\documentclass[a4paper,11pt]{article}
\pdfoutput=1
\usepackage{jcappub}

\usepackage{amsmath}
\usepackage{bm}
\usepackage{xcolor,color,colortbl}
\usepackage[utf8]{inputenc}
\usepackage{hyperref}
\usepackage{url}
\usepackage{graphicx}
\usepackage{multirow,bigdelim}
\usepackage{amssymb}
\usepackage[amssymb]{SIunits}
\usepackage{multirow}
\usepackage{bbold}
\usepackage{verbatim}

\definecolor{darkgreen}{HTML}{339933}

\title{Cosmology at high redshift - a probe of fundamental physics}

\author[a,b]{Noah Sailer,}
\author[d,e]{Emanuele Castorina,}
\author[c,b]{Simone Ferraro}
\author[a,b,c]{and Martin White}
\affiliation[a]{Department of Physics, University of California, Berkeley, CA 94720, USA}
\affiliation[b]{Berkeley Center for Cosmological Physics, UC Berkeley, CA 94720, USA}
\affiliation[c]{Lawrence Berkeley National Laboratory, One Cyclotron Road, Berkeley, CA 94720, USA}
\affiliation[d]{Dipartimento di Fisica `Aldo Pontremoli', Universita' degli Studi di Milano, Milan, Italy}
\affiliation[e]{Theoretical Physics Department, CERN, 1211 Geneva 23, Switzerland}
\emailAdd{nsailer@berkeley.edu}
\emailAdd{emanuele.castorina@unimi.it}
\emailAdd{sferraro@lbl.gov}
\emailAdd{mwhite@berkeley.edu}

\abstract{An observational program focused on the high redshift ($2<z<6$) Universe has the opportunity to dramatically improve over upcoming LSS and CMB surveys on measurements of both the standard cosmological model and its extensions. Using a Fisher matrix formalism that builds upon recent advances in Lagrangian perturbation theory, we forecast constraints for future spectroscopic and 21-cm surveys on the standard cosmological model, curvature, neutrino mass, relativistic species, primordial features, primordial non-Gaussianity, dynamical dark energy, and gravitational slip. We compare these constraints with those achievable by current or near-future surveys such as DESI and Euclid, all under the same forecasting formalism, and compare our formalism with traditional linear methods. Our Python code \href{https://github.com/NoahSailer/FishLSS}{FishLSS} $-$ used to calculate the Fisher information of the full shape power spectrum, CMB lensing, the cross-correlation of CMB lensing with galaxies, and combinations thereof $-$ is publicly available. 
}

\begin{document}
\maketitle
\flushbottom

\section{Introduction}

Large Scale Structure (LSS), as probed by fluctuations in the Cosmic Microwave Background (CMB) radiation or large galaxy surveys, provides one of our best windows on fundamental physics.
Topics include, but are not limited to, constraints on the expansion history and curvature \cite{Slosar19c,Takada:2015mma}, primordial non-Gaussianity \cite{Meerburg19,Alonso:2015uua}, features in the power spectrum (primordial \cite{Slosar19b} or induced \cite{Hill20,Ivanov20,DAmico20b,Klypin20,Chen20b}) or running of the spectral index \cite{ferraro2019inflation,Castorina20}, models of inflation \cite{PlanckInf18}, dark energy \cite{PDE15,Slosar19a}, dark matter interactions \cite{CMBS4}, light relics and neutrino mass \cite{Dvorkin19,Green19} and modified gravity \cite{Jain10,Joyce16,EUCLID18,Slosar19c}.
Unfortunately, while many of these extensions of the standard model are well motivated, we do not have strong guidance from theory on which of these many probes is most likely to turn up evidence of new, beyond standard model, physics.  However, well understood phenomenology allows us to forecast where our sensitivities to such physics will be highest and our inference cleanest.  To work where the inference is cleanest and the noise lowest we should push to high redshift.

Continuous advances in detector technology and experimental techniques are pushing us into a new regime, enabling mapping of large-scale structure in the redshift window $2<z<6$ using both relativistic and non-relativistic tracers.  This will allow us to probe the metric, particle content and \emph{both} epochs of accelerated expansion (Inflation and Dark Energy domination) with high precision.
Moving to higher redshift allows us to take advantage of four simultaneous trends.  First, a wider lever arm in redshift leads to rotating degeneracy directions among cosmological parameters and thus tighter constraints.  Second, the volume on the past lightcone increases dramatically, leading to much tighter constraints on sample-variance dominated modes.  Third, the degree of non-linearity is smaller, so the field is better correlated with the early Universe.  Fourth, high precision perturbative models built around principles familiar from high-energy particle physics become increasingly applicable.  Indeed at high redshift, we increasingly probe long wavelength modes which are linear or quasi-linear and carry a great deal of “unprocessed” information from the early Universe.

In this paper we forecast the constraints on cosmological parameters and extensions to the standard model that could be achieved by an observational program focused on the high redshift Universe.
We compare the reach of different experiments using a Fisher matrix formalism that builds upon recent advances in cosmological perturbation theory. Our Python code \verb|FishLSS|\footnote{\href{https://github.com/NoahSailer/FishLSS}{https://github.com/NoahSailer/FishLSS}} used to create these forecasts is publicly available, and is described in greater detail in Appendix \ref{app:code}. 

Since at high redshift the non-linear scale shifts to higher $k$, perturbative models provide a robust method for forecasting that allows for consideration of scale dependent bias, stochastic small-scale power and the decorrelation of the initial and observed modes (see also ref.~\cite{Castorina19}).
We begin in \S\ref{sec:surveys and probes}, where we list observational probes and discuss our assumptions about their number density and linear bias, and describe the experimental parameters of the (future) surveys for which we forecast.
Details of the forecasting formalism we employ are provided in \S\ref{sec:fisher} and our results are presented in \S\ref{sec:results}. We explore how these results depend on observational limitations in \S\ref{sec:issues}. We conclude in \S\ref{sec:conclusions}.

\section{Surveys and probes}
\label{sec:surveys and probes}

In this section we describe the various surveys and probes considered in our forecasts, as summarized in Table~\ref{tab:galaxy_surveys}. These include spectroscopic galaxy surveys, which observe some combination of emission line galaxies (ELGs), H$\alpha$ galaxies, or Lyman break galaxies (LBGs), along with interferometers dedicated to 21-cm intensity mapping, and CMB observatories. 

\begin{figure}[!h]
\centering
\includegraphics[width=0.8\linewidth]{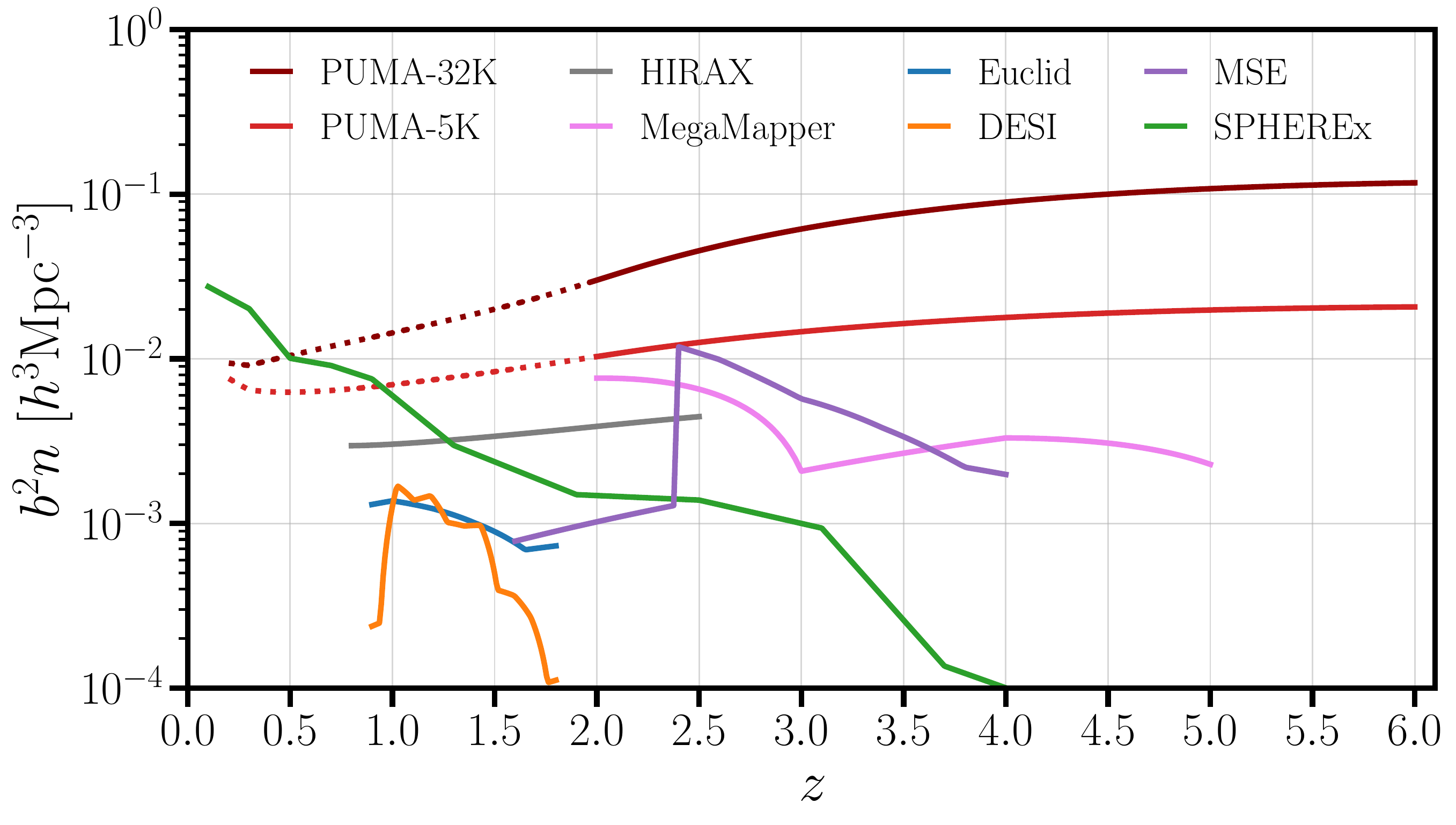}
\caption{
$b^2n$ for each of the LSS surveys considered in our forecasts.  This combination of linear bias and number density is proportional to the signal-to-noise ratio in linear theory, at fixed $\bm{k}$ (\S\ref{sec:fisher}).  
For both PUMA and HIRAX, the ``number density'' is taken to be the inverse of the thermal $+$ shot noise. We note that while the fiducial PUMA design would enable observations down to $z=0.2$ (dotted lines), we neglect this region in our forecasts, as it will be well measured by upcoming surveys such as DESI or Euclid which we already include.
}
\label{fig:b2n}
\end{figure}

\begin{figure}[!h]
    \centering
    \includegraphics[width=0.8\linewidth]{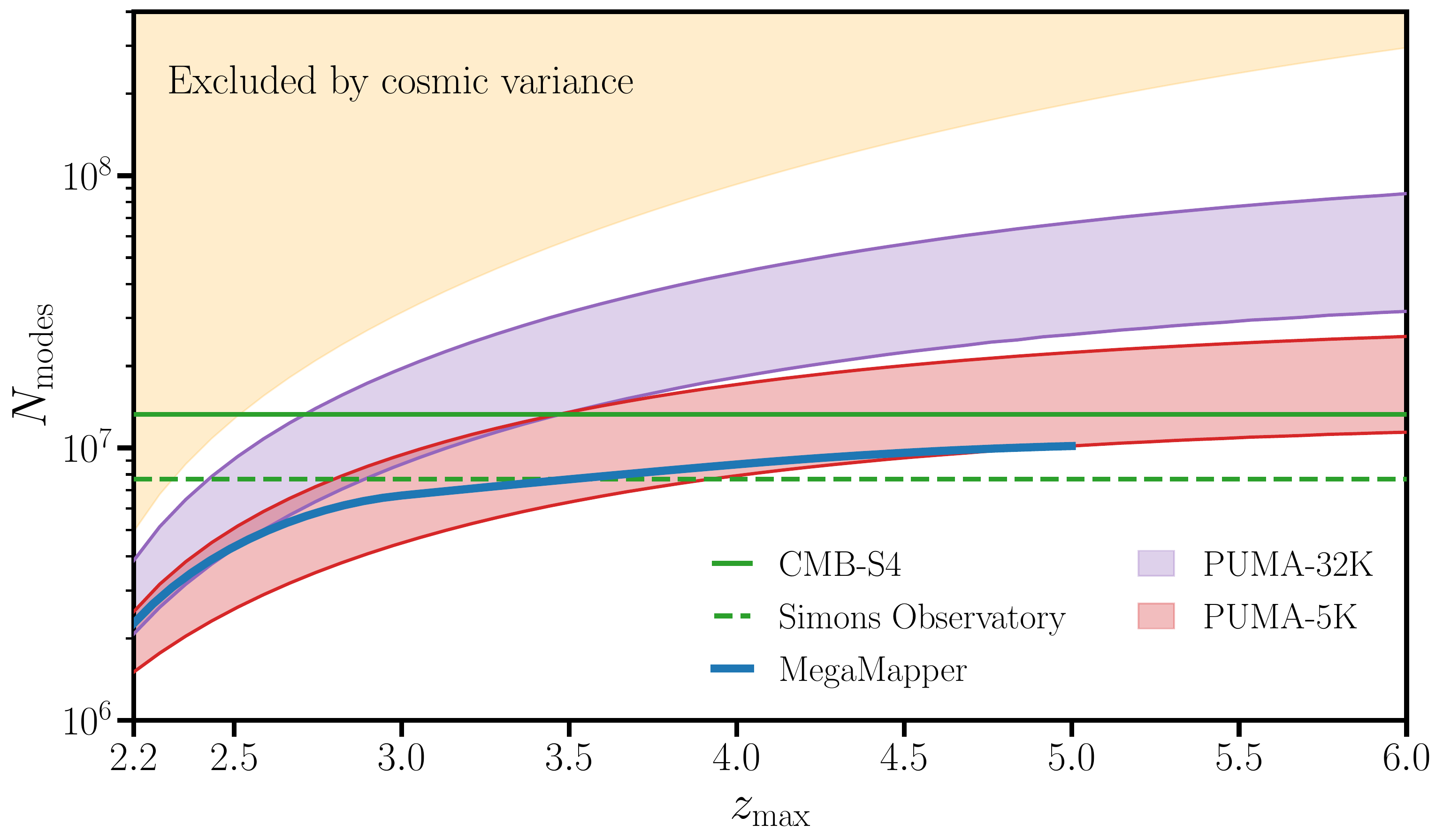}
    \caption{$N_\text{modes}$ as a function of $z_\text{max}$ for PUMA (-5K and -32K) and MegaMapper ($2<z<z_\text{max}$). For PUMA, we consider both optimistic and pessimistic foreground models (\S\ref{sec:full shape power spectrum}), which are the boundaries of the shaded regions. In green are the number of modes for Simons Obervatory and CMB-S4 like experiments with 7 $\mu$K-arcmin and 1.5 $\mu$K-arcmin noise levels respectively, both with a $1.4^\prime$ beam and $f_\text{sky}=0.4$.
    } 
\label{fig:figure_of_merit}
\end{figure}

\begin{figure}[!h]
    \centering
    \includegraphics[width=\linewidth]{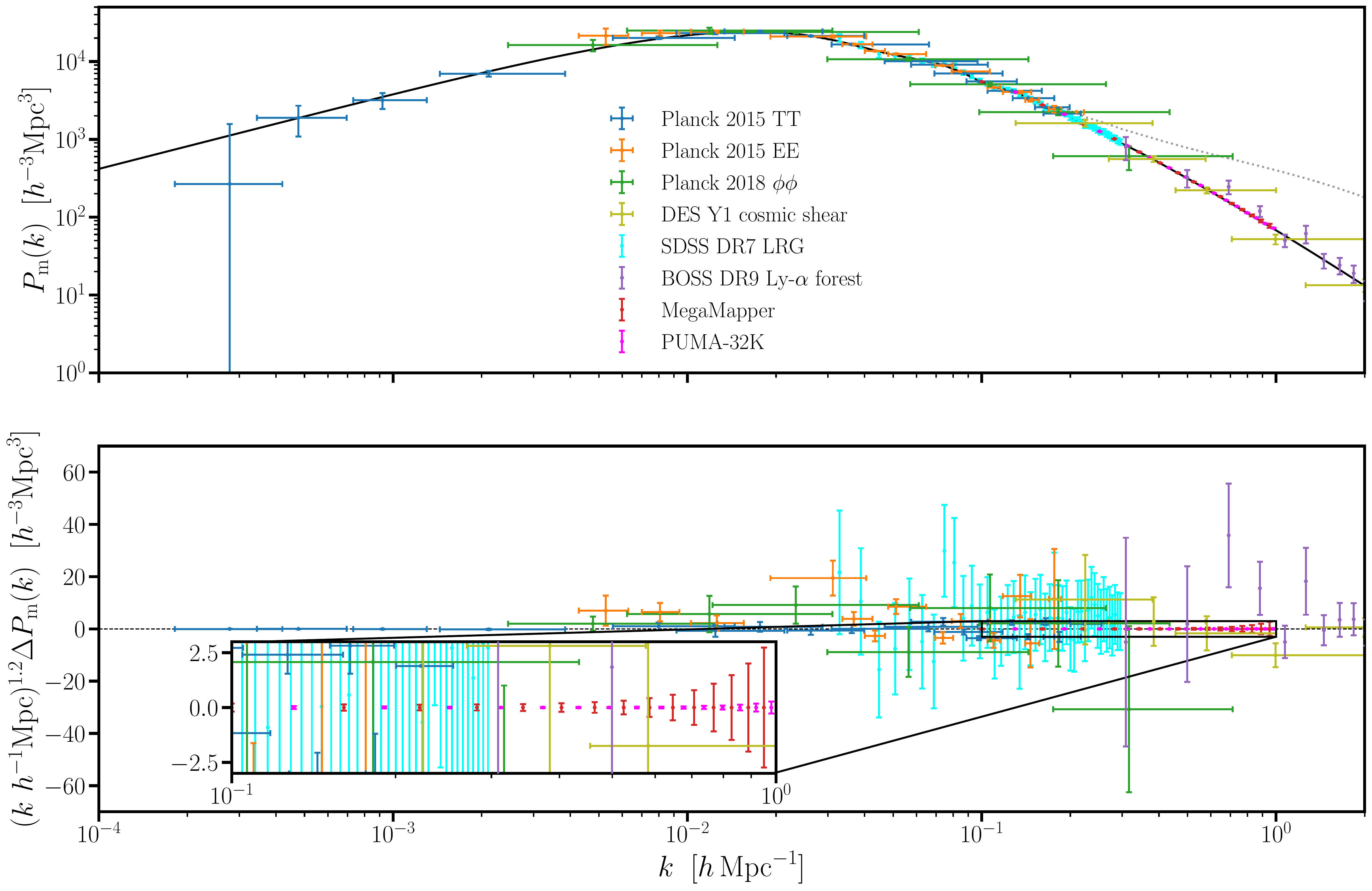}
    \caption{Measurements of the linear matter power spectrum at $z=0$. For both MegaMapper and PUMA-32K we show constraints for 15 linearly space $k$-bins between $0.1\,\,h\,\text{Mpc}^{-1}\lesssim k\lesssim 1\,\,h\,\text{Mpc}^{-1}$. This figure is adapted from refs. \cite{PlanckLegacy18,Chabanier_2019} using \href{https://github.com/marius311/mpk\_compilation/}{https://github.com/marius311/mpk\_compilation/}.
    } 
\label{fig:marius_plot}
\end{figure}

In the following subsections we discuss our assumptions about the linear bias and number density of each LSS probe. These values are summarized in Fig.~\ref{fig:b2n}, where we plot the effective number density $b^2 n$ of each survey. This figure already provides some baseline intuition for the results of the following forecasts. In particular, we see that MegaMapper has a comparable noise to the Euclid H$\alpha$ sample, but covers a much broader redshift range, surveying a volume $\sim 2.5$ times as large, as illustrated in Fig.~\ref{fig:nLBG}. Similarly, PUMA (-5K and -32K) has a lower noise than either MegaMapper, Euclid or DESI, and its survey volume is $\sim 3.2$ times that of Euclid\footnote{This rough factor ignores differences in sky coverage.}. In addition to lower shot noise and more volume, high redshift $(z>2)$ experiments have access to more ``linear'' modes, i.e.\ modes that can be modeled very accurately using perturbation theory. In Fig.~\ref{fig:figure_of_merit} we plot the ``effective number of linear (primordial) modes'' $(N_\text{modes})$ that MegaMapper, PUMA, and two upcoming CMB observatories will have access to. For a precise definition of $N_\text{modes}$, as well as a discussion of its interpretation and caveats on the comparison between CMB and LSS experiments using this metric, see Appendix \ref{sec:fom}. In particular, we see that MegaMapper will have access to roughly $25\%$ more modes than SO, and that PUMA-32K will have access to $\gtrsim 4$ times more modes than CMB-S4, assuming optimistic foregrounds. Our definition of $N_\text{modes}$ can be easily modified to forecast the error on a measurement of the linear matter power spectrum $P_m(k,z=0)$, as shown in Fig.~\ref{fig:marius_plot}. In particular we see that MegaMapper and PUMA will be capable of measuring the power spectrum out to $k= 0.9\,\,h\,\text{Mpc}^{-1}$ with an order of magnitude higher precision than Planck at $k\sim0.2\,\,h\,\text{Mpc}^{-1}$, despite having finer-grained $k$-bins. These estimates suggest that cosmological constraints from future, high-redshift LSS surveys will be competitive with CMB measurements. As we show in \S\ref{sec:results}, combining the two dramatically improves constraints on $\Lambda$CDM parameters, $N_{\rm eff}$, the neutrino masses and other extensions of the standard cosmological model.

We note that the forecasts presented in this work do not include constraints on cosmological parameters coming from cosmic shear or galaxy-galaxy lensing. Both quantities will be measured to exquisite precision by upcoming surveys such as Euclid, the Rubin Observatory and the Roman Space Telescope. However, most of their cosmological constraints from lensing will come from $z < 2$, due to the difficulty in defining high number density samples with well-characterized photometric redshifts and shape measurements at high redshift. Moreover, due to the relatively large redshift uncertainties compared to a spectroscopic survey, detailed features in the power spectrum (such as the BAO) are suppressed, together with the majority of the modes in the radial direction. Therefore, the analysis is often performed in a small number (order 10) of 2-dimensional tomographic bins. For this reason, photometric surveys are very effective at measuring the amplitude of structure and/or lensing at low redshift, while spectroscopic surveys can measure the 3-dimensional power spectrum and hence obtain growth and expansion measurements directly. Thus the high-redshift spectroscopic surveys studied here are highly complementary to other upcoming and planned photometric experiments.

\subsection{Near-term galaxy surveys}

The next five years will see a massive influx of LSS data, with spectroscopic galaxy surveys including DESI, Euclid, and SPHEREx coming online. Together these surveys will map out the low redshift universe, taking spectra of hundreds of millions of galaxies. 

The Dark Energy Spectroscopic Instrument (DESI) \cite{DESI} is actively taking data on the $4\,$m Mayall Telescope. Using its five thousand fibers and 10 spetrographs, DESI will take spectra of tens of millions of Emission Line Galaxies (ELGs), Luminous Red Galaxies (LRGs), and quasars over the next $\sim5$ years. Since ELGs comprise the largest sample of galaxies in the DESI catalogue, providing a significant fraction of the constraining power for most cosmological parameters, we only include the ELG sample in our forecasts. ELGs are active star-forming galaxies which exhibit strong nebular emission lines that originate from gas in the ionized regions around massive stars. This nebular emission includes the [O{\sc ii}] doublet, a distinct spectral signature that can be used to accurately determine a galaxy's redshift. DESI's spectrographs cover a spectral range of $360\,$nm to $980\,$nm, enabling [O{\sc ii}] measurements out to $z\sim1.7$. We convert the mean values of $dN_{\rm ELG}/dz$ (per square degree) in Table 2.3 of ref.~\cite{DESI} to comoving number densities using $n_{\rm ELG} = (180/\pi)^2 \chi^{-2} (d\chi/dz)^{-1} dN_{\rm ELG}/dz$. These values assume a [O{\sc ii}] flux cut of approximately $8\times 10^{-17}\,\,{\rm erg}\,\,{\rm cm}^{-2}\,{\rm s}^{-1}$, and account for both selection cuts and target success rates. We follow ref.~\cite{DESI} and approximate the ELG linear bias as $b_{\rm ELG} = 0.84/D(z)$, where $D(z)$ is the linear growth factor normalized to unity at $z=0$. 

The Euclid satellite \cite{EUCLID18} is a near-infrared space telescope that is currently under developement, with a nominal launch date in 2022. Over its six-year long mission Euclid will measure spectra of tens of millions of galaxies out to $z\sim2$, mapping out roughly 15,000 square degrees. Euclid consists of a $1.2\,$m mirror coupled to three imaging and spectroscopic instruments (with wavelength ranges 500-800 nm, 920-1250 nm, and 1250-1850 nm), enabling observations of H$\alpha$-emitters from $0.9<z<1.8$. 
For our forecasts we use the reference case number densities $n_{\text{H}\alpha}$ and linear biases $b_{\text{H}\alpha}$ listed in Table 3 of ref.~\cite{Euclid2020prep}(see also \cite{Ilic:2021dwx}).

The Spectro-Photometer for the History of the Universe, Epoch of Reionization and Ices Explorer (SPHEREx) is a space-based, low resolution spectral survey, with a current launch date in 2024 \cite{Dore14}. It will image $\sim 450$ million galaxies in both the optical and near-infrared, primarily targeting low redshift ($z<1$) objects and searching for evidence of primordial non-Gaussianity. We take the number densities and biases from ref. \cite{Dore14}, which were split into five samples according to their maximum redshift uncertainty. While the redshift uncertainty can be relatively large for a subset of the combined sample, the key science goal for SPHEREx (i.e.\ primordial non-Gaussianity) is relatively insensitive to photo-$z$ uncertainties. Moreover SPHEREx was designed to have a very low fraction of redshift outliers, and therefore is especially well-suited to this kind of measurement. Thus we simply add the number densities of the individual samples to get the number density of the combined sample. For the bias of the combined sample we take the appropriate weighted average: $\sum_i b_n n_i/\sum_i n_i$, where $b_i$ and $n_i$ are the bias and number density of the $i$'th sample. 

\subsection{Future galaxy surveys}

While the construction and deployment of various low reshift ($z<2$) facilities are currently underway, designs for new high-redshift surveys are actively being developed. In this work we consider two of these futuristic surveys in detail: MegaMapper and MSE. These surveys would exploit dropout selection to measure galaxies out to $z\sim5$, providing cosmological constraints that are competitive with upcoming CMB surveys (see Fig.~\ref{fig:figure_of_merit}). 

MegaMapper \cite{MegaMapper} is a proposed highly-multiplexed spectroscopic instrument that is designed to survey galaxies at high redshift ($2<z<5$), covering 14,000 square degrees. Its $6.5\,$m telescope aperture, $\sim20$K fibers, and five year observation time would yield galaxy number densities $> 10^{-4} h^3\,{\rm Mpc}^{-3}$ across its redshift range. 
MegaMapper's primary target would be Lyman Break Galaxies (LBGs), which are abundant, well studied and well understood, representing massive, actively star-forming galaxies with a luminosity approximately proportional to their stellar mass. A sizeable fraction have bright emission lines which could facilitate obtaining a redshift~\cite{Wilson19}.

\begin{figure}[!h] 
\centering
\includegraphics[width=\linewidth]{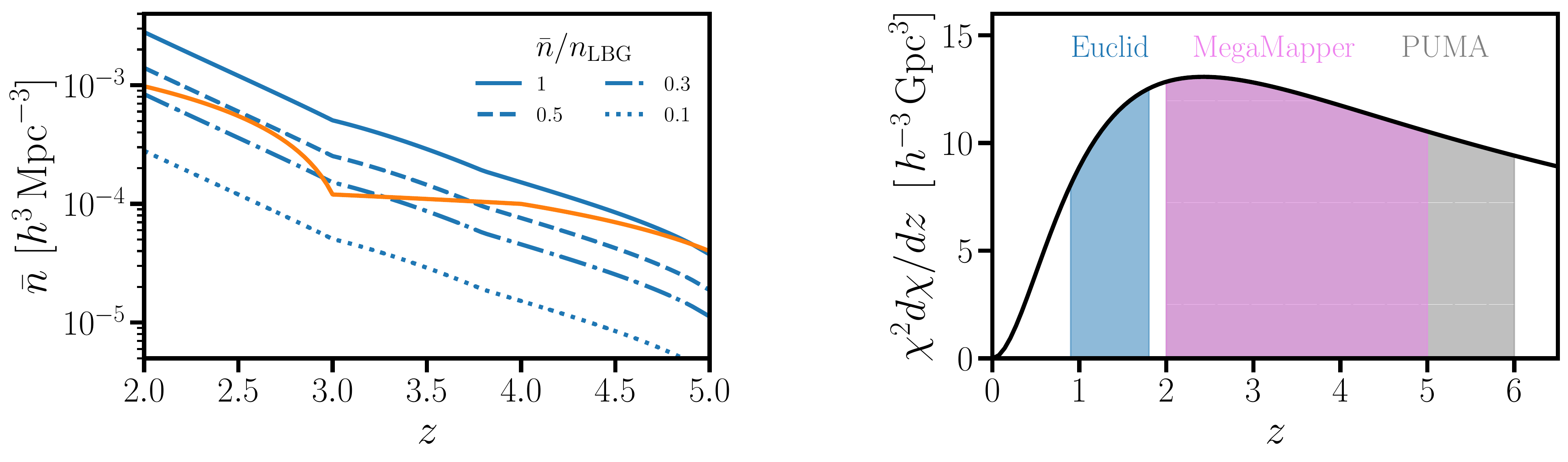}
\caption{\textit{Left:} In blue we plot several values of $\bar{n}/n_{\rm LBG}$, where $n_{\rm LBG}$ is the idealized LBG number density with $m<24.5$. For comparison, we plot the number density of the fiducial MegaMapper survey in orange.
\textit{Right:} Comoving volume per $dz$ per sky area ($\chi^2 d\chi /dz$) as a function of redshift. The shaded blue, purple, and (hatched) grey regions represent the volume covered by Euclid, MegaMapper and PUMA respectively, excluding sky coverage. 
}
\label{fig:nLBG}
\end{figure}

For the fiducial MegaMapper experiment we use the LBG number densities listed in Table 2 of ref.~\cite{ferraro2019inflation}, which are shown in Fig.~\ref{fig:nLBG}.
In \S\ref{sec:results} we also consider an idealized MegaMapper-like experiment that detects some redshift-independent fraction of LBGs $\bar{n}/n_{\rm LBG}$ with apparent magnitude $m<24.5$. This is slightly unrealistic, but provides an easy way to see in what manner our forecasts are sensitive to noise levels.  The ``idealized'' number densities are calculated using a Schechter function of absolute magnitude:
\begin{equation}
        \Phi(M_{\rm UV})
        =
        \left(\frac{\ln 10}{2.5}\right)
        \phi^\star
        10^{-0.4(1+\alpha) (M_{\rm UV}-M_{\rm UV}^\star)}
        \exp\left(
        -10^{-0.4 (M_{\rm UV}-M_{\rm UV}^\star)}
        \right).
\label{eq:schechter}
\end{equation}
We use the best fit values to the UV luminosity function for $\alpha,M_{\rm UV}^\star,$ and $\phi^*$, which are listed in Table 3 of ref.~\cite{Wilson19}. The number density is obtained via $n_\text{LBG} = \int^{M_c}_{-\infty} \Phi(M) dM$, where the cutoff $M_c$ corresponds to galaxies with apparent magnitude $m=24.5$:  
\begin{equation}
        M_c
        \simeq
        24.5 
        -
        5 \log_{10}\left(\frac{D_L(z)}{10\,{\rm pc}}\right)
        +2.5 \log_{10}(1+z).
\label{eq:mc}
\end{equation}
For both the fiducial and idealized MegaMapper surveys, we adopt the bias model of ref.~\cite{Wilson19}, in which the bias of galaxies with apparent magnitude $m$ is approximated to linear order in $m$ and to second order in the scale factor:
\begin{equation}
        b(m,z) = A(m) (1+z) + B(m) (1+z)^2
\end{equation}
with $A(m) = -0.98 (m-25) + 0.11$ and $B(m) = 0.12(m-25) + 0.17$. We assume the LBG bias is dominated by the faintest (and most numerous) galaxies, approximating $b_\text{LBG}(z) = b(m=24.5,z)$. These approximations to the number density and bias are consistent with the idealised version of MegaMapper considered in ref.~\cite{ferraro2019inflation}. As shown in Fig.~\ref{fig:nLBG}, the fiducial MegaMapper number density is within the range $0.3 \lesssim \bar{n}/n_{\rm LBG} < 1$, depending upon the redshift. In \S\ref{sec:results} we show that taking $\bar{n}/n_{\rm LBG}=0.5$ yields constraints similar to those from the fiducial MegaMapper survey.

The Maunakea Spectroscopic Explorer (MSE) \cite{MSE}, planned for first light in 2029, will cover over 10,000 square degrees. MSE will couple a $11.25\,$m mirror with a 1.5 square degree field of view to $\sim 4000$ fibers, feeding to spectrographs that cover 360 to $1300\,$nm. This design enables the detection of ELGs out to $z\sim 2.4$, and LBGs between $2.4<z<4$. For the ELG sample we use the number densities listed in Table 1 of ref.~\cite{percival2019cosmology}, and take $b_\text{ELG}(z) = D(0)/D(z)$, where $D(z)$ is the linear growth factor. To be consistent with our assumptions regarding LBGs, we use the same linear bias $b(m=24.5,z)$ for both MegaMapper and the MSE LBG sample. For the LBG sample we estimate the number density via $\int_{-\infty}^{M_c} \Phi(M)E(M)dM$, where $\Phi$ and $M_c$ are defined in Eqs.~\eqref{eq:schechter} and \eqref{eq:mc} respectively, and $E(M)$ is MSE's LBG target efficiency. We take the efficiency to be an average over the templates shown in Figure 2 of ref.~\cite{percival2019cosmology}, assuming that $40\%$ of the LBGs have a negative equivalent width (EW), $30\%$ have $0<\text{EW}<20$ and $30\%$ have $\text{EW}>20$.

In addition to the aforementioned surveys, several high redshift spectroscopic surveys have been proposed or are set to launch within the coming decade, including GAUSS \cite{Blanchard:2021ffq}, SpecTel \cite{SpecTel,dawson2018cosmic} and the Roman telescope \cite{Roman}. We provide a brief description of these surveys below, however, we do not attempt forecasts for these experiments as their survey designs are still in flux. 

The Gravitation And the Universe from large
Scale-Structures (GAUSS) \cite{Blanchard:2021ffq} telescope is a proposed large-scale structure focused space-based mission. GAUSS would consist of a $4\,$m aperture telescope coupled to both imaging and spectroscopic instruments, covering both the optical and infrared ($\sim$ 500 to $5000\,$nm). The survey would measure spectra of tens of billions of galaxies out to $z\sim5$, achieving number densities larger than Euclid's across its entire redshift range.

SpecTel \cite{SpecTel,dawson2018cosmic} is a proposed spectroscopic survey in the southern hemisphere, making it ideally situated for cross-correlations with SO or CMB-S4. SpecTel would couple an $11.4\,$m dish (with a 5 square degree FoV) to 60,000 fibers $-$ enabling 120 million fiber exposures (each $\geq 4,000$ seconds long) over its survey period $-$ which feed to spectrographs covering 360 to $1330\,$nm. Its design would permit observations of LBGs and Lyman Alpha Emitters (LAEs) out to redshift $\sim 5$, with number densities $2$ to $5$ times those achievable by a MegaMapper-like survey, depending upon the redshift bin. In \S\ref{sec:results} we explore how MegaMapper's parameter constraints improve with increased LBG number density, which can be interpreted as a rough estimate for SpecTel constraints.

The Roman telescope \cite{Roman} is a space-based NASA mission set to launch in 2025, carrying a spectrometer that covers the visible to near-infrared spectrum. Under the straw-man proposal for a $2,000\,\mathrm{deg}^2$ survey, constraints from Roman would not be competitive with those from DESI or the other proposed surveys we investigate in detail.  Study of trade-offs which would improve Roman's performance in these forecasts is beyond the scope of this paper and would require a better understanding of achievable depth and area for the galaxy redshift survey given its current constraints.

\subsection{21-cm intensity mapping}

Proof-of-concept interferometers dedicated to 21-cm mapping are currently being deployed \cite{CHIME,Tianlai,CHORD}, while designs for future larger-scale facilities are actively under development \cite{Newburgh_2016,Slosar19a,SKACosmo}.
We consider two 21-cm surveys in detail: the Hydrogen Intensity and Real-time Analysis eXperiment (HIRAX) \cite{Newburgh_2016} and the Packed Ultra-Wideband Mapping array (PUMA) \cite{Slosar19a}. 
HIRAX is a  $400–800\,$MHz radio interferometer that is currently under development for deployment in South Africa, while PUMA is a proposed ultra-wideband, low-resolution and transit interferometric radio telescope operating at $200–1100\,$MHz. PUMA's fiducial telescope design is a hexagonal array of 32K $6\,$m dishes, designed to survey half of the sky ($f_{\rm sky}=0.5$) over five years.
We also consider a smaller, proof-of-concept design with only 5K dishes (PUMA-5K). In our noise model (see below), we assume a hex-packed array with a $50\%$ fill factor. This fill factor is accounted for by doubling the number of detectors and setting the observation time to $t_{\rm int}=5/4$ years. All other couplings, efficiencies, temperatures etc are set to the values quoted in ref.~\cite{collaboration2018inflation}. We use the same noise model for HIRAX, but with a 1K, fully filled array, and $t_{\rm int} = 4$ years. 

Below $z\simeq 6$ the majority of the neutral Hydrogen in the Universe resides in dense, self-shielded regions within galaxies \cite{VN18}.  While the details of how H{\sc i} traces the dark matter are not well understood, current observations and numerical simulations suggest that we can 
approximate the H{\sc i} mass within a given halo with $M_{\rm HI}(M_h;z) = A(z) M_h e^{-M_\text{min}/M_h}$, where $M_\text{min} = 5\times 10^9 h^{-1} M_\odot$~\cite{Chen19a}. Using this relation, along with a halo number density and bias function, we can then approximate the linear bias of H{\sc i} by weighting the halo bias with the halo energy density:
\begin{equation}
       b_{\rm HI}(z)
       =
       \frac{
       \int_0^\infty b(M_h;z) n(M_h;z) M_{\rm HI}(M_h;z) dM_h
       }
       {
       \int_0^\infty n(M_h;z) M_{\rm HI}(M_h;z) dM_h
       }.
\end{equation}
We adopt the same conventions as ref.~\cite{Chen19a} and use the Sheth-Tormen \cite{Sheth_2001} halo bias and mass function. 

There are two sources of noise in H{\sc i} observations: a shot noise due to the finite sampling of the continuous H{\sc i} field, and the thermal noise from the instrument.
For all H{\sc i} surveys we adopt the thermal noise model described in the appendix of ref.~\cite{collaboration2018inflation} (see also 
\cite{Bull2014,Seo2010}), along with the shot noise from ref.~\cite{Castorina_2017,Obuljen2017}.  Further discussion of observational and data-analysis issues related to interferometric, 21-cm intensity mapping can be found in ref.~\cite{Liu20}.

\begin{table}[!h]
\centering
\resizebox{\textwidth}{!}{\begin{tabular}{cc} 
% Table 1
\begin{tabular}{cccc} 
\hline
Experiment & $z_{\rm min},\,z_{\rm max}$ & Sample & $f_{\rm sky}$\\
\hline
\hline
DESI \cite{DESI} &  0.6, 1.7 & ELG & 0.34 \\
Euclid \cite{EUCLID18,Bagley20,Euclid2020prep} & 0.9, 1.8 & H$\alpha$ & 0.36\\
SPHEREx \cite{Dore14,Dore16} & 0.1,4.3 & low-$z$ galaxies& 0.65 \\
MegaMapper \cite{MegaMapper} & 2, 5 & LBG & 0.34\\
MSE \cite{MSE} &  1.6, 4 & ELG $+$ LBG & 0.24 \\
\hline
\end{tabular} &  
% Table 2
\begin{tabular}{cccc} 
\hline
Experiment & $z_{\rm min},\,z_{\rm max}$ & Sample & $f_{\rm sky}$ \\
\hline
\hline
PUMA \cite{Slosar19a}& 2, 6 & 21-cm & 0.50 \\
HIRAX \cite{Newburgh_2016}& 0.8, 2.5 & 21-cm & 0.36\\
\hline\\
\end{tabular} 
\end{tabular}}
\caption{\textit{Left:} A list of the spectroscopic galaxy surveys considered in our forecasts, along with their redshift ranges, samples and sky coverage. \textit{Right:} A list of 21-cm experiments, their redshift range and sky coverage.
}
\label{tab:galaxy_surveys}
\end{table}

\subsection{CMB experiments}
\label{sec:CMB_experiments}

In addition to LSS probes, we consider external priors from CMB surveys, with configurations similar to the Planck \cite{PlanckLegacy18}, Simons Observatory (SO; \cite{SimonsObs}) CMB-S4 \cite{CMBS4} and LiteBIRD \cite{LiteBIRD} surveys.
To obtain the Planck Fisher matrix, we use the bands, beams and sensitivities from Table 4 of ref.~\cite{PlanckLegacy18}. We include both temperature and polarization anisotropies, and we adjust the minimum $\ell_{\rm min, P}$ in polarization to match the published value of uncertainty on the optical depth $\sigma(\tau) = 0.007$. We have checked that this method gives good agreement with the published uncertainties on the standard cosmological parameters, while noting that the main purpose of this paper is not forecasting parameters for the next generation of CMB experiments, but rather to highlight the improvements from high-redshift LSS surveys. Because of this, we consider a simplified setup in which detector noise is treated as white noise uncorrelated between different frequency channels, and we do not include any effects from the atmosphere on large scales (beyond the scale cut $\ell_{\rm min} = 30$), and do not include any instrumental or foreground nuisance parameters.
For SO and CMB-S4 we take $\ell_{\rm max, T} = 3000$ in temperature and $\ell_{\rm max, P} = 5000$ in polarization, and $\ell_{\rm min, T, P} = 30$ for both. Since SO and CMB-S4 are ground based and not expected to measure the very largest scales due to contamination from the atmosphere, we always add a low-$\ell$ Planck (or LiteBIRD) Fisher matrix to SO or CMB-S4. This is accomplished by splitting the SO/S4/Planck forecasts into high- and low-$\ell$ Fisher matrices (with cutoff $\ell = 30$), and adding the low-$\ell$ Planck Fisher matrix to the high-$\ell$ SO/S4 Fisher matrix. This simplistic addition is justified since the two Fisher matrices use a disjoint set of multipoles, and are therefore largely uncorrelated (we expect any mode coupling from lensing or non-Gaussian foregrounds to be small). A fiducial model for the foregrounds is added to the total power spectrum in temperature, following the model of ref.~\cite{Dunkley:2013vu} (with the modifications listed in ref.~\cite{SimonsObs}). We neglect any foregrounds in polarization, noting that point source masking should be very effective at mitigating their effect on small scales.
Unless stated otherwise, we use the 2-point function of the unlensed primary CMB temperature and polarization fluctuations for the derivatives with respect to the cosmological parameters, and do not include CMB lensing reconstruction.
This allows us to simply add the Fisher matrices for CMB and LSS probes. Another feature of using the unlensed CMB power spectrum is that it should provide accurate forecasts from high-$\ell$ temperature. If we included lensing the damping tail would be proportional to the lensing amplitude, and therefore largely degenerate with the foreground amplitude and shape, as well as baryon effects in the lensing power spectrum which is not marginalized over here.  This can lead to artificially tight constraints \cite{McCarthy:2021lfp}. 
We also consider a futuristic LiteBIRD-like space based CMB experiment, capable of reaching the cosmic variance limit $\sigma(\tau) \approx 0.002$
from $\ell < 30$. For this we take a $1.4 \mu$K-arcmin experiment in polarization with a 33 arcmin Gaussian beam covering 65\% of the sky.
Although not considered in this work, we note that an instrument like PICO \cite{Pico} would also be highly complementary to LSS probes.

\section{Forecasting formalism}
\label{sec:fisher}

In this section we outline our forecasting procedure, based on the Fisher formalism. We examine three primary observables of interest: primary anisotropies in the CMB (as described in \S\ref{sec:CMB_experiments}), the redshift space galaxy power spectrum (\S\ref{sec:full shape power spectrum}), and CMB lensing (\S\ref{sec:lensing}). We take $\Lambda$CDM+$M_\nu$ as our fiducial cosmology, with $\Lambda$CDM values given in Table~\ref{tab:marg_params} and $N_\text{eff}=3.046$, consistent with refs.~\cite{PlanckLegacy18,PCP18}. We assume a normal neutrino hierarchy with minimal mass: $m_{\nu_1}=0$, $m_{\nu_2} = 0.01$ eV and $m_{\nu_3}=0.05$ eV. Unless stated otherwise the neutrino mass $M_\nu=\sum_i m_{\nu_i}$ and $N_\text{eff}$ are held fixed in all forecasts. A list of all of the free parameters in our base model is given in Table~\ref{tab:marg_params}.
\subsection{Full shape measurements}
\label{sec:full shape power spectrum}

Both historically and currently, most forecasts use the linear power spectrum and scale-independent bias, or the non-linear matter power spectrum computed using fitting formulae, to model the power spectrum of a biased tracer \cite{Euclid2020prep,DESI2013,DESI,Schaan2020,Yu2018,Schmittfull17,Obuljen2017,Brinckmann2018,Vagnozzi2018,LSST,Slosar19a}.  In this work, we use more sophisticated models based on perturbation theory and a general bias expansion (see refs.~\cite{Castorina19,Chudaykin19,Boyle20,chen2021precise} for related work on forecasts and ref.~\cite{Mead20} for inclusion of scale-dependent biases in the halo model). 
We consider a general quadratic bias model and non-linear evolution up to one-loop order in perturbation theory.  On large scales and at high redshifts, perturbation theory provides an accurate model that allows us to take into account scale-dependent bias, broadening of the acoustic peaks and mode-coupling due to non-linear evolution.  Including these effects should improve the reliability of the forecasts. We further discuss the differences in linear vs.\ non-linear forecasting in \S\ref{sec:lin vs non lin}.

To include the non-linear effects and more complex bias model we use the \verb|velocileptors| \verb|LPT_RSD| module \cite{Chen19a,Chen20b,Chen20c}, which takes the linear power spectrum and biases as inputs, and self-consistently calculates the non-linear redshift-space power spectrum $P^\text{1-loop}_{gg}(k,\mu)$ to one-loop order in perturbation theory. We use the linear CDM+baryon power spectrum  $P^\text{lin}_{cb}(k,z)$ as our input to \verb|velocileptors|, calculated using \verb|CLASS| \cite{CLASS}, along with a second order biasing scheme: 
\begin{equation}
    \delta_g(\bm{x},z) 
    = 
    b\, \delta_{cb}(\bm{x},z)
    +
    \frac{1}{2}b_2\, \delta^2_{cb}(\bm{x},z)
    +
    b_s\, s^2(\bm{x},z)
\end{equation}
where $\delta_g$ is the non-linear tracer density contrast, $\delta_{cb}$ is the non-linear CDM+baryon density contrast, and $s_{ij} = (\partial_i\partial_j/\partial^2 - \delta_{ij}/3) \delta_{cb}$ is the shear field.  To better connect our formalism to previous literature here, and elsewhere in the text, these biases are in the Eulerian frame\footnote{We assume cubic bias operators are fixed and generated by time evolution in Lagrangian perturbation theory.}.  However, the \verb|LPT_RSD| module computes $P^\text{1-loop}_{gg}(\bm{k})$ using Lagrangian perturbation theory, which has been shown to perform well in these situations \cite{Chen20c} and allows for a simple treatment of models with complex features in the power spectrum \cite{Chen20b}.  To compute the Lagrangian biases, which are inputs to \verb|velocileptors|, we perform the following rotation \cite{Chan2012,Baldauf2012}:
\begin{equation}
    b^{\rm L} = b-1
    \quad
    \quad
    b_2^{\rm L} = b_2 - \frac{8}{21}(b-1)
    \quad
    \quad
    b_s^{\rm L} = b_s + \frac{2}{7}(b-1).
\end{equation}
We use the linear biases given in \S\ref{sec:surveys and probes}. The fiducial values of $b_2$ and $b_s$ are set by assuming $b^\text{L}_2=b^\text{L}_s=0$.
This choice is only a rough approximation whose validity is pending more data. In all forecasts we marginalize over these non-linear biases, making our results mostly insensitive to their fiducial values (see Appendix \ref{app:checks} for a more detailed discussion).

To model the effects of small-scale physics, we include a handful of counterterms $(\alpha_n)$ and stochastic contributions ($N_n$) in our fiducial power spectra:
\begin{equation}
\label{eqn:stochastic}
    P_{gg}(\bm{k})
    =
    P_{gg}^\text{1-loop}(\bm{k})+
    \big(\alpha_0 +\alpha_2 \mu^2 + \alpha_4 \mu^4\big) \left(k/k_*\right)^2 P^\text{Zel}_{cb}(\bm{k})
    + N_0 + N_2(\mu k)^2 + N_4(\mu k)^4,
\end{equation}
where $N_2$ and $N_4$ encode small-scale velocities (i.e.\ FoG effects and redshift errors, if appropriate), $N_0$ is the shot noise, $k_* = 1\,h\,{\rm Mpc}^{-1}$, $P^\text{Zel}_{cb}$ is the CDM+baryon power spectrum in the Zel'dovich approximation, and $\mu=\hat{\bm{k}}\cdot\hat{\bm{n}}$ is the cosine of the angle between the wavevector and the line of sight. Throughout we approximate  $P_{gg}(k,\mu) \simeq \sum_{\ell=0}^4 \mathcal{L}_\ell(\mu)  P_{gg,\ell}(k)$ by its first three non-vanishing multiples, where $\mathcal{L}_\ell(\mu)$ is the Legendre polynomial of degree $\ell$. For the surveys, redshifts and $k$-ranges considered in this work we find that this approximation is good to $2\%$. 

We determine the fiducial value of $\alpha_0$ by fitting the \verb|velocileptors| output to \verb|CLASS|' default version of \verb|HaloFit| \cite{Takahashi12}.  The fiducial values for the higher order stochastic terms ($\alpha_2$, $\alpha_4$) are set to zero purely for convenience.  We take the Poisson value $\bar{n}_{\rm gal}^{-1}$ for the fiducial value of $N_0$ and set $N_2 = -N_0 \sigma_v^2$, where $\sigma_v$ is a typical velocity dispersion. For galaxy surveys we set the velocity dispersion to $100\,$km/s, converted to comoving length units: $\sigma_v = (1+z)(100\,\text{km/s})/H(z)$. Finger-of-god effects in 21-cm maps are much smaller, driven by a small number of satellites (occupying $\sim$10\% of the sample) with typical rms velocities $\approx 40\,$km/s \cite{Villaescusa-Navarro2018}. Thus for 21-cm surveys, we choose a fiducial velocity dispersion of $10\,$km/s ($\sim \sqrt{0.1}\,40\,$km/s). 
The fiducial value of $N_4$ is set to zero purely for convenience. In all forecasts we marginalize over these counterterms and stochastic contributions, making our results largely insensitive to their fiducial values, as discussed in Appendix \ref{app:checks}.

The modeling of the power spectrum for 21-cm surveys is slightly more complex than for galaxy surveys. Radio interferometers measure the 21-cm signal in intensity units, so that the power spectrum has units of $(\text{temperature})^2\times \text{volume}$:
\begin{equation}
    P^\text{obs}_\text{HI}(\bm{k})
    =
    T_b^2
    \left[
    P_\text{HI}(\bm{k})
    +
    P_\text{SN}
    \right]
    +
    P_\text{N},
\end{equation}
where $P_\text{HI}(\bm{k})$ is the non-linear HI power spectrum (taking the same form as Eq.~\ref{eqn:stochastic}, with $P_{\rm SN}$ replacing $N_0$), $P_\text{SN}$ is the shot-noise coming from halos that contribute to the HI signal, $P_N$ is the instrumental noise \cite{collaboration2018inflation}, and $T_b$ is the mean 21-cm brightness temperature \cite{Field59}:
\begin{align}
\label{eqn:Tb}
    T_b(z) = 188(1+z)^2 E(z) \Omega_{\rm HI}(z) \,\text{mK}\,,
\end{align}
where $E(z)\equiv H(z)/H_0$ and $\Omega_{\rm HI}(z)\equiv \rho_{\rm HI}(z)/\rho_c(z)$ is the cosmic density of neutral hydrogen, taken from the fitting formula of ref.~\cite{Crighton2015}. In all forecasts that include 21-cm surveys we marginalize over the 21-cm brightness temperature by default as its true value is still largely uncertain.

Because our fiducial model includes massive neutrinos, we need to specify their perturbative treatment in our setup.
It is by now well established that halos and galaxies can be well approximated as biased tracers of the dark matter and baryon fluids only \cite{Villaescusa-Navarro2013,Castorina2013,Castorina2015,Villaescusa-Navarro2017,LoVerde2014,Munoz2018,Fidler2018}.  For this reason our real-space power spectrum calculation goes through as for the no-neutrino case if we make the trivial replacement of the total matter field with $P_{cb}$, the dark matter plus baryon spectrum, as described above. We also assume the Green's function associated with higher order perturbative solutions is very well approximated by the Einstein-de-Sitter one, which implies the structure of the loops remains unchanged. This is an excellent approximation for the small fiducial value of neutrino masses considered in this work \cite{Aviles:2020cax}. In principle one needs to consider the scale-dependence of the growth rate, which we shall neglect for the same reason mentioned above. 

For our full shape forecasts we work at the power spectrum level. As in Eq.~\eqref{eqn:stochastic}, let $P_{gg}(k,\mu,z)$ denote the observed nonlinear redshift-space power spectrum of the matter tracer. Let $i$ index a $(\mu_i,k_i)$ pair, e.g.\ $P_{gg,i}(z) = P_{gg}(k_i,\mu_i,z)$. We divide each survey into linearly spaced redshift bins $\{B_m;\,m=1,2,\cdots\}$, and calculate the corresponding Fisher matrix in each bin as\footnote{The shot noise $N_0(z) = 1/n_\text{gal}(z)$ is included in our definition of the observed power spectrum, so it does not appear explicitly in our covariance.}:
\begin{equation}
\begin{aligned}
    F_{ab}(B_m)
    =
    \sum_{ij}
    \frac{\partial P_{gg,i}(z)}{\partial \theta_a}
    C^{-1}_{ij}(z)
    \frac{\partial P_{gg,j}(z)}{\partial \theta_b}
    \bigg|_{z=z_m}
    \\
    \text{ where }
    \quad
    C_{ij}(z)
    = 
    \delta^K_{ij}
    \frac{4\pi^2}{k^3_i V_m\,d\mu\,d\ln k}
    \left[
    P_{gg,i}(z)
    \right]^2.
\label{eqn:partial_fisher}
\end{aligned}
\end{equation}
Here $V_m$ is the comoving volume of bin $B_m$, which implicitly includes a factor of $f_\text{sky}$, $z_m = (z_\text{max}+z_\text{min})/2$ is the central redshift in bin $B_m$, and $\theta_a$ represents either a cosmological or a nuisance parameter. Throughout we assume that $P_{gg}(k,\mu,z)$ is even in $\mu$, and use Simpson's rule to evaluate Eq.~\eqref{eqn:partial_fisher} with 2,000 logarithmically spaced $k$-bins (ranging from $5\times 10^{-4}$ to $1\,\,h\,\text{Mpc}^{-1}$) and 100 linearly spaced $\mu$-bins (ranging from $0$ to $1$). We limit the domain of integration by multiplying the summand by an appropriate window function. To calculate the Fisher matrix of the full survey we make the standard approximation that correlations between non-overlapping redshift bins are negligible, so that the final Fisher matrix is obtained by summing over redshift bins:
\begin{equation}
    F_{ab} =  \sum_m F_{ab}(B_m).
\end{equation}
In Eq.~\eqref{eqn:partial_fisher} we neglect any temporal or spatial variation in the number density within each bin. This effect can be included analytically \cite{Wadekar20}, and can lead to significant error in the forecast for sufficiently large redshift bins, or for sufficiently low number densities. However, as we show in Appendix \ref{sec:cov_approx}, for the surveys and redshift bins considered in this work these effects are negligibly small. We thus ignore this correction, and take Eq.~\eqref{eqn:partial_fisher} at face value.

\begin{figure}[!h]
\centering
\includegraphics[width=0.47\linewidth]{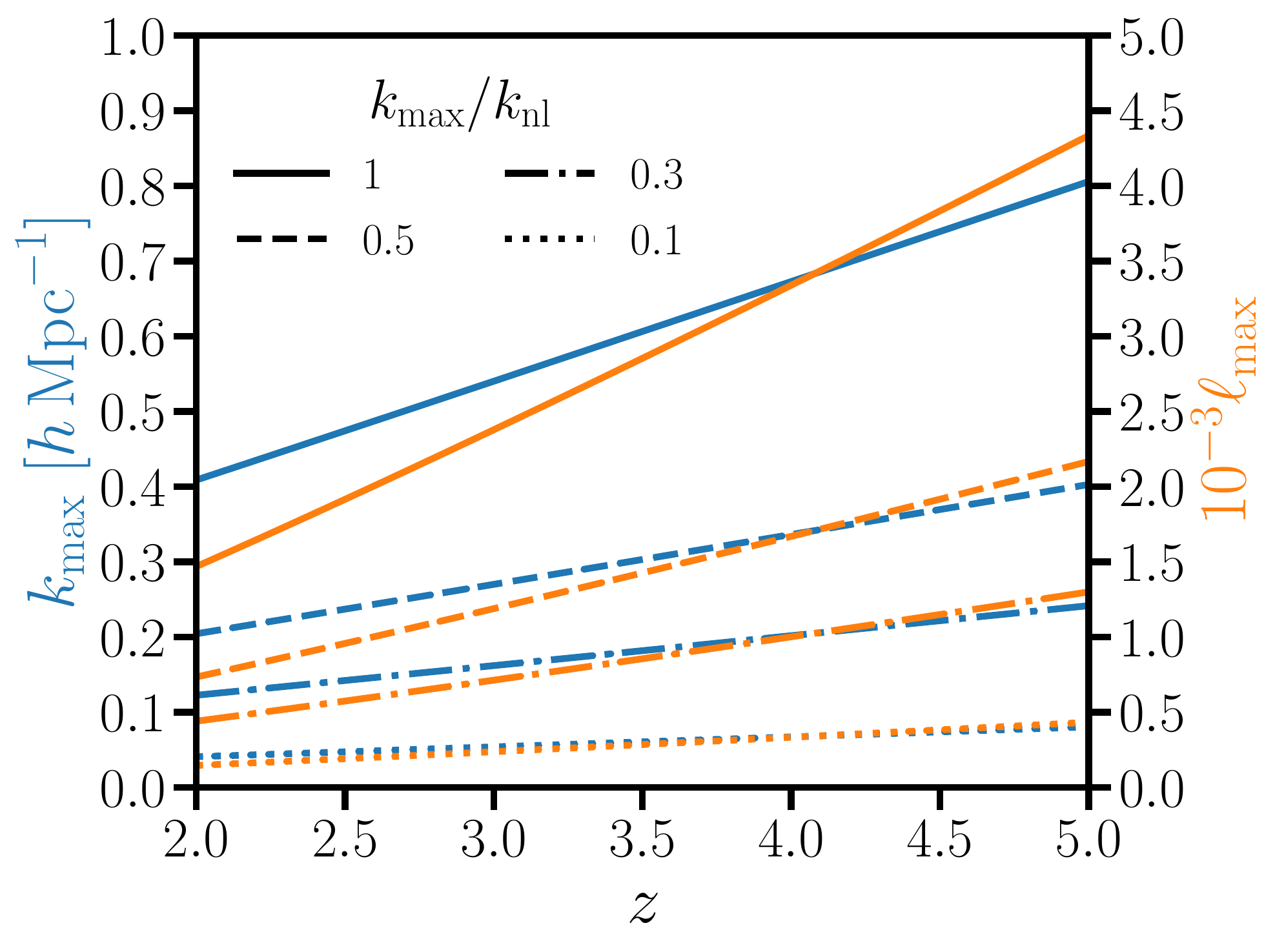}
\hspace{10mm}
\includegraphics[width=0.39\linewidth]{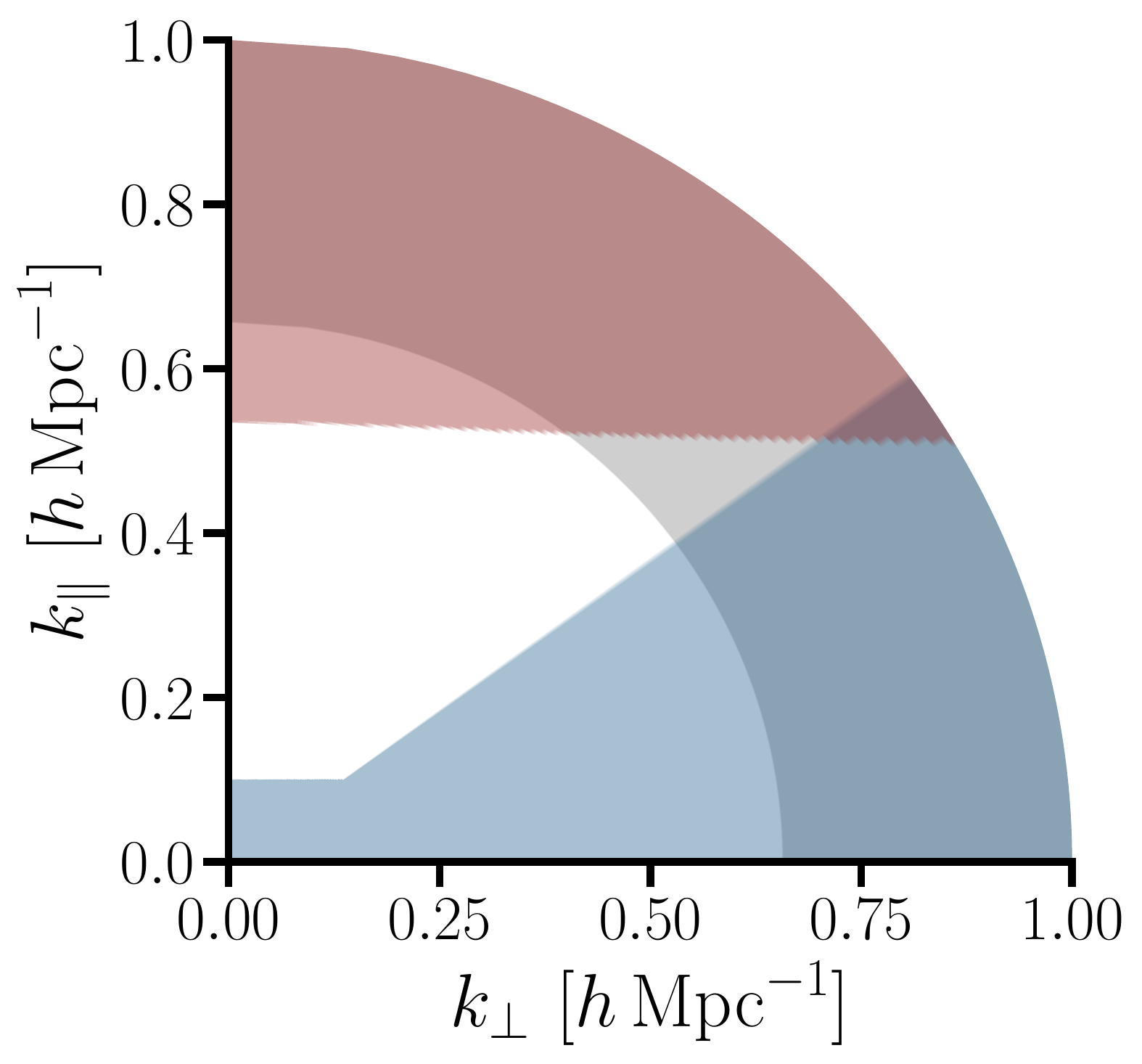}
\caption{\textit{Left}: $k_\text{max}(z)$ and $\ell_\text{max}(z) = k_\text{max}(z)\chi(z)$, assuming that $k_\text{max}$ is some constant fraction of $k_\text{nl}(z)=\Sigma^{-1}(z)$. \textit{Right}: Domain of integration in the $k_\parallel-k_\perp$ plane at $z\simeq 3.9$. The grey region is excluded by the $k<k_\text{nl}$ constraint. For 21-cm surveys the blue region is excluded by the (pessimistic) foreground wedge. The red region is excluded by FoG-like effects, chosen so that $N_0 \sigma_v^2(k\mu)^2$ is less than 20\% of the fiducial power. In this figure we plot the red region for a MegaMapper-like survey (it would shift to higher $k_\parallel$ for a 21-cm survey).
}
\label{fig:integration_region}
\end{figure}

In all forecasts we take $k_\text{min} = \text{max}\big[ 0.003\,h\,\text{Mpc}^{-1}, 2\pi/V^{1/3}_m \big]$ for each redshift bin. Since the clustering signal is low for $k<k_{\rm min}$, our results would not change significantly if a slightly different $k_{\rm min}$ were assumed, with the exception of local primordial non-Gaussianity, as discussed in Sec.~\ref{sec:PNG}.
Unless explicitly stated otherwise we take $k_\text{max}(z) = k_{\rm nl}(z)$, where $k_{\rm nl}(z) = \Sigma^{-1}(z)$ is the non-linear scale set by the RMS displacement in the Zel'dovich approximation:
\begin{equation}
    \Sigma^2(z)
    =
    \int
    \frac{dk}{6\pi^2}
    P^\text{lin}(k,z),
\label{eqn:Sigma2}
\end{equation}
where $P^\text{lin}(k,z)$ is the linear matter power spectrum. For 21-cm surveys, we also include a foreground wedge \cite{collaboration2018inflation} that further constrains the limits of integration:
\begin{equation}
    k_\parallel 
    >
    \text{max}\left[
    k_\perp
    \frac{\chi(z)H(z)}{c(1+z)} \sin(\theta_w(z))
    \,\,,\,\,k_\parallel^\text{min}\right]
    \quad
    \text{where}
    \quad 
    \theta_w(z) =
    N_w
    \frac{1.22}{2 \sqrt{0.7} } 
    \frac{\lambda_{\rm obs}}{D_\text{phys}}
\end{equation}
where $\lambda_{\rm obs}=(1+z)\,21\,$cm is the observed H{\sc i} wavelength, $D_\text{phys}$ is the physical diameter of the dish, $k_\parallel = k\mu$ and $k_\perp = k\sqrt{1-\mu^2}$. We take $D_\text{phys}=6$ m for both HIRAX and PUMA. For each 21-cm survey we consider ``optimistic'' and ``pessimistic'' foreground cases, in which $N_w=1$ , $k_\parallel^\text{min}=0.01\,h\,{\rm Mpc}^{-1}$ and $N_w=3$ , $k_\parallel^\text{min}=0.1\,h\,{\rm Mpc}^{-1}$ respectively. This is consistent with the definitions in ref.~\cite{collaboration2018inflation}. Finally, we note that non-perturbative effects like Fingers-of-God (FoG) limit the information that we can extract from high values of $\mu$. The importance of FoG is not known a priori, but roughly speaking they can be accounted for by a stochastic term $N_0 \sigma_v^2(k\mu)^2$ (see Eq.~\ref{eqn:stochastic}) on sufficiently large scales. We therefore make the conservative choice of integrating only over modes where this term is less than 20\% of the fiducial power $P_{gg}(k,\mu)$. We have checked that this choice in cutoff has an insignificant impact on our forecasts, as discussed in Appendix \ref{app:checks} and shown in Fig.~\ref{fig:N2_cutoff}. In Fig.~\ref{fig:integration_region} we show both the foreground wedge (for the pessimistic case) and the region excluded by FoG-like effects for our fiducial MegaMapper survey.

\subsection{CMB lensing and cross-correlations}
\label{sec:lensing}

Weak lensing of the CMB is a tracer of the
matter distribution integrated along the line of sight \cite{LEWIS_2006}, and is one of the central science cases for upcoming CMB surveys \cite{SimonsObs,LiteBIRD,CMBS4,Bermejo-Climent:2021jxf}. Throughout we work under both the Born and Limber approximations so that $C_\ell^{XY}$ may be expressed as a single integral:
\begin{equation}
    C_\ell^{XY}
    =
    \int_{0}^{\chi_*} d\chi
    \frac{W^X(\chi)W^Y(\chi)}{\chi^2}
    P_{\delta_X\delta_Y}\left(k_\perp = \frac{\ell+1/2}{\chi},k_\parallel=0\right),
\label{eqn:Limber}
\end{equation}
where $X$ and $Y$ denote either the CMB lensing convergence $\kappa$ or the galaxy sample $g$, and $\delta_\kappa = \delta_m$. In Eq.~\eqref{eqn:Limber} we have included the lowest order correction to the Limber approximation, replacing $\ell\to\ell+1/2$, which increases the accuracy to $\mathcal{O}(\ell^{-2})$ \cite{LoVerde_2008}. In principle both approximations could be relaxed, but in practice we do not expect them to have a large impact on the parameters constraints \cite{LovAfs08,Krause2010,Pratten2016,Marozzi2016,Fabbian2017,Fabbian2019,Boehm2019}.
The projection kernels take the form:
\begin{equation}
    W^\kappa(\chi)
    =
    \frac{3}{2}\Omega_m H_0^2 (1+z)
    \frac{\chi(\chi_*-\chi)}{\chi_*}
    \quad
    \text{and}
    \quad
    W^g(\chi) \propto H(z) \frac{dN}{dz},
\end{equation}
where $W^g$ is normalized so that $\int d\chi W^g(\chi)=1$, and $\chi_*$ is the comoving distance to the surface of last scattering. For simplicity we neglect any contribution from magnification bias, as this will be measured by future LSS surveys to high enough accuracy to be treated as an essentially fixed parameter \cite{Chen19a}. 

The nonlinear real space power spectra ($P_{gg}$ and $P_{gm}$) are calculated to one-loop in LPT using \verb|velocileptors|' \verb|cleft_fftw.py|, which includes a second order biasing scheme ($b, b_2, b_s$), two counterterms ($\alpha_0$ and $\alpha_x$) for the auto- and cross-correlations, and a shot noise contribution to $P_{gg}$. Note that lensing cross-correlations introduces only one new nuisance term ($\alpha_x$), and that $P_{gm}$ and $P_{gg}$ manifestly have different dependencies on the bias parameters. Thus lensing cross-correlations play a role in breaking degeneracies, improving the constraints on parameters which are highly degenerate with bias (e.g.\ $M_\nu$), as discussed in Appendix~\ref{app:lensing_confusion} (see Fig.~\ref{fig:kk_vs_kg_gg}). The fiducial values for the biases, counterterms and shot noise are identical to those for the full shape power spectrum, and are listed in Table~\ref{tab:marg_params}. 

For both the cross-correlation $C^{\kappa g}_\ell$ and the galaxy auto-correlation $C^{gg}_\ell$ we approximate Eq.~\eqref{eqn:Limber} with 100 integration points that are linearly spaced in the redshift interval corresponding to the galaxy sample $g$. In principle the power spectra should be evaluated at each integration point, but this is computationally expensive to do in practice. To speed up our calculations we split the redshift interval into $m$ linearly-spaced bins, each of which contain $100/m$ integration points. Within each bin the power spectrum is evaluated at a single effective redshift \cite{Modi_2017}:
\begin{equation}
    z^{XY}_\text{eff} 
    =
    \frac{\int d\chi W^X(\chi)W^Y(\chi)z(\chi)/\chi^2}
    {\int d\chi W^X(\chi)W^Y(\chi)/\chi^2},
\end{equation}
which is chosen to cancel the linear order correction in the expansion of $P_{\delta_X\delta_Y}(k,z)$ about $z^{XY}_\text{eff}$. We choose $m$ to be the smallest integer that yields bins with widths $\Delta z \leq 0.2$, resulting in $\mathcal{O}(0.1\%)$ accuracy for $30<\ell<500$ and $1<z<6$. Achieving a similar accuracy for the convergence power spectrum via direct integration is computationally inefficient. We instead calculate $C^{\kappa \kappa}_\ell$ with \verb|CLASS|, using its default version of $\verb|HaloFit|$ to model nonlinearities.

\begin{figure}[!h]
\centering
\includegraphics[width=\linewidth]{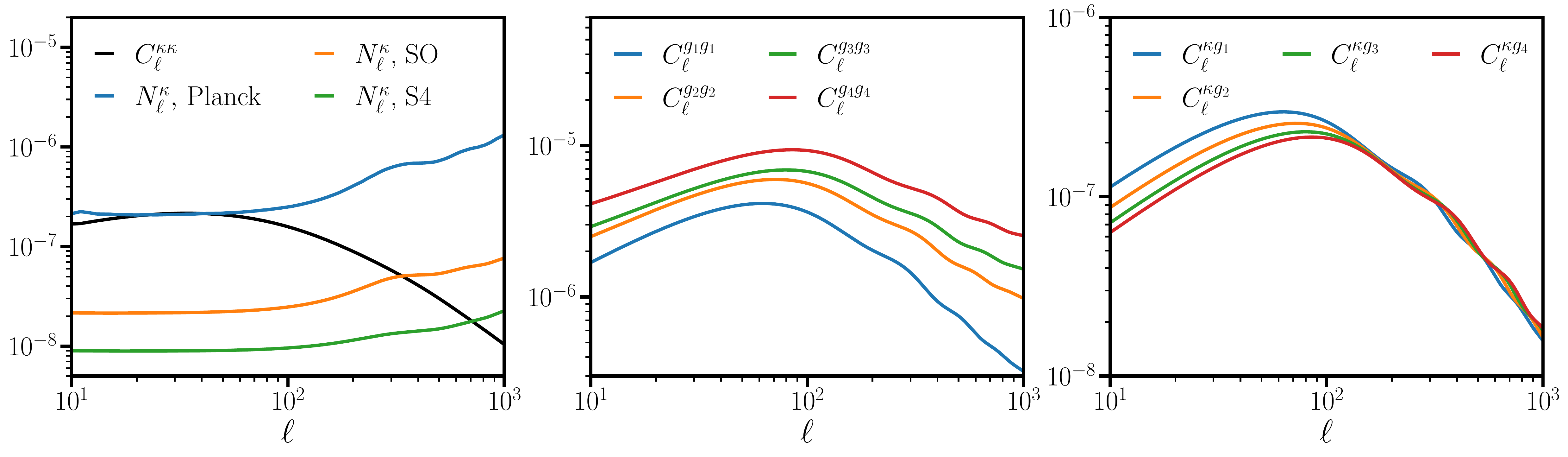}
\caption{
\textit{Left:} Lensing noise $N^\kappa_\ell$ for Planck, Simons Observatory and CMB-S4. The convergence power spectrum is shown in black. \textit{Middle:} The galaxy auto-correlation $C^{g _ig_i}_\ell$ for MegaMapper, where $g_i$ is the LBG sample in the $i^{\rm th}$ redshift bin: $z\in [2.0,2.8], [2.8,3.5], [3.5,4.3], \text{ or }[4.3,5.0]$. \textit{Right:} The convergence-galaxy cross-spectra in each redshift bin.
}
\label{fig:lensing_noise}
\end{figure}

We consider both SO and S4-like experiments when providing lensing constraints. For SO we use the noise $N^\kappa_\ell$ from the \href{https://github.com/simonsobs/so_noise_models/blob/master/LAT_lensing_noise/lensing_v3_1_0/nlkk_v3_1_0deproj0_SENS2_fsky0p4_it_lT30-3000_lP30-5000.dat}{SO noise calculator} while for S4, we use the lensing noise curve described on the \href{https://cmb-s4.uchicago.edu/wiki/index.php/Survey_Performance_Expectations}{CMB-S4 Wiki}. These curves are plotted in Fig.~\ref{fig:lensing_noise}. 
Both of these noises include atmospheric contributions, and are calculated with an iterated minimum variance (MV) quadratic estimator and MV Internal Linear Combination (ILC) of both CMB temperature ($30< \ell_\text{T} < 3000$) and polarization ($30< \ell_\text{P} < 5000$) data. 

In our forecasts we work at the power spectrum level and split our galaxy sample $g$ into $n$ non-overlapping redshift bins, denoted by $g_1, \cdots, g_n$, so that our averaged data vector is $\bm{\mu}^T_\ell = (C^{\kappa\kappa}_\ell,C^{\kappa g_1}_\ell,\cdots,C^{\kappa g_n}_\ell,C^{g_1 g_1}_\ell,\cdots,C^{g_n g_n}_\ell)$. The corresponding Fisher matrix is
\begin{equation}
F_{ij} \simeq \frac{f_\text{sky}}{2}\sum_{\ell=\ell_\text{min}}^{\ell_\text{max}} (2\ell+1)
\frac{\partial \bm{\mu}^T_\ell}{\partial \theta_i} \bm{C}^{-1}_\ell \frac{\partial \bm{\mu}_\ell}{\partial \theta_j},
\end{equation}
where the elements of the covariance $\bm{C}_\ell$ are given by\footnote{As in \S\ref{sec:full shape power spectrum}, we include the shot noise in our fiducial $C^{g_i g_i}_\ell$, so it does not explicitly appear in our covariance.}
\begin{equation}
(\bm{C}_\ell)_{XY} =
\begin{cases}
(C^{\kappa\kappa}_\ell+N^\kappa_\ell)^2 f_\text{sky}/f^\text{CMB}_\text{sky} & X=Y=\kappa\kappa\\
\frac{1}{2}\left[(C^{\kappa\kappa}_\ell+N^\kappa_\ell)C^{g_ig_i}_\ell\delta^K_{ij} + C^{\kappa g_i}_\ell C^{\kappa g_j}_\ell\right]& X=\kappa g_i,\,Y=\kappa g_j\\
(C^{g_i g_i}_\ell)^2 \delta^K_{ij}f_\text{sky}/f^\text{LSS}_\text{sky} & X=g_i g_i,\,Y=g_j g_j\\
(C^{\kappa\kappa}_\ell+N^\kappa_\ell)C^{\kappa g_i}_\ell & X=\kappa\kappa,\,Y=\kappa g_i\\
(C^{\kappa g_i}_\ell)^2& X=\kappa \kappa,\,Y= g_i g_i\\
C^{g_i g_i}_\ell C^{\kappa g_i}_\ell \delta^K_{ij}& X=g_i g_i,\,Y=\kappa g_j\\
\end{cases}
\label{eq:angular_covariance}
\end{equation}

In all forecasts we take $\ell_\text{min}=30$. To remove information from modes that are too small to be accurately modeled using PT, we multiply the appropriate elements of the covariance matrix ($C^{g_i g_i}_\ell$ and $C^{\kappa g_i}_\ell$) by $10^{20}$ whenever $\ell > k_\text{nl}(z_i)\chi(z_i)$, where $z_i$ is central redshift in the $i$'th redshift bin. Throughout we assume full spatial overlap of CMB and LSS surveys for cross-correlations: $f_\text{sky}=\text{min}(f^\text{LSS}_\text{sky}, f^\text{CMB}_\text{sky})$, with $f^\text{CMB}_\text{sky}=0.4$ for a SO-like experiment.  

Due to projection effects, the lensing auto-power spectrum receives substantial contributions from small scales, beyond the reach of perturbation theory. For such modes, baryonic effects cannot be neglected either, further complicating the model predictions. 
To be conservative we impose a cut ($\ell<500$) on the lensing angular modes included in our forecasts, but note that this is active area of research and improvements are likely to possible by the time the data arrive \cite{Foreman2015,Braganca2020,McCarth2020,Chung2019}. For $\ell<500$, baryonic feedback contributes at the subpercent level to the convergence power spectrum \cite{Chung_2020}, while at $\ell=500$, only $\sim 15\%$ of the power comes from $k>0.2 \,h\,{\rm Mpc}^{-1}$.

\begin{table}
\centering
\begin{tabular}{ccc}
\hline
Parameter & Definition & Fiducial value \\
\hline
\hline
$h$ & Hubble parameter: $H_0/$100 km$/$s$/$Mpc & 0.677 \\
$\ln(A_s)$ & $\ln($primordial amplitude$)$ & -19.97 \\
$n_s$ & spectral index & 0.96824 \\
$\omega_c$ & fractional dark matter density: $\Omega_c h^2$ & 0.11923 \\
$\omega_b$ & fractional baryon density: $\Omega_b h^2$ & 0.02247 \\
$\tau$ & optical depth to reionization & 0.0568 \\
\hline
$b$ & linear Eulerian bias: $\delta_g \supset b \delta_{cb}$ & survey dependent\\
$b_2$ & quadratic Eulerian bias: $\delta_g\supset \frac{1}{2}b_2\delta^2_{cb}$ & $(8/21)\times(b-1)$\\
$b_s$ & shear bias: $\delta_g \supset b_s s^2$ & $-(2/7)\times(b-1)$\\
$\alpha_0$ & $P_{gg} \supset \alpha_{0} (k/k_*)^2 P^\text{Zel}_{cb}$ & $1.22 +0.24\,b^2\,(z-5.96)$ \\
$\alpha_{2n}$ & $P_{gg} \supset \alpha_{2n} \mu^{2n}(k/k_*)^2 P^\text{Zel}_{cb},\,n=1,2$ & 0 \\
$N_0$ & Poisson white noise: $P_{gg}\supset N_0$ & survey dependent\\
$N_2$ & FoG-like contribution: $P_{gg}\supset N_2 (k\mu)^2$ & survey dependent\\
$N_4$ & FoG-like contribution: $P_{gg}\supset N_4 (k\mu)^4$ & 0\\
$T_b$ & 21-cm brightness temperature & Eq.~(\ref{eqn:Tb})\\
\hline
$\alpha_x$ & $P_{gm}\supset \alpha_x (k/k_*)^2 P^\text{Zel}_{cb}$ & 0\\
\hline
\end{tabular}
\caption{
A list of the free parameters in our base model, which are marginalized over in all forecasts unless explicitly stated otherwise. In the first set of rows we list the $\Lambda$CDM parameters and their fiducial values. In the second set of rows we list the nuisance terms that we use to model redshift-space clustering. The last row shows the counterterm introduced when including lensing cross-correlations. 
Here $\delta_g$ is the non-linear density contrast of the tracer with power spectrum $P_{gg}$, $\delta_{cb}$ is the non-linear density contrast of CDM+baryons, $P_{cb}^\text{Zel}$ is the power spectrum of the CDM+baryons within the Zeldovich approximation, $s_{ij}\equiv (\partial_i\partial_j/\partial^2-\delta_{ij}/3)\delta_{cb}$ is the shear field, $s^2\equiv s_{ij}s_{ij}$, and $k_*=1\,h\,{\rm Mpc}^{-1}$. 
}
\label{tab:marg_params}
\end{table}

\subsection{Combined forecasts}
\label{sec:combined fisher}

In our forecasts we treat the CMB surveys described in \S\ref{sec:CMB_experiments} as priors on $\Lambda$CDM, $M_\nu$, $N_\text{eff}$ and $\Omega_k$. Since our CMB priors are calculated using \textit{unlensed} power spectra, we neglect any covariances between these priors and lensing or full shape data. Thus to combine a CMB prior with a LSS Fisher matrix $\bm{F}_\text{LSS}$ with basis $\{\Lambda\text{CDM},M_\nu,N_\text{eff},\Omega_k,\cdots\}$, we simply add the CMB Fisher matrix to the appropriate submatrix of $\bm{F}_\text{LSS}$.

When combining lensing\footnote{From here on out we abbreviate $(C^{\kappa\kappa}_\ell, C^{\kappa g_i}_\ell, C^{g_i g_i}_\ell)$ as ``lensing data'', ``$+$ lensing'' or simply ``lensing'', unless stated otherwise.} $(C^{\kappa\kappa}_\ell, C^{\kappa g_i}_\ell, C^{g_i g_i}_\ell)$ and full shape clustering data ($P_{gg}(\bm{k})$) from the same survey, we impose a $k_\parallel > 10^{-3}\,h\,\text{Mpc}^{-1}$ cut on the full shape Fisher matrix. Since lensing probes modes with $k_\parallel\sim 0$, this cut nulls any covariance between the two observables in linear theory. We therefore neglect any covariance between the lensing and full shape data, and simply add the respective Fisher matrices when showing combined constraints. 

We also neglect any (negligibly small) covariances between two non-overlapping LSS surveys when combining their full shape data. The same cannot be done when combining lensing data from two surveys, however, since the covariance between $C^{\kappa g_i}_\ell$ and $C^{\kappa g_j}_\ell$ is nonzero even when the tracers $g_i$ and $g_j$ are at different redshifts. To avoid this complication we never combine lensing data from two different surveys $-$ when combining full shape information from two spectroscopic surveys with lensing data, we only include the lensing data from the higher redshift survey. Since the high-$z$ survey (e.g.\  MegaMapper) typically dominates the constraining power, neglecting the lensing-galaxy cross-correlations of the low-$z$ survey (e.g.\ DESI) has a negligible impact on our results. Since the foreground wedge (\S\ref{sec:full shape power spectrum}) restricts access to modes with $k_\parallel\sim0$, we never include 21-cm $\times$ CMB lensing in our results. Thus when adding lensing to (low-$z$ spectroscopic survey) $+$ (21-cm) clustering data, we only include the lensing data from the low-$z$ survey. 

The full set of free parameters in our base model is given in Table~\ref{tab:marg_params}. In all forecasts that include full shape information we marginalize over $\{\Lambda\text{CDM}, b, b_2, b_s, \alpha_0, \alpha_2, \alpha_4, N_0, N_2, N_4\}$, while in all forecasts that include lensing we marginalize over $\{\Lambda\text{CDM}, b, b_2, b_s, \alpha_0, N_0, \alpha_x\}$. We marginalize over the 21-cm brightness temperature $T_b$ in any forecast that includes 21-cm full shape data. We treat all time-dependent nuisance terms (biases, counterterms, stochastic terms, $T_b$) in different redshift bins as separate, independent variables which are individually marginalized over. We do the same for nuisance terms of separate surveys when showing combined constraints.

\begin{figure}[!h]
    \centering
    \includegraphics[width=\linewidth]{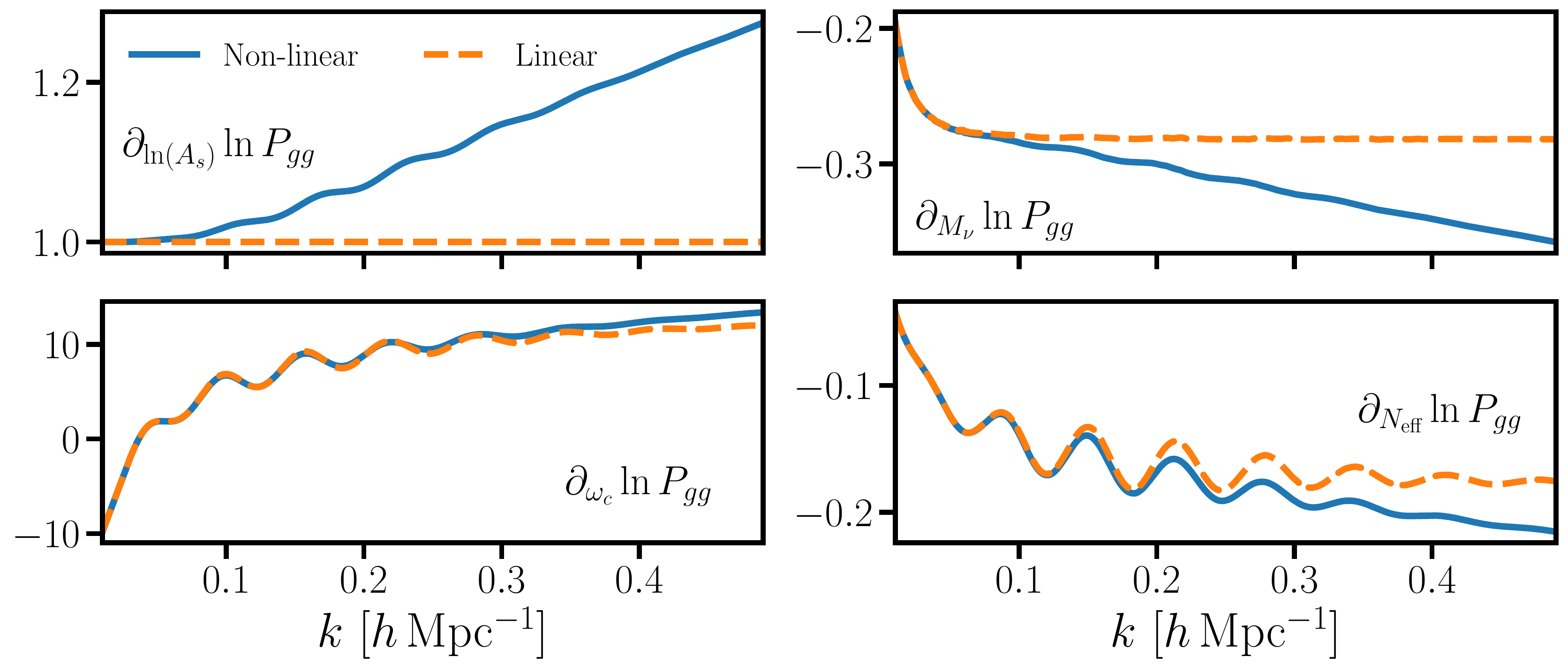}
    \caption{Some exemplary logarithmic derivatives at $z=2.5$, $\mu=0$. The solid curves are calculated using Lagrangian perturbation theory, the dashed curves are calculated using linear theory. These derivatives do not account for the A-P (\S\ref{sec:APeffect}) effect.
    }
\label{fig:lin vs nonlin derivatives}
\end{figure}

\subsection{Derivatives}
With the exception of derivatives with respect to the three stochastic terms ($N_0,N_2,N_4$) and the 21-cm brightness temperature, all of our derivatives are calculated numerically. For parameters whose fiducial value, $\theta_{\rm fid}$, is within the physically allowed region we use a five-point, two-sided derivative with relative step size $\delta \theta/\theta_\text{fid}$, which takes the form: 
\begin{equation}
\frac{\partial P}{\partial \theta} \bigg|_{\theta_\text{fid}}
=
\frac{
-P_{+2} + 8 P_{+1} - 8 P_{-1} + P_{-2}
}{
12\,\delta\theta
}
+\mathcal{O}\left(\delta \theta^4\right).
\end{equation}
where $P_{\pm n} \equiv P(\theta_\text{fid}\pm n \delta\theta)$. When the parameter has its fiducial value at a boundary of the allowed region we instead use a five-point, one-sided derivative: 
\begin{equation}
\begin{aligned}
\frac{\partial P}{\partial \theta} \bigg|_{\theta_\text{fid}}
= 
\frac{
-3P_{+4}+16P_{+3}-36P_{+2}+48P_{+1}-25P_{+0}
}{
12\,\delta\theta
}
+\mathcal{O}\left(\delta \theta^4\right).
\end{aligned}
\end{equation}
We take $\delta \theta/\theta_\text{fid}=0.01$ for most parameters.  When the fiducial value of the parameter is zero, we assume $\delta \theta=0.01$ by default. $\tau$, $M_\nu$ and $A_\text{lin}$ (see \S\ref{sec:primordial_features}) require larger (or smaller) step sizes for numerical convergence: $\delta \tau = 0.3 \tau$, $\delta M_\nu = 0.05 M_\nu$ and $\delta A_\text{lin} = 0.002$ respectively. For these choices the convergence of the derivatives is better than 0.5\% for all parameters.  We show some illustrative examples in Fig.~\ref{fig:lin vs nonlin derivatives}.  

When taking derivatives with respect to $M_\nu$, we fix $m_{\nu_1}=0$ to avoid numerical error, and take steps of the form $m_{\nu_2},m_{\nu_3}\to m_{\nu_2}+\delta M_\nu/2, m_{\nu_3}+\delta M_\nu/2$ to preserve the atmospheric mass splitting.

\subsection{Alcock-Paczynski effect}
\label{sec:APeffect}

Power spectra, being dimensionfull quantities expressed in terms of physical distances, are usually computed from observations by assuming a 
``fiducial model'' to convert redshifts and angular positions to 3D comoving coordinates.  This fixed fiducial cosmology must be accounted for when comparing the observations to a theoretical model that predicts a different conversion to comoving coordinates.  The distortion this introduces is known as the Alcock-Paczynski (A-P) effect \cite{AP}. The wavevector measured using the fiducial cosmology $\bm{k}_\text{fid}$ is related to the true wavevector $\bm{k}$ by:
\begin{equation}
\bm{k}_\text{fid} = \bm{k}_\parallel \alpha_\parallel(z)+\bm{k}_\perp \alpha_\perp(z).
\label{eqn:rescaling_AP}
\end{equation}
where $\alpha_\parallel(z) = H(z)_\text{fid}/H(z)_\text{true}$ and $\alpha_\perp(z) = D_A(z)_\text{true}/D_A(z)_\text{fid}$. 
The correlation function, being a dimensionless quantity, is left unchanged by this rescaling. Thus $P^\text{obs}_{gg}(\bm{k}_\text{fid},z)d^3\bm{k}_\text{fid} = P^\text{true}_{gg}(\bm{k},z)d^3 \bm{k}$, which gives \cite{Seo_2003}:
\begin{equation}
P^\text{obs}_{gg}(\bm{k}_\text{fid},z) = 
\alpha_\parallel^{-1}(z)\alpha_\perp^{-2}(z)
P^\text{true}_{gg}(\bm{k},z).
\label{eqn:AP_power_spectrum_level}
\end{equation}
This effect is included in our derivatives via the chain rule: 
\begin{equation}
\begin{aligned}
    \frac{\partial P}{\partial \theta}
    \to
    \frac{\partial P}{\partial \theta}
    &-
    \underbrace{
    \left[\frac{\partial\alpha_\parallel}{\partial\theta} + 2 \frac{\partial\alpha_\perp}{\partial\theta}\right]}_\text{volume rescaling}P\\
    &-
    \underbrace{
    \mu (1-\mu^2)\left[\frac{\partial\alpha_\parallel}{\partial\theta} - \frac{\partial\alpha_\perp}{\partial\theta}\right]}_{-\partial_\theta\mu}\frac{dP}{d\mu}
    -
    \underbrace{
    k
    \left[\frac{\partial\alpha_\parallel}{\partial\theta} \mu^2 + \frac{\partial\alpha_\perp}{\partial\theta} (1-\mu^2)\right]}_{-\partial_\theta k}\frac{dP}{dk}.
\end{aligned}
\end{equation}

\subsection{BAO modeling}
\label{sec:bao}

One of the major scientific goals of future redshift surveys is the mapping of the expansion history of the Universe using the baryon acoustic oscillation (BAO) technique \cite{Wei13}.  The BAO features in the power spectrum form a particularly robust standard ruler, and can benefit from a technique known as ``reconstruction'' \cite{ESSS07,Pad12}, that improves the distance fidelity.  Since BAO measurements are analyzed differently than many of the other parameters we've discussed, we forecast constraints on the distance scale differently, as we now discuss.

We hold the shape of the linear theory power spectrum fixed at a fiducial value and compute the post-reconstruction power spectrum $(P_\text{recon})$ with the ``Rec-Sym'' convention within the Zeldovich approximation \cite{Chen19b}. We use the same quadratic bias model of \S\ref{sec:full shape power spectrum}, with fiducial bias values listed in Table~\ref{tab:marg_params}, and approximate the reconstructed power spectrum by its first three non-zero multipoles. We include the A-P parameters, $\alpha_\parallel$ and $\alpha_\perp$, that are defined to include a factor of the sound horizon at the drag epoch ($r_d$), and marginalize over a linear bias and 15 ``broad band'' polynomials as is usually done experimentally (e.g.\ refs.\ \cite{BOSS_DR12,eBOSS21}):
\begin{equation}
    P_{\rm obs}(k_\perp,k_\parallel;\bm{b}) = \alpha_\parallel^{-1}\alpha_\perp^{-2}\, P_{\rm recon}(k_\perp/\alpha_\perp,k_\parallel/\alpha_\parallel,\bm{b}) + \sum_{n=0}^4 \sum_{m=0}^2 c_{nm} k^n \mu^{2m}
\end{equation}
with $\mu=k_\parallel/k$ and $\bm{b}=(b,b_2,b_s)$. We set the fiducial value of $c_{00}$ coefficient to the Poisson shot noise $1/\bar{n}_\text{gal}$, and the fiducial values of all higher order coefficients to zero. 

The Fisher matrix for the reconstructed power spectrum takes the same functional form as Eq.~\eqref{eqn:partial_fisher}. In our forecasts (\S\ref{sec:distance_measures}) we interpret the marginalized errors on the tangential and radial distance measures as $\sigma_{\alpha_\perp} D_A/r_d$ and $\sigma_{\alpha_\parallel} H r_d$ respectively. In Appendix~\ref{sec:bao_comparison} we compare this approach to the traditional method of ref.~\cite{Seo_2007}.

\begin{figure}[!h]
    \centering
    \includegraphics[width=0.9\linewidth]{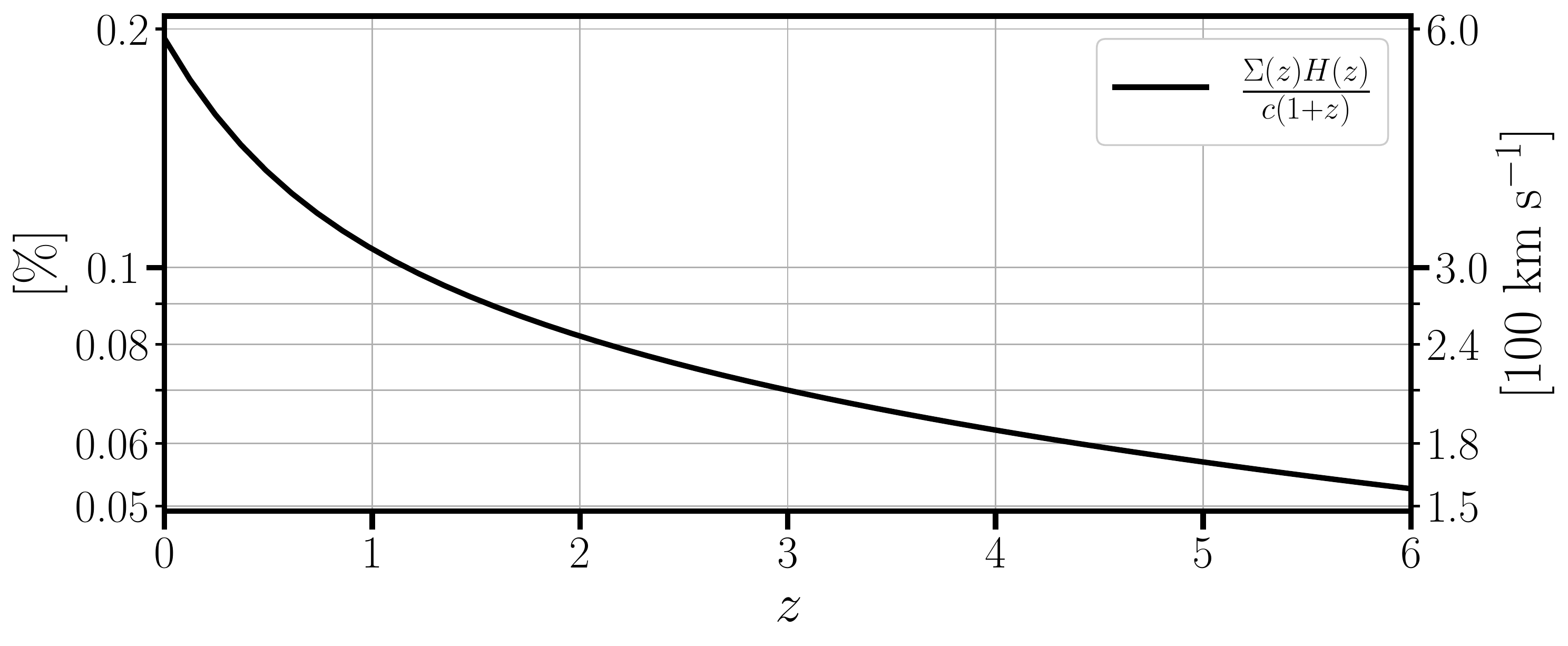}
    \caption{$\Sigma(z)H(z)/[c(1+z)]$ as a function of redshift. If the redshift uncertainties $\sigma_z/(1+z)$ are near or above this curve, then they will suppress modes that could carry valuable cosmological information.  The left $y$-axis shows the requirement on $\sigma_z/(1+z)$ in percent, while the right $y$-axis converts that to a velocity scale, in $100\,\mathrm{km}\,\mathrm{s}^{-1}$.
    }
    \label{fig:redshift_uncertainty}
\end{figure}

\subsection{Redshift uncertainties}
\label{sec:redshift_uncertainties}

An error $\sigma_z/(1+z)$ in the measurement of the redshift of an object corresponds to an error in the radial distance to said object $\sigma_\chi = \sigma_z d\chi/dz = [c(1+z)/H(z)]\sigma_z/(1+z)$. As a result of this error, the observed density field at Euleurian position $\bm{x}$ is related to the true density field by a simple translation 
$\delta^\text{obs}(\bm{x}) = \delta(\bm{x}-\epsilon(\bm{x})\hat{\bm{n}})$, where $\hat{\bm{n}}$ is the direction along the line of sight, and $\epsilon(\bm{x})$ is a stochastic variable which we assume is drawn from a Gaussian distribution with mean zero and standard deviation $\sigma_\chi$. Thus on average, the observed density field is damped in the $\hat{\bm{n}}$ direction:
\begin{equation}
\begin{aligned}
    \langle\delta^\text{obs}(\boldsymbol{k})\rangle_\epsilon &= 
    \int  d^3 \boldsymbol{y} 
    e^{-i\boldsymbol{k}\cdot\boldsymbol{y}}
    \delta(\boldsymbol{y})
    \frac{1}{\sqrt{2\pi}\sigma_\chi}
    \int d[\epsilon(\boldsymbol{x})]
    e^{-\epsilon(\boldsymbol{x})^2 / 2\sigma^2_\chi}
    e^{-ik_\parallel\epsilon(\boldsymbol{x})}\\
    &=
    e^{-k^2_\parallel \sigma^2_{\chi}/2}
    \delta(\boldsymbol{k}),
\end{aligned}
\end{equation}
which in turn damps the observed power spectrum by a factor of $\exp[-k^2\mu^2 \sigma_\chi^2]$. In our forecasts we assume that the redshift uncertainty can be well characterized and ignore this extra factor in the derivatives.  Therefore, at the Fisher matrix level (Eqn.~\eqref{eqn:partial_fisher}), we can in practice absorb the effect of redshift uncertainties in the noise, so that $N_0\to N_0\exp[k^2\mu^2 \sigma_\chi^2]$.

These uncertainties become increasingly important as an experiment pushes to higher redshifts. If we want to accurately reconstruct all modes with $k<k_\text{nl}$, we require that the uncertainty in radial length measurements is smaller than the corresponding wavelength at the non-linear scale: $\sigma_\chi < k_\text{nl}^{-1} = \Sigma(z)$. This in turn implies $\sigma_z/(1+z) < \Sigma(z) H(z)/[c(1+z)]$. The RHS of this inequality is plotted in Fig.~\ref{fig:redshift_uncertainty}.  Obtaining small redshift uncertainties at high redshift is challenging, because the tracers are intrinsically fainter, but also because some of the broader and easier lines to measure provide intrinsically less accurate redshifts (due to asymmetric profiles, or because they are affected by winds and outflows for example) \cite{Shapley03}.  While it may be possible to use the Ly$\alpha$ line profile to recover the systemic velocity to $\lesssim 100\,$km/s \cite{2018MNRAS.478L..60V} --- which would be more than sufficient for our science cases --- how well this works on large samples with achievable integration times remains an open question. We delay discussing the impact of redshift uncertainties on our forecasts to \S\ref{sec:issues}.

\section{Results}
\label{sec:results}

In sections \ref{sec:surveys and probes} and \ref{sec:fisher} we outlined the surveys, samples and methods used to create our forecasts. In this section we summarize our results, estimating the constraints on base $\Lambda$CDM, curvature, neutrino mass, relativistic species, primordial features, primordial non-Gaussianity, dynamical dark energy, and gravitational slip. 

\begin{figure}[!h]
\centering
\includegraphics[width=\linewidth]{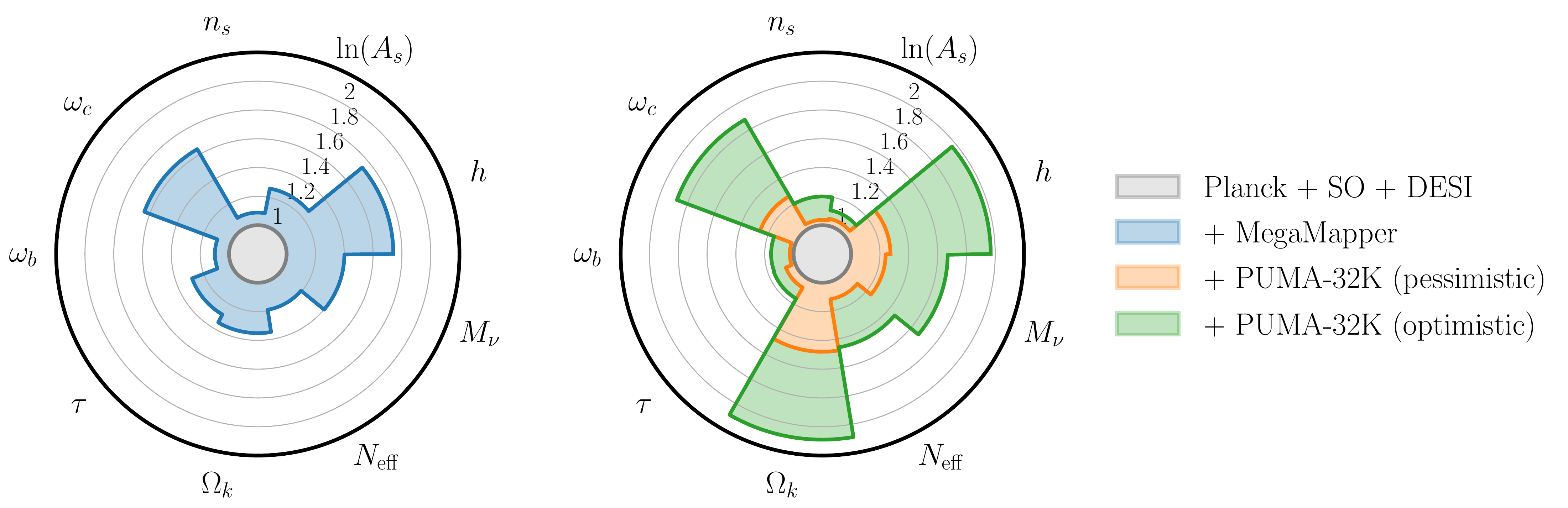}
\caption{
For each parameter $X$ on the perimeter, the corresponding radius is the ratio of the uncertainty $\sigma_X$ from Planck$+$SO$+$DESI (PSD) alone to the uncertainty from Planck$+$SO$+$DESI$+$high redshift ($z>2$) data.
The constraints on $\Lambda$CDM are calculated assuming our base model (see \S\ref{sec:fisher}) while each extension to our base model ($M_\nu,\,N_\text{eff},\,\Omega_k$) is treated as a single parameter extension. 
These constraints include both full shape and lensing data (see \S\ref{sec:combined fisher} for our conventions when combining surveys). For PSD$+$MegaMapper we include MegaMapper$\times$SO lensing, while for all other configurations DESI$\times$SO lensing is included. In green (orange) we show constraints for optimistic (pessimistic) foreground assumptions (\S\ref{sec:full shape power spectrum}).
}
\label{fig:polar_plot_HiZ}
\end{figure}

Because we are interested in the performances of proposed future experiments in the context of current state of the art surveys, we compute the Fisher information for DESI and Euclid as well. This way all current and future surveys are studied using the same perturbation theory scheme and share the same model assumptions, allowing for a fairer comparison.
In light of existing tensions between LSS datasets and Planck (e.g. the $H_0$ and $S_8$ tensions), we also present forecasted uncertainties on relevant parameters from individual surveys without any CMB external information.
In doing so we obtain an estimate for the expected level of agreement (or disagreement) between disjoint LSS and CMB analyses. 

We note that by taking $\Lambda$CDM as our fiducial cosmology, we are implicitly assuming that these tensions are either statistical flukes or due to systematic errors which will ultimately be resolved in the future. 
More generally, we implicitly assume that our base model (or our base model $+$ the extensions considered in the following subsections) is capable of accurately fitting future LSS and CMB data to the required precision. Should future tensions arise, or should current tensions worsen, it may be necessary to introduce extensions to this base model to accurately fit the data. Marginalizing over the free parameters associated with these extensions would inevitably result in larger errors than those derived assuming our base model. 

Such extensions could arise from new physics (e.g. early dark energy), new nuisance parameters which model systematic effects (e.g. relative velocities), or from freeing up parameters which had previously been held fixed. The latter is explored in detail in \S\ref{sec:Extending_the_base_model}. We note that we do not explore extending the base model to include the running of the spectral index $\alpha_s$. We expect a detection of the running to be well beyond the reach of the surveys considered in this work\footnote{We find $\sigma(\alpha_s)=0.005$ from our fiducial LiteBIRD + CMB-S4 + DESI + MegaMapper + lensing survey, while the expected value for the running is $\sim 0.001$.}, and it is therefore unlikely that one would need to extend the base model to fit for the running.

As partially summarized in Fig.~\ref{fig:polar_plot_HiZ}, high redshift surveys would yield significant gains in constraining power over current LSS and CMB surveys. For example, the addition of PUMA to the combination of Planck, SO and DESI tightens the constraints on $\omega_c,\,h$ and $\Omega_k$ by a factor of $\sim2$, while the constraints on $M_\nu$ improve by more than $60\%$. The addition of MegaMapper also yields to a large reduction of the uncertainties. Other extensions of the base model, such as primordial features (\S\ref{sec:primordial_features}) and EDE (\S\ref{sec:EDE}), show similar improvements, which we explore in detail in the following subsections.

\subsection{Distance measures}
\label{sec:distance_measures}

\begin{figure}[!h]
\centering
\includegraphics[width=\linewidth]{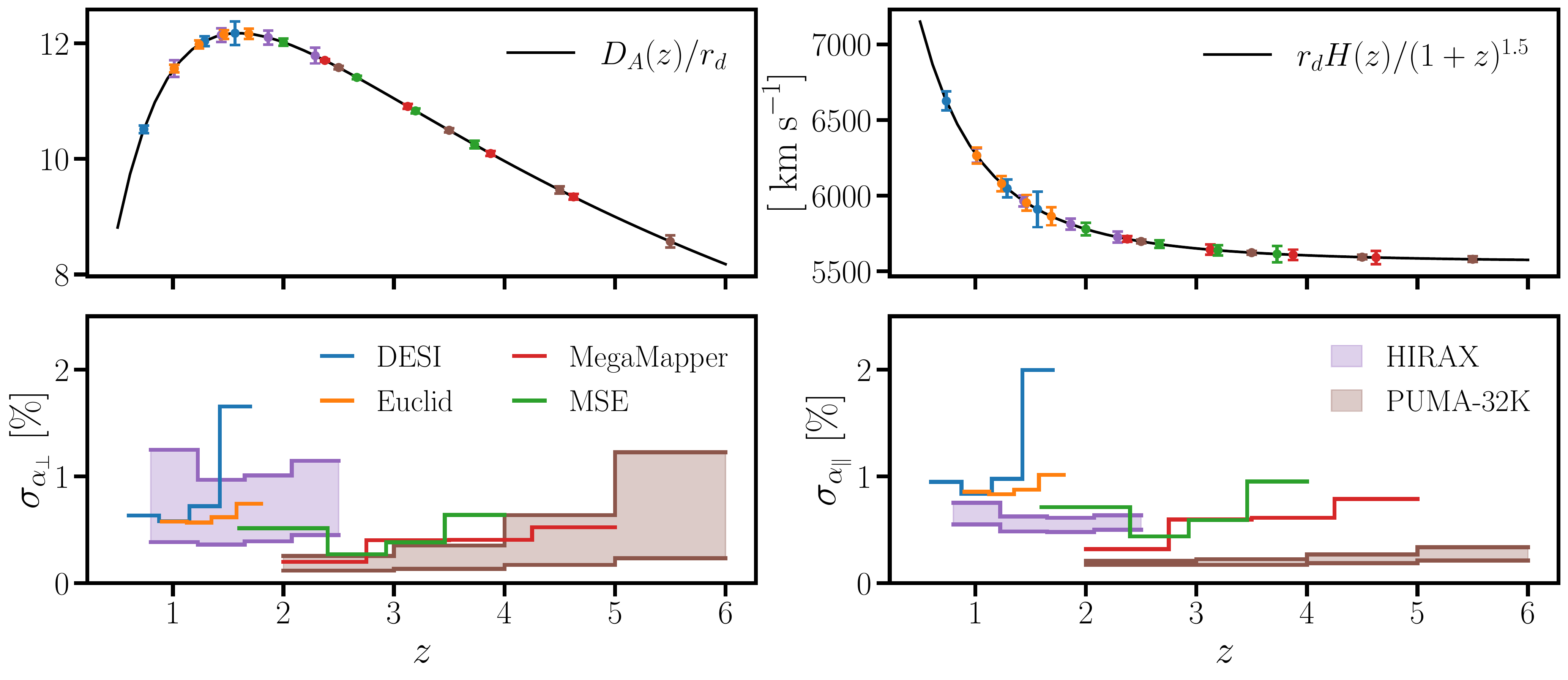}
\caption{
Error on the A-P parameters from the reconstructed power spectrum, which can be interpreted as relative errors on $D_A(z)/r_d$ and $r_d H(z)$ respectively. The boundaries of the shaded regions denote optimistic/pessimistic foreground assumptions for the 21-cm surveys. In the top panels we show the errorbars for the optimistic case.
}
\label{fig:distance_constraints}
\end{figure}

The first results we show concern the uncertainty on the distance measures obtained from measurements of the BAO feature (\S\ref{sec:bao}). Fig.~\ref{fig:distance_constraints} shows the relative error on $\alpha_\perp$ and $\alpha_\parallel$, which can be interpreted as relative errors on $D_A/r_d$ and $r_d H $, for different spectroscopic and 21-cm surveys.
Near-term surveys (DESI, Euclid and HIRAX) could measure the angular diameter distance to sub-percent precision all the way to $z\sim 2$, and the Hubble parameter to percent precision in the same redshift range. 
For 21-cm instruments we always show a shaded region encompassing optimistic and pessimistic foreground assumptions. 
We find that an MSE-like survey, shown in green, nicely extends the current spectroscopic program to higher redshifts, achieving the same accuracy in the redshift range $2\le z\le4$ as DESI and Euclid at low redshift. An instrument like MegaMapper would further improve these constraints, yielding 0.5\% measurements of both $D_A/r_d$ and $r_d H $ at $z\lesssim3.5$ and 1\% measurements at $z\sim5$. A Stage-II 21-cm experiment like PUMA-32K, shown in brown, could in principle outperform spectroscopic surveys, providing almost cosmic variance limited measurements of the Hubble parameter out to $z\sim 6$. For the measurement of distances perpendicular to the line of sight, 21-cm instruments are severely limited by the foreground contamination, but even in our pessimistic scenario PUMA-32K would reduce the uncertainties on $D_A/r_d$ by roughly 30\% compared to MegaMapper.

We would like to stress that for 21-cm instruments it is not yet clear how to perform BAO reconstruction and therefore enhance the constraints on the distance parameters. In auto-correlation experiments new approaches have been considered \cite{Obuljen2016}, but no practical implementation has been proposed for interferometers. On the other hand we notice that the main purpose of BAO reconstruction is to undo large scale flows and sharpen the position of the BAO feature in configuration space. The effect of such large scale motions is smaller at high redshifts, and becomes irrelevant at $z\gtrsim3$, after which we expect BAO reconstruction to yield diminishing returns.

Finally, it is possible to combine BAO information with the full shape analysis presented in the next few sections \cite{Philcox2020}. This would however require a model for the covariance between the two measurements, which is currently best obtained using mock catalogs. For this reason we do not include any prior on the distance measures in the forecasts discussed in the next sections.

\begin{table}
\centering
\resizebox{\textwidth}{!}{\begin{tabular}{c|ccccccccc}
\hline
Configuration & $10^4 h$ & $10^4\omega_c$ & $10^3 \tau$ &  $10^3 \Omega_k$ & $10^3N_{\rm eff}$ & $M_\nu\,\,[{\rm meV}]$\\
\hline
\hline
Planck & $62.14$ & $13.66$ & $7.87$ & $17.19$ & $187.09$ & $83.39$\\
DESI & $63.83$ & $42.29$ & $>10^5$ & $20.38$ & $599.83$ & $311.98$ \\
Euclid & $58.02$ & $35.55$ & $\cdots$ & $17.0$ & $476.77$ & $245.8$ \\
MSE & $49.84$ & $21.36$ & $\cdots$ & $10.79$ & $214.09$ & $129.18$ \\
MegaMapper & $45.75$ & $19.42$ & $\cdots$ & $8.01$ & $173.57$ & $104.58$ \\
HIRAX & $48.9/98.2$ & $23.2/42.5$ & $\cdots$ & $10.7/29.4$ & $267.5/707.0$ & $157.0/516.9$ \\
PUMA-5K & $39.4/64.9$ & $14.6/16.8$ & $\cdots$ & $5.5/12.1$ & $101.9/204.1$ & $67.4/154.2$ \\
PUMA-32K & $32.8/49.7$ & $12.2/14.3$ & $\cdots$ & $3.8/8.8$ & $81.4/150.7$ & $52.4/116.1$ \\
\hline
\hline
Planck $+$ DESI & $31.83$ & $7.28$ & $5.86$ & $1.24$ & $112.8$ & $41.49$ \\
$+$ lensing& $28.72$ & $6.58$ & $5.17$ & $1.19$ & $107.32$ & $38.01$ \\
\hline
\rowcolor{gray!20}
Planck $+$ SO $+$ DESI (PSD) & $23.42$ & $6.13$ & $5.83$ & $0.68$ & $43.93$ & $39.76$ \\
\rowcolor{gray!20}
$+$ lensing& $18.68$ & $4.9$ & $4.2$ & $0.68$ & $42.84$ & $35.39$ \\
\hline
Planck $+$ SO $+$ Euclid & $23.04$ & $6.05$ & $5.64$ & $0.66$ & $43.64$ & $37.17$ \\
$+$ lensing & $17.23$ & $4.54$ & $3.97$ & $0.65$ & $42.62$ & $33.67$ \\
\hline
Planck $+$ SO $+$ MSE & $17.63$ & $4.76$ & $5.5$ & $0.58$ & $41.51$ & $31.57$ \\
$+$ lensing & $12.47$ & $3.41$ & $3.46$ & $0.57$ & $39.27$ & $29.2$ \\
\hline
\hline
\rowcolor{gray!20}
PSD $+$ MegaMapper & $13.77$ & $3.77$ & $4.67$ & $0.51$ & $38.1$ & $26.33$ \\
\rowcolor{gray!20}
$+$ lensing & $10.65$ & $2.96$ & $3.23$ & $0.5$ & $35.91$ & $25.29$ \\
\hline
PSD $+$ HIRAX & $16.9/22.4$ & $4.4/5.9$ & $5.8/5.8$ & $0.6/0.6$ & $41.1/43.6$ & $34.1/38.9$ \\
$+$ lensing & $14.9/18.2$ & $3.9/4.8$ & $3.9/4.2$ & $0.6/0.6$ & $40.3/42.6$ & $30.0/34.5$ \\
\hline
PSD $+$ PUMA-5K & $12.9/19.7$ & $3.5/5.2$ & $5.8/5.8$ & $0.4/0.5$ & $34.2/41.5$ & $27.1/35.3$ \\
$+$ lensing & $11.9/16.7$ & $3.2/4.4$ & $3.7/4.0$ & $0.4/0.5$ & $33.8/40.6$ & $25.2/31.6$ \\
\hline
\rowcolor{gray!20}
PSD $+$ PUMA-32K & $10.0/16.6$ & $2.8/4.4$ & $5.7/5.8$ & $0.3/0.5$ & $29.6/39.0$ & $22.4/31.6$ \\
\rowcolor{gray!20}
$+$ lensing & $9.6/14.7$ & $2.6/3.9$ & $3.6/3.9$ & $0.3/0.5$ & $29.4/38.3$ & $21.2/28.5$ \\
\hline
\hline
LiteBIRD $+$ S4 $+$ DESI (LSD) & $21.39$ & $5.74$ & $2.02$ & $0.34$ & $24.57$ & $34.78$ \\
$+$ lensing & $12.86$ & $3.41$ & $1.93$ & $0.34$ & $24.02$ & $24.38$ \\
\hline
LSD $+$ MegaMapper & $12.76$ & $3.52$ & $1.94$ & $0.31$ & $22.63$ & $20.02$ \\
$+$ lensing & $8.86$ & $2.43$ & $1.74$ & $0.31$ & $21.62$ & $14.92$ \\
\hline
LSD $+$ HIRAX & $15.9/20.6$ & $4.3/5.5$ & $2.0/2.0$ & $0.3/0.3$ & $23.7/24.5$ & $30.2/34.1$ \\
$+$ lensing & $11.3/12.7$ & $3.0/3.4$ & $1.9/1.9$ & $0.3/0.3$ & $23.3/24.0$ & $19.5/23.6$ \\
\hline
LSD $+$ PUMA-5K & $12.4/18.4$ & $3.4/4.9$ & $2.0/2.0$ & $0.3/0.3$ & $21.3/24.1$ & $24.7/31.1$ \\
$+$ lensing & $9.7/12.1$ & $2.6/3.2$ & $1.8/1.9$ & $0.3/0.3$ & $21.0/23.6$ & $17.4/21.4$ \\
\hline
LSD $+$ PUMA-32K & $9.7/15.8$ & $2.7/4.2$ & $2.0/2.0$ & $0.3/0.3$ & $19.4/23.4$ & $20.3/28.0$ \\
$+$ lensing & $8.3/11.3$ & $2.2/3.0$ & $1.8/1.9$ & $0.3/0.3$ & $19.3/23.0$ & $14.5/18.9$ \\
\rowcolor{gray!20}
Everything bagel & $6.95$ & $1.94$ & $1.69$ & $0.24$ & $18.36$ & $12.24$ \\
\hline
\end{tabular}}
\caption{
$1\sigma$ uncertainties on selected $\Lambda$CDM parameters and extensions from a combination of primary CMB, large-scale clustering ($P_{gg}(\bm{k})$), and CMB$\times$LSS lensing data. For all configurations that include 21-cm data, we give constraints for optimistic/pessimistic foreground assumptions. The Planck constraints in the first row are the diagonal components of the measured TT,TE,EE+lowE covariance \cite{PCP18}. These would be the true constraints if the Planck likelihood were a perfect Gaussian, and agree with the constraints quoted in ref.~\cite{PCP18} to within $5\%$ (with the exception of $M_\nu$, where non-Gaussianities in the likelihood increase the true $\sigma_{M_\nu}$ to $130$ meV). Everywhere else in this paper, Planck refers to the CMB prior described in $\S\ref{sec:CMB_experiments}$. In rows $2-8$ (LSS alone) we include a conservative $\sigma(\omega_b)=0.0005$ prior from BBN \cite{PCP18}. PSD is short for Planck + SO + DESI, while LSD is short for LiteBIRD + S4 + DESI. ``Everything bagel'' is short for LSD $+$ PUMA-32K (with optimistic foreground assumptions) $+$ MegaMapper $+$ MegaMapper$\times$S4 lensing, and neglects any correlations between 21-cm and LBG clustering.
}
\label{tab:big_summary}
\end{table}

\subsection{Base $\Lambda$CDM}
\label{sec:base_lcdm}

Historically, the standard ($\Lambda$CDM) model has been extremely successful in modeling cosmological observables, ranging from the Universe's expansion history to the statistics of CMB anisotropies and the evolution of galaxy clustering (e.g.\ Fig.~\ref{fig:marius_plot}). Future LSS and CMB surveys will significantly improve over current constraints, providing the most stringent test of $\Lambda$CDM to date.

Within the next $\sim5$ years, our forecasts indicate that the combination of DESI and Planck will improve upon current CMB measurements of the Hubble constant and $\omega_c$ by more than a factor of $2$. By themselves, DESI and Euclid will provide similar constraints to Planck on several $\Lambda$CDM parameters, but LSS and CMB data combined are larger than the sum of their parts.
In Table~\ref{tab:big_summary} we also show how the inclusion of CMB lensing (and its cross-correlation with galaxies) improves the constraints. For Planck lensing and current or upcoming surveys like DESI and Euclid, the addition of lensing does not yield significant improvements. Given the time frame of future LSS surveys, we expect to (in principle) be able to combine them with Planck$+$SO$+$DESI (PSD) data. The addition of MegaMapper to PSD improves the constraint on the Hubble constant by almost a factor of 2, and by $\sim40$\% on $\omega_c$. We find that cross-correlations with CMB lensing maps return larger gains for a MegaMapper-like survey, presumably due to the increased overlap with the CMB lensing kernel. For 21-cm surveys, foregrounds will likely make the cross correlation with CMB lensing maps not possible. Thus, in Table~\ref{tab:big_summary}, the ``$+$lensing" rows below any PSD$+$21 cm surveys only include the cross-correlation of DESI with SO (or S4) lensing maps. Nonetheless PSD$+$PUMA-5K or -32K could yield comparable constraints to PSD$+$MegaMapper on most $\Lambda$CDM parameters. In the future, density reconstruction methods could help 21-cm surveys to recover the modes lost in the foreground wedge, allowing for a non-zero correlation with CMB lensing maps \cite{Modi:2019hnu, Modi:2021okf}.

On a similar timescale to future LSS surveys, CMB experiments like LiteBIRD and CMB-S4 will also be taking data. It is therefore interesting to consider their combination with DESI as the baseline, which we dub LSD. As shown in Table~\ref{tab:big_summary}, clustering data at high redshifts enable significant improvements in $\Lambda$CDM constraints over a nearly cosmic variance limited CMB survey. The uncertainties on $h$ and $\omega_c$ from LSD only are $>50\%$ larger than those from  LSD$+$MegaMapper. Improvements from cross-correlations with CMB lensing are also of the same order.  
We note that constraints from LSD+PUMA-32K are comparable to those from LSD+MegaMapper, despite 21-cm surveys having a lower effective shot noise. 
This is ultimately due to degeneracies between the brightness temperature $T_b$, $A_s$ and the linear bias $b$, which in turn worsen the constraints on any cosmological parameters that the latter are degenerate with.

\subsection{Structure growth}
\label{sec:structure_growth}

Future high-redshift surveys will enable $1\%$ measurements of the amplitude of structure growth $\sigma_8(z)$ out to $z\sim 5$. We consider two different methodologies for structure growth measurements: one in which $\sigma_8$ is treated as a derived parameter from $\Lambda$CDM, and another, more cosmology-agnostic approach in which the shape of the power spectrum is held fixed.

\begin{figure}[!h]
    \centering
    \includegraphics[width=\linewidth]{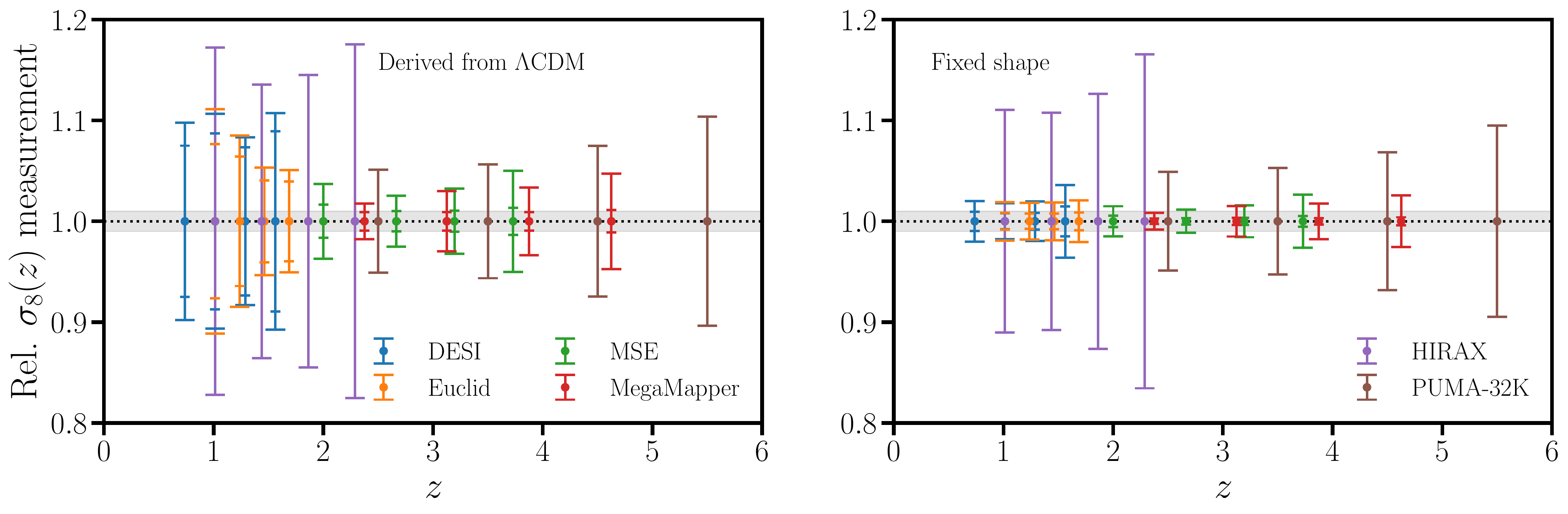}
    \caption{
    \textit{Left:} Errors on $\sigma_8(z)$ when derived from $\Lambda$CDM, assuming a $\sigma(\omega_b)=0.0005$ prior from BBN. \textit{Right:} Errors on $\sigma_8(z)$ from full shape data when calculated using the fixed shape procedure, without any external priors. The errorbars with smaller caps include cross-correlations with SO lensing, but do not include any information from the convergence auto-spectrum. All 21-cm measurements assume optimistic foregrounds. The gray bands extend from $1\pm0.01$.
    }
\label{fig:fs8}
\end{figure}

In the derived-from-$\Lambda$CDM procedure we calculate the Fisher matrix from full shape clustering data with basis $\{\theta_i\} = \{\Lambda\text{CDM},b,b_2,b_s,N_0,N_2,N_4,\alpha_0,\alpha_2,\alpha_4\}$ in each redshift bin. Our choice of $\Lambda$CDM parameters is listed in Table~\ref{tab:marg_params}. We then transform to a basis that includes $\sigma_8(z)$, where $z$ is the central redshift of some redshift bin. The Fisher matrix transforms as a two-form under a change of basis: $\bm{F}\to \bm{R}^T \bm{F}\bm{R}$, where
\begin{equation}
    \bm{R}^{-1}
    =
    \begin{pmatrix}
    \frac{\partial \sigma_8(z)}{\partial \theta_1} & \frac{\partial \sigma_8(z)}{\partial \theta_2} & \cdots & \frac{\partial \sigma_8(z)}{\partial \theta_n} \\
    0 & 1 & \cdots & 0\\
    \vdots & \vdots & \ddots & \vdots \\
    0 & 0 & \cdots & 1
    \end{pmatrix}.
\end{equation}
We then add a conservative $\sigma(\omega_b) = 0.0005$ \cite{PCP18} prior from BBN, and invert this matrix to yield the marginalized constraint on $\sigma_8(z)$. Thus the measurements of $\sigma_8(z)$ are local, derived only from data within the redshift bin of interest, apart from the shared $\omega_b$ prior. As shown in the left panel of Fig.~\ref{fig:fs8}, DESI and Euclid will measure $\sigma_8(z)$ to $5-10\%$ across the range $0.6<z<2$, while MSE and MegaMapper will extend these measurements out to $z=5$ with less than $5\%$ error. Due to a degeneracy between $\sigma_8$ and $T_b$, the constraints from HIRAX are more modest than those from DESI or Euclid. Since $\sigma_8$ and $T_b$ are perfectly degenerate in linear theory, 21-cm measurements of $\sigma_8$ heavily depend on how accurately one can model and measure non-linear effects \cite{Castorina19}. While PUMA also suffers from the same $\sigma_8-T_b$ degeneracy, it has a much lower shot noise than HIRAX, and is thus more sensitive to non-linear scales that differentiate the effects of the two parameters, enabling $5-10\%$ measurements across its redshift range. We find that PUMA-32K (with a BBN prior on $\omega_b$) can measure $T_b(z)$ to the same relative accuracy as $\sigma_8(z)$, or $5-10\%$ depending on the redshift bin. Thus a $\approx\,1\%$ prior on $T_b(z)$ would be required for 21-cm measurements of $\sigma_8(z)$ to be competitive with MegaMapper, which is likely beyond the reach of astrophysical determinations from high-column-density systems. 

In Fig.~\ref{fig:fs8} we also show the improvement from adding SO$\times$LSS lensing to full shape data. When adding lensing to this figure we only include the cross-correlation, and not the convergence power spectrum. This ensures that the constraints on $\sigma_8(z)$ remain local apart from the $\omega_b$ prior. Thus these constraints make no assumptions about the expansion history outside of the redshift bin of interest. As was true in the previous section, lensing cross-correlations offer small improvements for DESI and Euclid, while for MegaMapper, the improvement can be more than a factor of $2$. With lensing cross-correlations, MegaMapper can measure $\sigma_8(z)$ with a $1\%$ error across its entire redshift range.

While the derived-from-$\Lambda$CDM constraints are only valid within the $\Lambda$CDM framework, fixed shape constraints assume very little about the underlying cosmology. For any cosmology whose free parameters include an amplitude of the matter power spectrum, one can obtain a fixed shape measurement by fitting to the amplitude with all other parameters fixed at some fiducial value. Thus fixed-shape forecasts give a sense of a ``best-case'' scenario, since the error on the amplitude from the fixed shape procedure is always smaller than the error derived from the full cosmological model. In our case we fix $\{h,\omega_c,\omega_b,n_s,\tau\}$ at their fiducial $\Lambda$CDM values, and treat the amplitude of the matter power spectrum as an independent variable in each redshift bin. In each bin we marginalize over the nuisance terms listed in Table~\ref{tab:marg_params}. With this procedure, we find that DESI and Euclid yield $2\%$ measurements across $0.6<z<2$, while MSE and MegaMapper extend these measurements out to $z=5$ with comparable uncertainty. When cross-correlations with SO lensing are included, these errors decrease to sub-percent values across the entire redshift range covered by spectroscopic surveys $(0.6<z<5)$. Again, we see that 21-cm measurements suffer from a $\sigma_8-T_b$ degeneracy, resulting in much more modest $\mathcal{O}(10\%)$ constraints. 

We find that fixed shape measurements from spectroscopic surveys have $2$ to $5$ times smaller error than when derived from $\Lambda$CDM (without lensing cross-correlations). For 21-cm surveys, however, the two methods yield similar constraints, presumably due to the $\sigma_8-T_b$ degeneracy dominating over any other degeneracy between $\sigma_8$ and the remaining $\Lambda$CDM parameters, such as $\tau$. 

\subsection{Massive neutrinos}
\label{sec:neutrinos}

One of the best-motivated extensions beyond the $\Lambda$CDM model described above is allowing the neutrino masses to be non-minimal \cite{PDG20}. Measurements of solar and atmospheric neutrino oscillations probe $m_{\nu_2}^2-m_{\nu_1}^2$ and $m_{\nu_2}^2-m_{\nu_3}^2$ respectively, which together imply a minimum value for the sum of the neutrino masses $M_\nu = \sum_i m_{\nu_i} > 60\,$meV. Oscillation measurements have yet to determine if neutrinos follow a normal 
$(m_{\nu_2} < m_{\nu_3})$ 
or inverted 
$(m_{\nu_2} > m_{\nu_3})$ 
mass hierarchy, which is one of the major goals of upcoming oscillation measurements and neutrino-less double beta decay experiments. Cosmological constraints of $M_\nu$ are complementary to these experiments. CMB and BAO data currently provide our tightest upper limits on $M_\nu$ \cite{planckcollaboration2019planck}, and future surveys should tighten these constraints significantly. 

Neutrinos are relativistic at early times, washing out their primordial fluctuations at length scales smaller than the horizon. However, once the neutrino fluid cools and becomes non-relativistic (occurring at a scale factor  $a_{\rm nr}$), so that its energy density is dominated by the neutrino mass $M_\nu$, neutrinos cluster in the same way as the CDM+baryon fluid. The comoving length scale associated with this transition is $k_\text{nr} = a_\text{nr} H(a_\text{nr}) \sim 10^{-3}h$Mpc$^{-1}$ \cite{CMBS4}. Thus, another way of saying the above is that neutrinos only cluster at large scales ($k\ll k_\text{nr}$). At small scales, the neutrino energy density is nearly spatially homogeneous, contributing to the Hubble drag associated with the evolution of the CDM+baryon fluid, but not to the gravitational potential, which drives the clustering. Thus a nonzero neutrino mass slows the growth of structure, damping the matter power spectrum at scales smaller than the horizon at $a_\text{nr}$. 

\begin{figure}[!h]
    \centering
    \includegraphics[width=\linewidth]{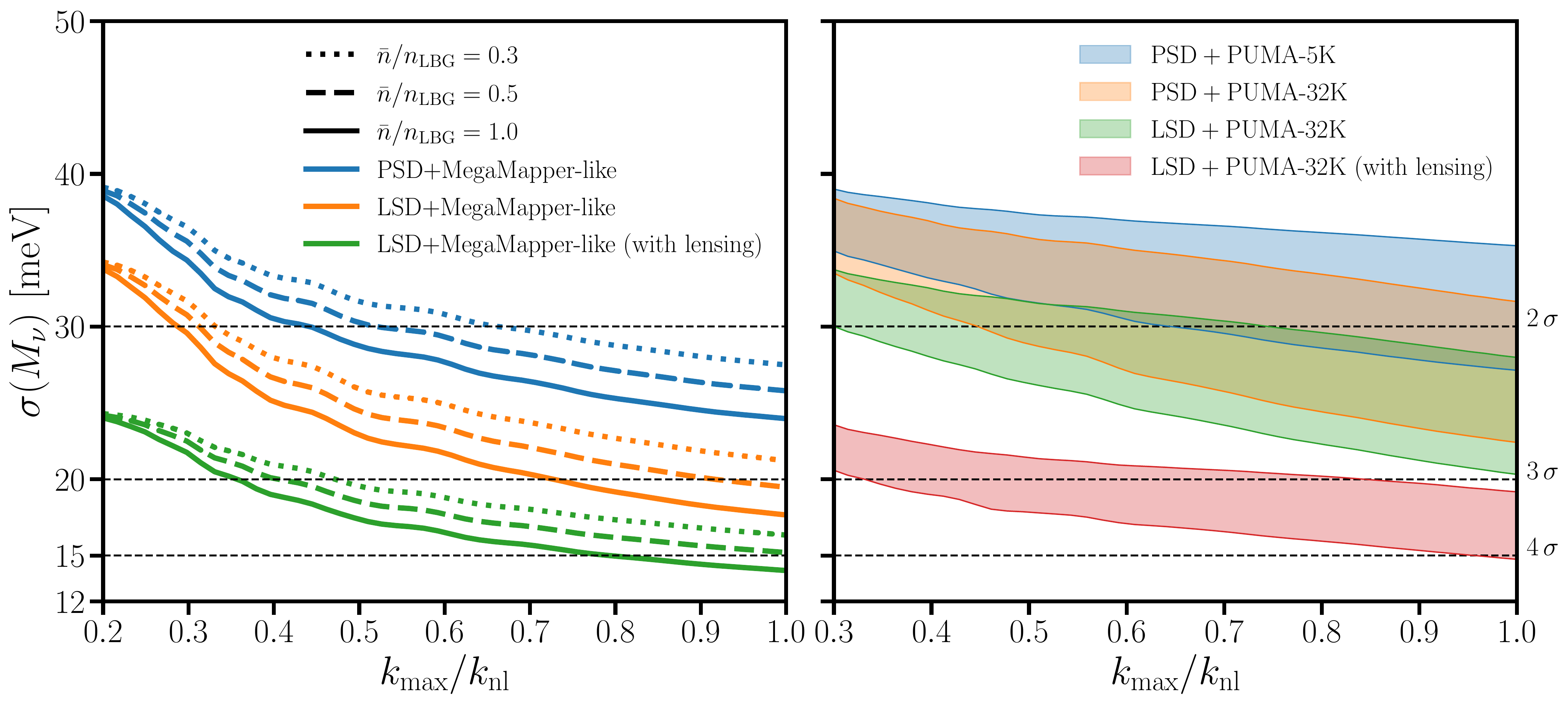}
    \caption{
    Constraints on $M_\nu$ as a function of $k_\text{max}/k_\text{nl}$ from the primary CMB and full shape data. The curves that are labeled ``(with lensing)'' include information from DESI$\times$S4 lensing ($\kappa\kappa$, $\kappa g$, $gg$). In both plots we set $k_\text{max}=k_\text{nl}$ for DESI, and only vary $k_\text{max}/k_\text{nl}$ for the high redshift survey of interest. 
    \textit{Left:} Constraints for a MegaMapper-like survey that observes a redshift-independent fraction $\bar{n}/n_{\rm LBG}$ of LBGs from $2<z<5$ with $f_\text{sky}=0.34$. 
    \textit{Right}: Constraints for PUMA. The edges of the shaded regions represent optimistic and pessimistic foreground assumptions. 
    } 
\label{fig:mnu_vs_kmax}
\end{figure}

Since this effect is largest at small scales and at late times, high redshift LSS surveys $-$ which will measure high $k$ modes in the late-time Universe $-$ are ideally situated for $M_\nu$ measurements. As shown in Table~\ref{tab:big_summary}, we see that $\sigma(M_\nu)$ drops dramatically with the addition of high redshift data: the constraint from PSD is 35 (40) meV with(out) lensing, which drops to 25 (26) meV for PSD+MegaMapper, or 21 (22) meV for PSD+PUMA-32K. As shown in Fig.~\ref{fig:mnu_vs_kmax}, these results depend significantly on the $k_\text{max}$ used in the forecast. The constraints change by nearly a factor of two for a LSD+MegaMapper-like survey between $0.2 < k_\text{max}/k_\text{nl}< 1.0$. Likewise, the 21-cm foreground assumptions have an equally significant effect on the forecasted neutrino constraints, impacting the results by up to $50\%$ for the highest values of $k_\text{max}$ for PSD+PUMA-32K\footnote{For parameters with a physical priors, \textit{e.g.}  $M_{\nu}>0$, the Fisher matrix approach does not necessarily return trustworthy error bars if the fiducial value chosen is close to the physical boundary. For this reason we caution to reader about the interpretation of our results when $\sigma(M_{\nu})$ is large.}. However, even with a more conservative cutoff of $k_\text{max}/k_\text{nl}=0.5$, and with mildly conservative foreground assumptions, a next generation CMB survey coupled to either MegaMapper or PUMA-32K could comfortably achieve $\sigma(M_\nu)\lesssim 20$ meV. 

Currently in the literature there is a significant bifurcation in $M_\nu$ forecasts for the upcoming Euclid satellite. Ref.\ \cite{Boyle20} found $34$ meV for Planck+SO+Euclid+Euclid$\times$SO lensing using a PT-based method in which nine nuisance terms were marginalized over, including four bias parameters ($b$, $b_2$, $b_s$, $b_3$), three counterterms ($\alpha_0$, $\alpha_2$, $\alpha_4$) and two stochastic contributions ($N_0$, $N_2$). When we fix $N_4$ and match their $k_\text{max}=0.2\,h\,\text{Mpc}^{-1}$ cut we find 40 meV for Planck+SO+Euclid+Euclid$\times$SO lensing. A $\sim15\%$ discrepancy is entirely reasonable given that their Euclid sample spans a broader redshift range ($0.6<z<2.1$) than ours ($0.9<z<1.8$), uses slightly different number densities and biases, and also marginalizes over a $b_3$ term. By contrast, ref. \cite{Chudaykin19} finds 17 meV for Planck+Euclid+lensing (PS only) using a PT-based approach in which seven nuisance parameters were marginalized over, including three bias terms, three counterterms, and the shot noise. When we fix $N_2$ and $N_4$ and set $k_\text{max}=0.5\,h\,\text{Mpc}^{-1}$, so that our set of nuisance terms and $k_\text{max}$ match ref. \cite{Chudaykin19}, we find 30 meV for Planck+Euclid+lensing+BAO.  If we fix $\tau$ our constraint drops to 20 meV, in good agreement with ref. \cite{Chudaykin19}. The residual difference could be due to the fact that the authors in \cite{Chudaykin19} employ a full MCMC approach as opposed to the Fisher matrix presented in this work.  

Similar results for the cross-correlation of galaxies with CMB lensing were also recently obtained by  \cite{chen2021precise}, who consider a non-linear bias expansion similar to ours.  
Our results are largely consistent with recent forecasts for upcoming high redshift surveys. Ref.\ \cite{ferraro2019inflation} found 26 meV for PSD+MegaMapper\footnote{This includes BAO only and no lensing, uses a different $k_\text{max}$, and does not marginalize over non-linear terms, so the comparison is not entirely fair.}, while we find 25 meV. Recent PUMA forecasts \cite{collaboration2018inflation} show $M_\nu$ improving from 38 meV with Planck+LSST+DESI to 20 meV with CMB-S4+PUMA-32K, while we get 40 meV for Planck+SO+DESI and 20 meV for LSD+PUMA-32K. 

We note that the inclusion of a future cosmic-variance limited measurement of $\tau$ from large scale CMB polarization such as from the proposed LiteBIRD satellite provides significant improvements on $M_{\nu}$ by removing the $\tau-A_s$ degeneracy present in the primary CMB. The very large-scale polarization signal is challenging to measure from the ground, potentially requiring a future satellite mission\footnote{The CLASS experiment \cite{2014SPIE.9153E..1IE} aims to measure these very large-scale polarization fluctuations from the ground, and depending on performance, it may lead to competitive constraints on $\tau$ compared to a future space mission.}. However, small scale information in both the CMB temperature (through fluctuations in the kinematic Sunyaev-Zel'dovich effect \cite{Ferraro:2018izc, Alvarez:2020gvl}), CMB lensing \cite{Meerburg:2017lfh}, or the 21cm signal from reionization \cite{Liu:2015txa, billings2021extracting} can potentially constrain $\tau$ to a similar level as LiteBIRD ($\sigma(\tau) \approx 0.002$), if astrophysical systematics can be kept under control. Including such a prior on $\tau$ reduces the PSD$+$MegaMapper neutrino mass constraint from 25 meV to 16 meV (including lensing) while for PSD$+$PUMA-32K it drops from 21 meV to 16 meV. As shown in Table~\ref{tab:big_summary} and Fig.~\ref{fig:extensions_to_base_model}, these improvements are similar to those obtained by upgrading from PSD to LSD, suggesting that the improvement from these future CMB data is almost entirely dominated by the better $\tau$ measurement.

Finally we note that with a LiteBIRD-like prior on $\tau$ and $k_\text{max}/k_\text{nl}=1$, MegaMapper (which corresponds to $\bar{n}/n_\text{LBG}\approx 0.5$) is primarily limited by its shot noise, and that small improvements are possible with a higher LBG target efficiency. Increasing the number density to  $\bar{n}/n_\text{LBG}=1$ decreases the LSD + MegaMapper (without lensing) constraint from $20$ to $18$ meV.
 
\subsection{Light relics}
\label{sec:relics}

Measurements of the total amount of radiation in the Universe are a powerful way to search for new physics. Any light particle that was once in thermal equilibrium with the Standard Model, and hence had a cosmological abundance, leads to well known modifications in the power spectrum of the CMB and LSS.
These effects are commonly parameterized through $N_\text{eff}$, defined such that
\begin{equation}
    \rho_r
    =
    \frac{\pi^2}{15}
    \left[
    1
    +
    \frac{7}{8}
    \left(\frac{4}{11}\right)^{4/3}
    N_\text{eff}
    \right]
    T^4_\gamma
\end{equation}
during the radiation dominated era, with $N_\text{eff}=3.046$ \cite{Mangano_2005} in the Standard Model (or $N_\text{eff}=3$ in the limit of instantaneous neutrino decoupling). The presence of new light relics would shift this value to $N_\text{eff}=3.046+\Delta N_\text{eff}$, where $\Delta N_\text{eff}$ depends on the number of independent spin states and decoupling temperature(s). By considering decoupling temperatures that are significantly larger than the top quark mass, one obtains lower limits to $\Delta N_\text{eff}$ for broad classes of particles: $\Delta N_\text{eff} > 0.027$ for scalars, $0.047$ for Weyl fermions, and $0.054$ for vector bosons~\cite{CMBS4}. In many examples of physics beyond the Standard Model, these new states are too weakly coupled to be probed by facilities on Earth, therefore highlighting the synergy between cosmological probes and traditional particle physics experiments.

As shown in Fig.~\ref{fig:lin vs nonlin derivatives}, the presence of extra radiation produces a change in the position of the BAO peaks and a damping of the matter power spectrum on small scales \cite{Bashinsky:2003tk}. The density perturbations of the new degrees of freedom also produce a phase shift in the oscillation scale of the baryon-photon fluid before recombination \cite{Bashinsky:2003tk}.
In the CMB power spectrum, the small-scale suppression due to extra relativistic species is largely determined by a shift in the damping scale $k_d$, which is sourced by a change in the early time Hubble parameter $H(z)$ through the change in $\rho_r$. As such, $N_\text{eff}$ is rather degenerate with the primordial helium abundance $Y_p$ in CMB data, and marginalizing over $Y_p$ can have a significant effect on $N_\text{eff}$ constraints, weakening them by up to a factor of three\footnote{In our forecasts $Y_p$ is treated as a derived parameter from the BBN consistency relation by default (see later in this section). The BBN calculation assumes a known value for the neutron lifetime and no other physics beyond the Standard Model. However, there is a current $4\,\sigma$ discrepancy between beam \cite{Nico:2004ie} and bottle \cite{Pattie:2017vsj} measurements of this lifetime. Thus it may be necessary to marginalize over $Y_p$ to obtain an unbiased measurement of $N_\text{eff}$.} \cite{Baumann:2015rya}. 
Note that this degradation is present despite the $N_\text{eff}$-induced phase shift in the CMB peaks location which is independent of $Y_p$ and therefore partly helps break this degeneracy, suggesting that the phase-shift alone is not playing a major role. 

By contrast, the growth of dark matter density fluctuations on scales probed by LSS experiments is sensitive to the modified expansion history introduced by new free streaming particles, and it is not affected by Silk damping. It is therefore largely insensitive to $Y_p$, which is thus not included in our analysis. The BAO phase shift is also not the major source of information on $N_{\rm eff}$ in galaxy clustering data \cite{Baumann:2017lmt}, whose constraining power is mostly driven by the change in the full shape of the power spectrum.

\begin{figure}[!h]
    \centering
    \includegraphics[width=0.95\linewidth]{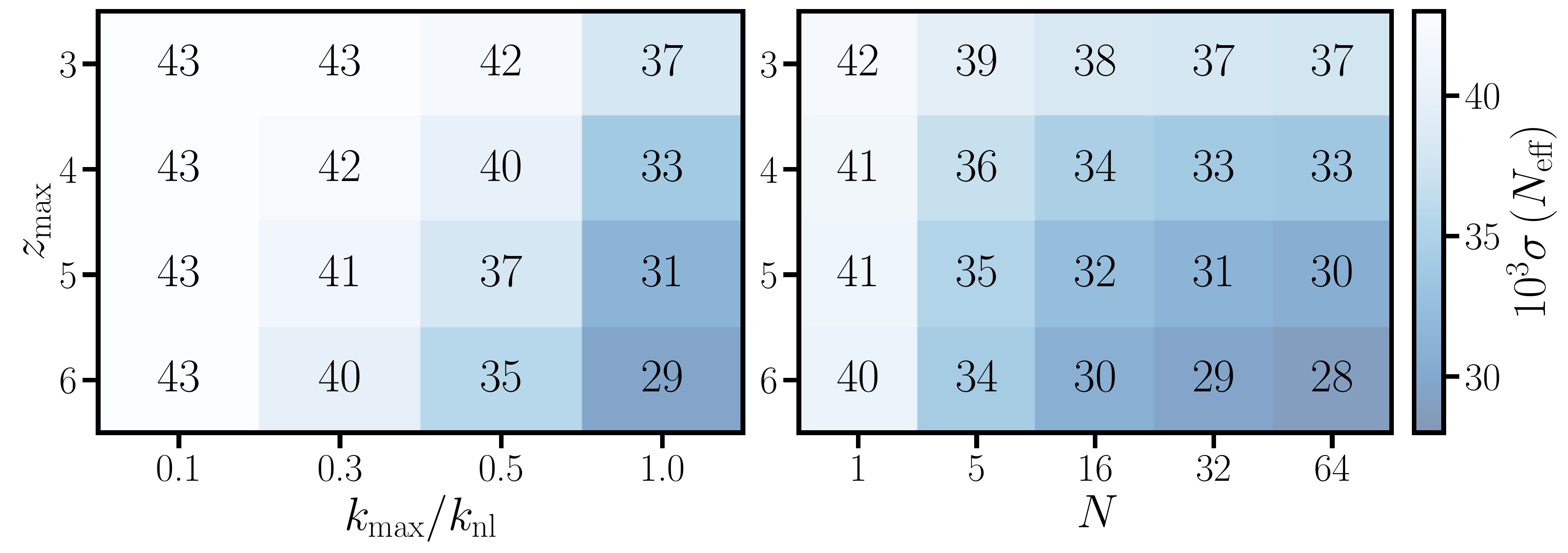}
    \caption{
    $N_\text{eff}$ constraints from Planck+SO (primary CMB) +DESI+PUMA-$N$K (full shape) for variable dish number $N$ and $z_\text{max}$. No lensing data is included in this figure. In the right plot we fix $k_\text{max}/k_\text{nl}=1$, while in the left plot we fix $N=32$. For reference, the constraint from Planck+SO (primary CMB) +DESI (full shape) is $0.044$.
    }
\label{fig:Neff}
\end{figure}

In our forecasts we marginalize over $\Lambda$CDM and the nuisance terms listed in Table~\ref{tab:marg_params}. Our results are summarized in Table~\ref{tab:big_summary} and Fig.~\ref{fig:Neff}. Near term LSS surveys (DESI, Euclid, MSE) will improve over the SO goal of $\sigma(N_\text{eff})=0.05$ by only $\sim 15\%$. The addition of high redshift data reduces this uncertainty to $0.03$, a $40\%$ improvement over Stage-III CMB measurements alone. With a Stage-IV CMB survey these constraints drop to $0.019$, or $0.018$ when a combination of high redshift LBG and 21-cm data are included. A measurement with uncertainty $\sigma(N_\text{eff})=0.018$ could detect light scalars (vectors) at $1.5\sigma$ ($3\sigma$) significance, regardless of their freeze-out temperature, and would be capable of detecting (or ruling out) all light relics with a freeze-out temperature smaller than the QCD phase transition ($\sim 0.2$ GeV, or $\Delta N_\text{eff}\sim 0.1$) at $>5\sigma$ significance \cite{CMBS4}. 

As shown in Fig.~\ref{fig:Neff} the constraints from a PUMA-$N$K survey plateau with detector number $N$ at around $N=32$, suggesting that PUMA-32K constraints are limited by the foreground removal of modes inside the wedge. From Table~\ref{tab:big_summary}, we see that both MegaMapper and PUMA-32K achieve similar constraints ($\sim 0.02$ with CMB-S4). Thus MegaMapper is also sufficient for achieving a nearly cosmic-variance limited $N_\text{eff}$ measurement from LSS. As one would expect, from Fig.~\ref{fig:Neff} we see that these constraints significantly depend on the $k_\text{max}$ assumed in the forecast, changing at the $\sim 25\%$ level between $0.1 < k_\text{max}/k_\text{nl}<1$. From Fig.~\ref{fig:Neff} we also see that decreasing PUMA's redshift range to $2<z<4$, which halves the survey volume, only increases the $N_\text{eff}$ constraint by $\sim 10\%$, which implies that most of the information is coming from the breaking of the degeneracies between cosmological parameters obtained by combining CMB with LSS data.

Our results are in good agreement with ref. \cite{collaboration2018inflation}, which show $N_{\rm eff}$ constraints dropping from 0.026 with CMB-S4 to 0.013 with CMB-S4 + PUMA-32K. Without lensing, our constraints are 0.019 for LSD + PUMA-32K. This apparent discrepancy can be entirely explained by differences in nuisance parameters. When non-linear nuisance parameters ($b_2,b_s,N_2,N_4,\alpha_2,\alpha_4$) are held fixed, we find $\sigma(N_\text{eff})\sim 0.012$, which is in excellent agreement with ref. \cite{collaboration2018inflation}. This suggests that improved models of galaxy formation, which could yield tight priors on non-linear nuisance parameters, would significantly improve a LSS measurement of $N_\text{eff}$, with the ``extra information'' primarily coming from small scales.

In our fiducial analysis we have fixed $Y_p$ as a function of $N_\text{eff}$ using the Big Bang Nucleosynthesis (BBN) consistency relation \cite{Simha:2008zj, Kneller:2004jz, Hou:2011ec}, rather than varying it independently. However, as an important check of the Standard Model of particle physics and cosmology, we can also treat them as independent quantities. In this case, marginalizing over $Y_p$ can significantly degrade the constraints on $N_\text{eff}$ from CMB surveys. 
For example, the constraints from Planck $+$ SO (LiteBIRD $+$ S4) increase from 0.047 (0.025) to 0.116 (0.077) with $Y_p$ marginalized over. 
When clustering data from high redshift surveys are included,
this degradation is reduced by $\sim 50\%$. 
For example, the constraint from PSD + PUMA-32K (LSD + PUMA-32K) only increases from 0.030 (0.019) to 0.048 (0.041). 
Since LSS is insensitive to $Y_p$, the resulting constraints with $Y_p$ marginalized over are primarily driven by LSS. 
These constraints are comparable to the SO goal $(0.05)$, suggesting that future high-redshift LSS surveys will have a similar constraining power to a Stage-III CMB survey, but with the added benefit of being immune to $Y_p$. 

\subsection{Primordial non-Gaussianity}
\label{sec:PNG}

In the simplest inflationary models one expands the action of a single scalar field minimally coupled to gravity about a spatially homogeneous solution. The leading order corrections to the action are quadratic in the small fluctuations about this solution. Thus these fluctuations are drawn from a Gaussian distribution to lowest order. Higher order terms in the action, arising from interactions of the inflaton with itself or other fields, result in trace amounts of Primordial non-Gaussianity (PNG). Of particular interest are local PNG, which vanish in the squeezed limit for slow-roll single field models of inflation \cite{Maldacena_2003,Creminelli04}.

The non-Gaussian contributions to the primordial potential $\Phi$ are commonly modeled as a perturbative expansion in a Gaussian random field $\phi$ \cite{Schmidt_2010}:
\begin{equation}
\Phi(\bm{x})
=
\phi(\bm{x})
+
f_\text{NL}
\int
d^3 y\, d^3 z
\ W(\bm{y},\bm{z})
\phi(\bm{x}+\bm{y})
\phi(\bm{x}+\bm{z})
+\mathcal{O}(\phi^3),
\label{eqn:png}
\end{equation}
where $W(\bm{y},\bm{z})$ is some kernel and $f_\text{NL}$ controls the amplitude of the non-Gaussianity. In this paper we specify to the local type of PNG in which $W(\bm{y},\bm{z}) = \delta^D(\bm{y})\delta^D(\bm{z})$. Currently the tightest constraints on the amplitude of local PNG $f^\text{Loc}_\text{NL} = -0.9 \pm 5.1$ come from the Planck satellite \cite{planckcollaboration2019planck}, which are within a factor of $3$ of a cosmic-variance limited CMB survey \cite{Dore14}. LSS surveys are not yet competitive with the CMB, with current constraints around  $\sigma(f^\text{Loc}_\text{NL})\sim 25$ \cite{Castorina2019}. Pushing beyond $\sigma(f^\text{Loc}_\text{NL})\sim1$ requires the addition of future 3D LSS data,  which has access to significantly more modes than the 2D CMB on large scales. 

\begin{figure}[!h]
\centering
\includegraphics[width=\linewidth]{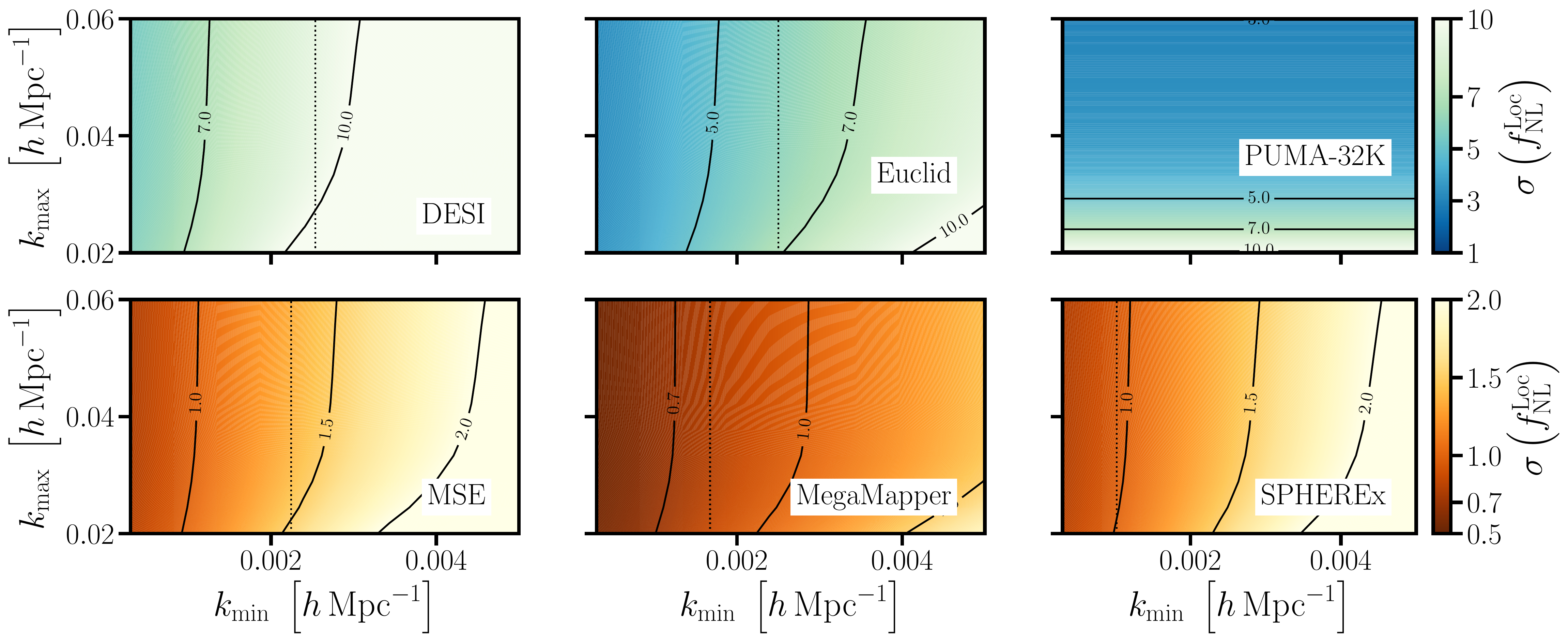}
\caption{ Constraints on local primordial non-Gaussianity as a function of $k_\text{min}$ and $k_\text{max}$, using the full shape power spectrum without any CMB priors. For SPHEREx we set $\sigma_z/(1+z)=0.05$, while for all other surveys we set $\sigma_z=0$. The vertical lines are the lowest possible $k_\text{min}$ for each survey: $2\pi / V^{1/3}$, where $V$ is the volume of the survey (including sky coverage).
}
\label{fig:png}
\end{figure}

In addition to modifying the Universe's initial conditions (e.g.\ producing a non-zero bispectrum) PNG modifies the non-linear evolution of structure growth (producing new mode-couplings), which can in principle be modeled using standard perturbative techniques. However, the development of a PT-based code which self-consistently models non-linearities (in redshift space) in the presence of PNG is beyond the scope of this paper. We instead focus on PNG's effect on the linear bias, and restrict our forecasts to $k_\text{max}<0.06\,\,h$Mpc$^{-1}$ where linear theory is a good  approximation for the high redshifts we consider. As seen in Eq.~\ref{eqn:png}, PNG couples short wavelength modes of $\Phi$ to long wavelength modes of $\phi$. This coupling induces an additional scale-dependent bias ($b\to b+\Delta b$) which takes the form \cite{Slosar:2008}:
\begin{equation}
    \Delta b 
    \equiv
    \frac{3\Omega_m H_0^2}{2k^2 D(z) T(k)} b_\phi f^\text{Loc}_\text{NL} =  \frac{3\Omega_m \delta_c H_0^2}{ k^2 D(z) T(k)}  (b-1)f^\text{Loc}_\text{NL},
\end{equation}
where $D(z)$ is normalized to $a$ in the matter dominated era and $\delta_c=1.686$. In the above equation we made the assumption that $b_\phi = 2 \delta_c(b-1)$ \cite{Slosar:2008,Biagetti:2019bnp}, which does not necessarily hold for the galaxies that will be observed by current and future LSS surveys \cite{Biagetti2016,Barreira2020,MoradinezhadDizgah2020}, and it cautions against combining the constraints from different instruments.

Shown in Fig.~\ref{fig:png} are our forecasted constraints on $f^\text{Loc}_\text{NL}$ from a variety of upcoming or proposed LSS surveys, using the power spectrum only. Since all information about $f^\text{Loc}_\text{NL}$ comes from low $k$ where linear theory is an excellent approximation, in our forecasts we only marginalize over $\Lambda$CDM, $b$ and $T_b$ (for 21-cm surveys\footnote{See ref.~\cite{Cunnington:2020wdu} for a detailed discussion of degeneracies between 21-cm foregrounds and $f^\text{Loc}_\text{NL}$.}). For all surveys we include a prior on $\Lambda$CDM from Planck. 
Even with just the power spectrum, we see that near-term LSS surveys (such as Euclid) have the capability of achieving similar constraints on PNG as Planck, while a future high-redshift LBG survey has the potential to achieve $\sigma(f^\text{Loc}_\text{NL})\sim 0.8$, more than a factor of $2$ improvement over a cosmic variance limited CMB survey. Similar constraints from the power spectrum should be obtainable with the SPHEREx mission \cite{Dore14}.

Here we have only focused on local PNG from the power spectrum, making use of the $f^\text{Loc}_\text{NL}$ induced scale-dependence of linear bias.  We note that adding bispectrum or higher point function information can potentially help \cite{dePutter:2018jqk,Karagiannis:2018jdt,MoradinezhadDizgah2020}. Extracting bispectrum information will be crucial for constraining shapes of PNG beyond the local type considered here (such as equilateral or orthogonal, for example), where the effects on the power spectrum are largely degenerate with non linear evolution \cite{Gleyzes_2017}. 

Finally we note that the signatures of a non-zero $f^\text{Loc}_\text{NL}$ appear on very large scales, which are often most affected by systematics (especially angular measurements) \cite{Kitanidis:2019rzi,2013PASP..125..705P,Leistedt:2014zqa,Agarwal:2013ajb,2013MNRAS.428.1116R,Weaverdyck:2020mff}.
In Section \ref{sec:issues} we explore deprojecting angular modes by introducing a minimum $k_\parallel$ cut. Alternatively, one can use the fact that most galaxy imaging systematics \cite{2013MNRAS.432.2945H} are uncorrelated with CMB lensing maps, and measure the scale dependent bias by cross-correlating the galaxy positions with CMB lensing. Ref.~\cite{Schmittfull18} found that $\sigma(f^\text{Loc}_\text{NL}) \lesssim 1$ can be achieved by future surveys using this technique, thus providing an important independent check on systematics.
\subsection{Primordial features}
\label{sec:primordial_features}

Current observations of the CMB are consistent with the primordial potential having a nearly scale-invariant power spectrum. This is a generic prediction of single-field slow-roll
inflation, sourced by the (near) flatness of the inflationary potential \cite{Baumann09}. 
However, attempts to connect this simple picture of inflation with more fundamental ultra-violet scenarios often result in deviation from scale invariance on some scales \cite{Slosar19c}. In this paper we focus on a subclass of these features (sharp features), which are commonly modeled as a step in the inflationary potential \cite{Adams_2001}. This step produces a scale-dependent oscillatory feature in the primordial power spectrum, which we model as a sinusoidal modulation of the linear power spectrum: $P_m(k)\to [1+A_\text{lin}\sin(\omega_\text{lin}k+\phi_\text{lin})]P_m(k)$ \cite{Ballardini:2016hpi,Beutler_2019,PlanckInf18}. 

\begin{figure}[!h]
\centering
\includegraphics[width=0.7\linewidth]{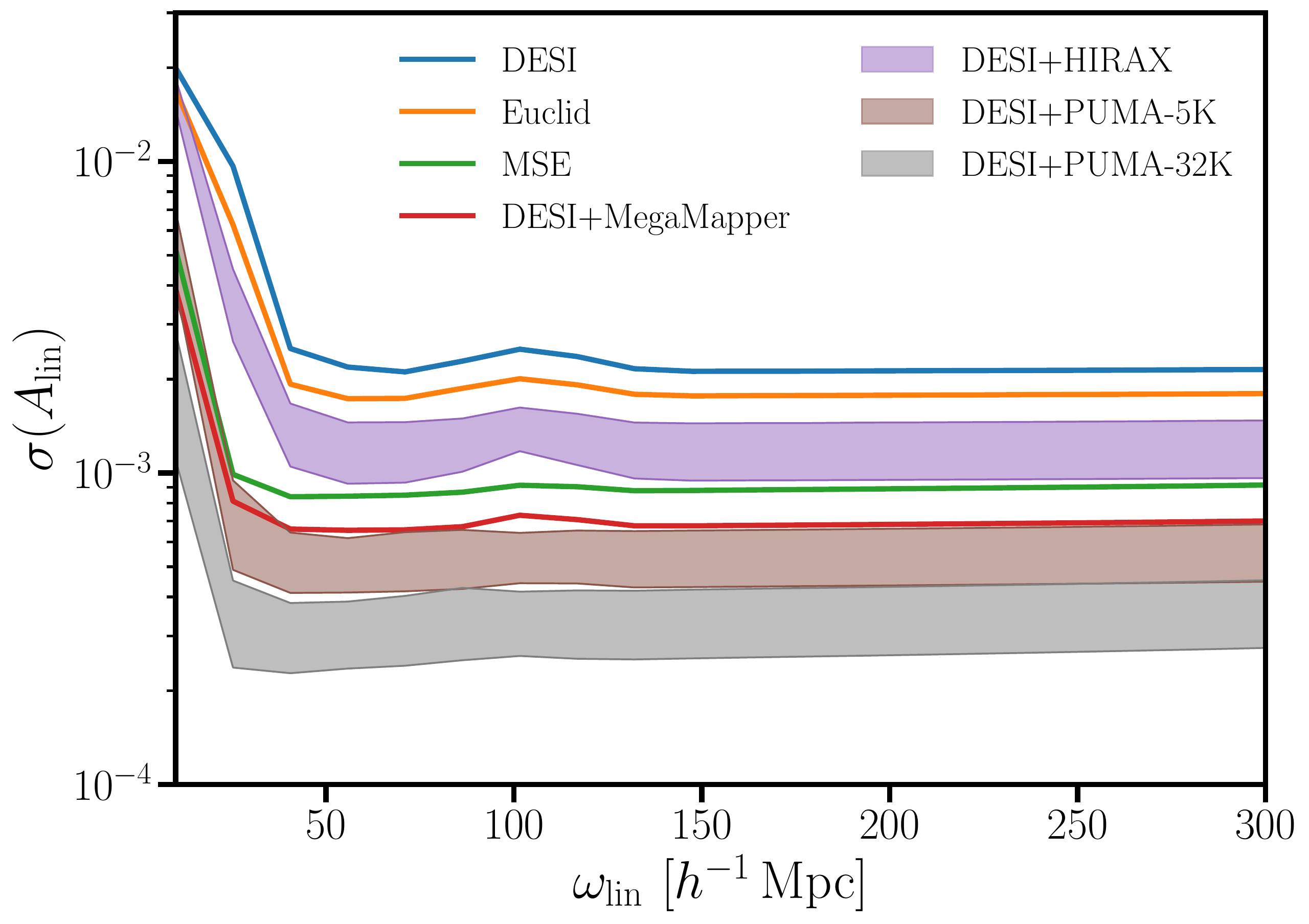}
\caption{
Constraints on $A_\text{lin}$ for various fiducial values of $\omega_\text{lin}$, assuming $\phi_\text{lin}=\pi/2$. For all surveys we include a Planck+SO prior on $\Lambda$CDM. The slight ``bump'' in the constraints at $\omega_\text{lin}\sim 100\,\,h^{-1}{\rm Mpc}$ is due to the degeneracy with the BAO frequency. 
}
\label{fig:primodial_features}
\end{figure}

In our forecasts we fix both $\omega_\text{lin}$ and $\phi_\text{lin}$, and find the corresponding uncertainty on $A_\text{lin}$ after marginalizing over the parameters in Table~\ref{tab:marg_params}. Our results are shown in Fig.~\ref{fig:primodial_features} for $10\,\,h^{-1}\text{Mpc}<\omega_\text{lin} < 300\,\,h^{-1}\text{Mpc}$, assuming $k_\text{max} = k_\text{nl}$\footnote{Since the high frequency oscillatory feature induced by primordial features is immune to small-scale non-linearities \cite{Beutler_2019} up to an overall damping of their amplitude \cite{Ballardini:2019tuc}, one could in principle extend the $k_\text{max}$ used in an analysis beyond the nonlinear scale. In practice, however, we find that both the MegaMapper and PUMA-32K constraints improve by at most $2\%$ when $k_\text{max}$ is increased from $k_\text{nl}$ to $1.4\,k_\text{nl}$. For MegaMapper this is mostly due to the large value of the shot noise, which limits the ability of reconstructing oscillatory features on small scales.}. We find that near-term spectroscopic surveys will be capable of achieving $\sigma(A_\text{lin}) \sim 2\times10^{-3}$, while HIRAX has the capability to improve upon these constraints by a factor of two. Order of magnitude improvement are possible by extending out to high redshift, with an optimistic configuration of PUMA-32K achieving $\sigma(A_\text{lin}) \sim 2\times10^{-4}$. 

These results are largely consistent with ref.~\cite{Beutler_2019}, who found $\sigma(A_\text{lin})\simeq 2\times10^{-3}, 1.5\times 10^{-3},$ and $10^{-4}$ for DESI, Euclid, and a cosmic variance limited LSS survey (out to $z=6$) respectively. Our results for DESI and Euclid are within $\sim25\%$ of these values, which is entirely reasonable given that ref.~\cite{Beutler_2019} uses a different forecasting approach in which $ P^\text{w}_{gg}(k)/P^\text{nw}_{gg}(k)$ is the observable of interest. This suggests that PUMA-32K (with optimistic foreground assumptions) is within a factor of two of a cosmic variance limited survey. We note that the slight ``bump'' in our constraints around $\omega_\text{lin}\sim 100\,\,h^{-1}{\rm Mpc}$ is smaller than the bump found by ref.~\cite{Beutler_2019}, which is due to a degeneracy with the BAO frequency. This suggests that the full shape information is sufficient to break the degeneracy between primordial features and the BAO feature. 

\subsection{Dynamical dark energy}
\label{sec:EDE}

Models of dark energy are typically invoked to explain the observed accelerated expansion of the Universe at the present epoch. Current observations are consistent with the existence of a cosmological constant $\Lambda$. In this section we quantify the sensitivity of future surveys to departures from the $\Lambda$-model, using both a ``model-independent'' approach\footnote{Ref. \cite{EricLinderRise} performs a similar analysis to estimate constraints on $\Omega_\text{DE}(z)$, focusing on the redshift interval $1.5<z<2.5$.}, and through constraints on a specific model of dynamical dark energy.

\begin{figure}[!h]
\centering
\includegraphics[width=\linewidth]{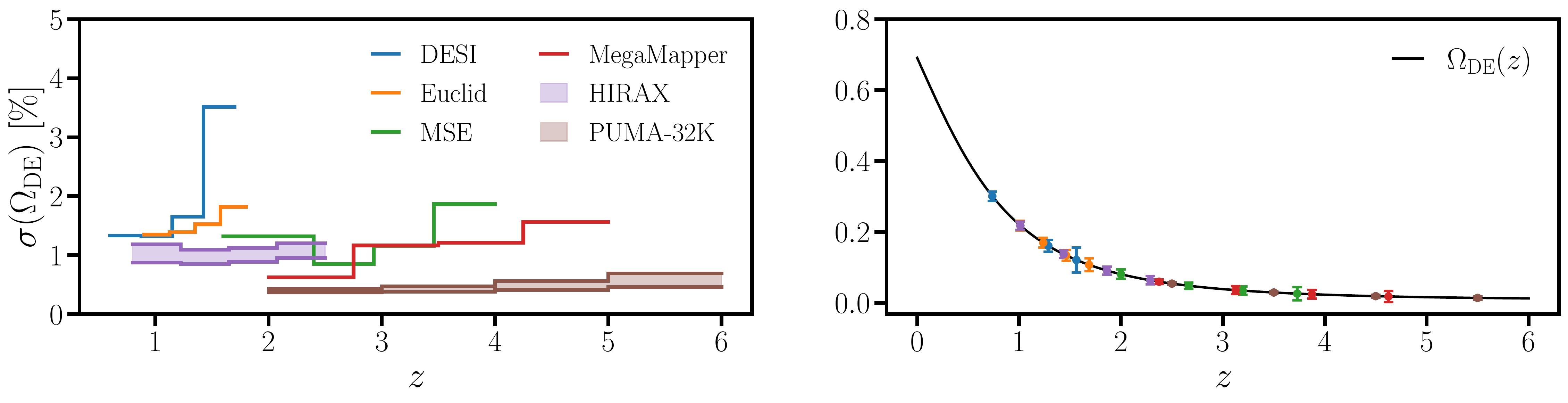}
\caption{The absolute error on the dark energy fraction ($\Omega_\text{DE}$) achievable by current and future LSS surveys.    
}
\label{fig:Omega_DE}
\end{figure}

Over the redshift range covered by our observations (typically up to $z \sim 5-6$), direct expansion and growth measurements can constrain any dynamical dark energy model by quantifying the fraction of dark energy in a number of redshift bins as we now explain. 
At redshifts $z<6$ the Universe is dominated by nonrelativstic matter and dark energy, so that the first Friedmann equation is approximately given by $\Omega_\text{DE}(z) + \Omega_m(z) = 1$, where $\Omega_m(z) = \rho_m(z)/\rho_\text{crit}(z) = \omega_m(1+z)^3/h^2(z)$ and $h(z)$ is defined through $H(z) = 100 \,h(z)$ km/s/Mpc. We estimate the error on the dark energy fraction via standard error propagation, and neglect the covariance between $h(z)$ and $\omega_m$, so that 
$\sigma^2_{\Omega_\text{DE}(z)} = \Omega^2_m(z)(4 [\sigma_{h(z)}/h(z)]^2  + [\sigma_{\omega_m}/\omega_m]^2)$. 
We take the relative error on $h(z)$ from the radial BAO measurements $(\sigma_{h(z)}/h(z) = \sigma_{\alpha_\parallel(z)})$, which are described in sections \ref{sec:bao} and \ref{sec:distance_measures}. We assume that the measurement of $\omega_m$ is driven primarily from CMB measurements, and take $\sigma_{\omega_m} = 2.4\times 10^{-4}$ for all surveys, motivated by the constraints achievable by LSD + MegaMapper (see Table~\ref{tab:big_summary}). Our results are shown in Fig.~\ref{fig:Omega_DE}. We find that MegaMapper will be able to constrain the dark energy fraction to better than $2\%$ across its redshift range, while PUMA-32K could improve these measurements to the sub-percent level. 

While the simplest models have $\Omega_{\rm DE}$ drop rapidly at higher redshift, so that dark energy is a late-time phenomenon, there is little guidance from theory as to how $\rho_{\rm DE}$ should evolve, and the presence of an inflationary epoch in the very early Universe calls into question whether there should be only one late-time epoch of dark energy domination.  Recently, early dark energy (EDE) has been invoked to decrease the CMB sound horizon via an increase in the expansion rate before recombination.  Doing so would increase the inferred value of $H_0$ from CMB observations, making EDE a possible candidate for resolving the $H_0$ tension \cite{Efstathiou20}.
Current constraints from a combination of CMB, SNe and LSS data tend to disfavor this explanation \cite{Hill20,Ivanov20,DAmico20b,Klypin20}, with much of the constraint coming from EDE's effects on the growth of large-scale structure.

\begin{figure}[!h]
\centering
\includegraphics[width=0.95\linewidth]{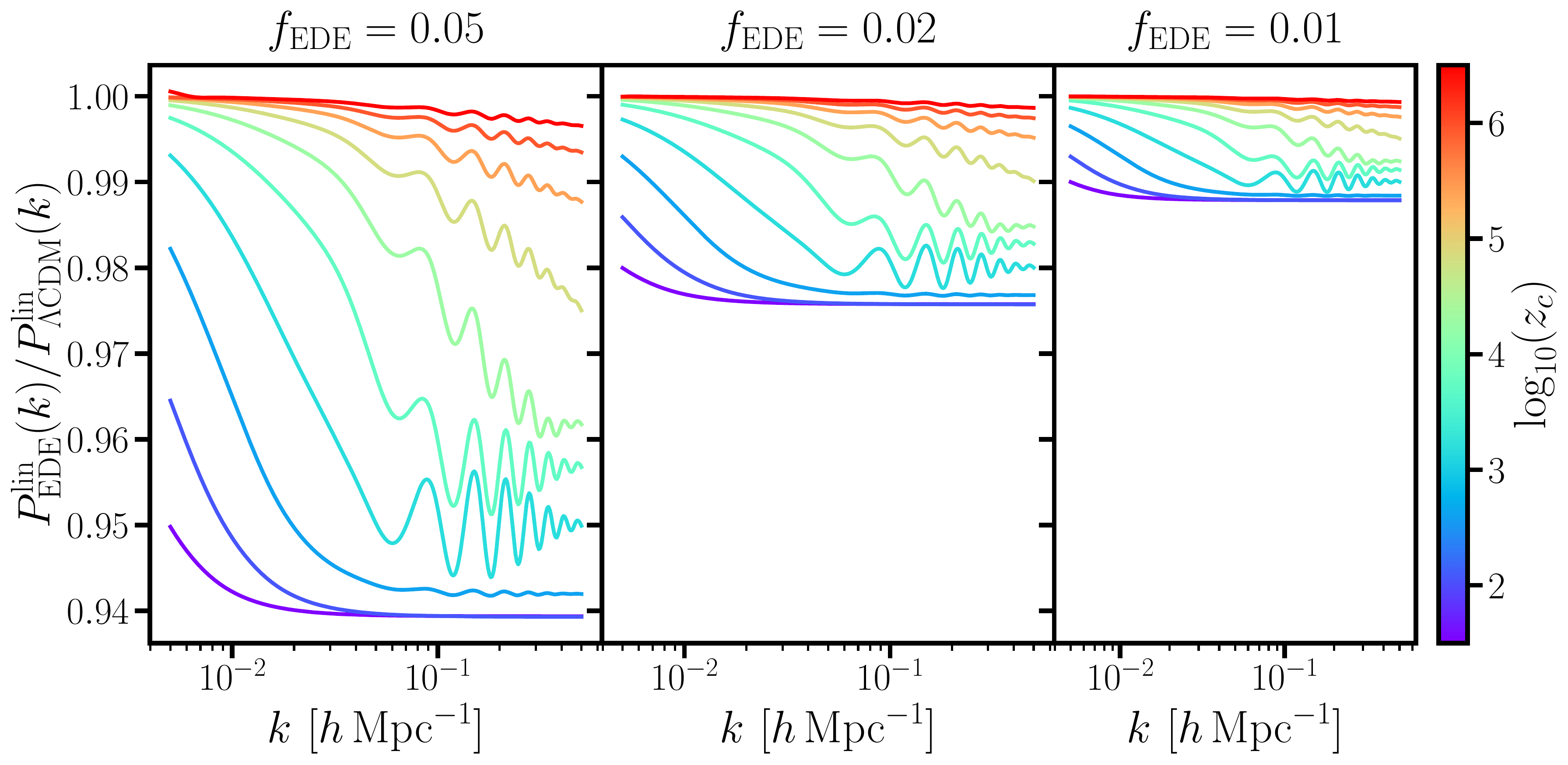}
\caption{Change in the linear matter power spectrum (at $z=2$) when including Early Dark Energy (EDE) for various values of $f_\text{EDE}$ and $1.5<\log_{10}(z_c)<6.5$. $\theta_i=2.83$, and all standard cosmological parameters are held at their fiducial $\Lambda$CDM values. For $z_c\lesssim 30$, the ratio $P^\text{lin}_\text{EDE}(k)/P^\text{lin}_{\Lambda\text{CDM}}(k)$ is roughly given by $1-f_\text{EDE}$.
}
\label{fig:percent_change_in_pk_from_EDE}
\end{figure}

This effect is much more general than the above-mentioned example however, and future LSS surveys can surprisingly put very tight constraints on EDE models at redshift much higher than the range covered by the observations, thanks to the exquisite precision with which they measure the shape of the power spectrum.  Since the growth of structure is damped by Hubble expansion, long periods where unclustered species dominate the expansion lead to suppression of large-scale structure that can be detected by comparing early- and late-time measures of the fluctuation amplitude.  A precise measurement of the power spectrum shape can also be used to place constraints on (or detect) short-term deviations from matter or radiation domination, since such periods will change the shape of the power spectrum due to differential growth of modes.  An accurate measurement of the power spectrum constrains deviations in the expansion history over a broad range of redshifts \cite{Chen20b}.

\begin{figure}[!h]
\centering
\includegraphics[width=0.9\linewidth]{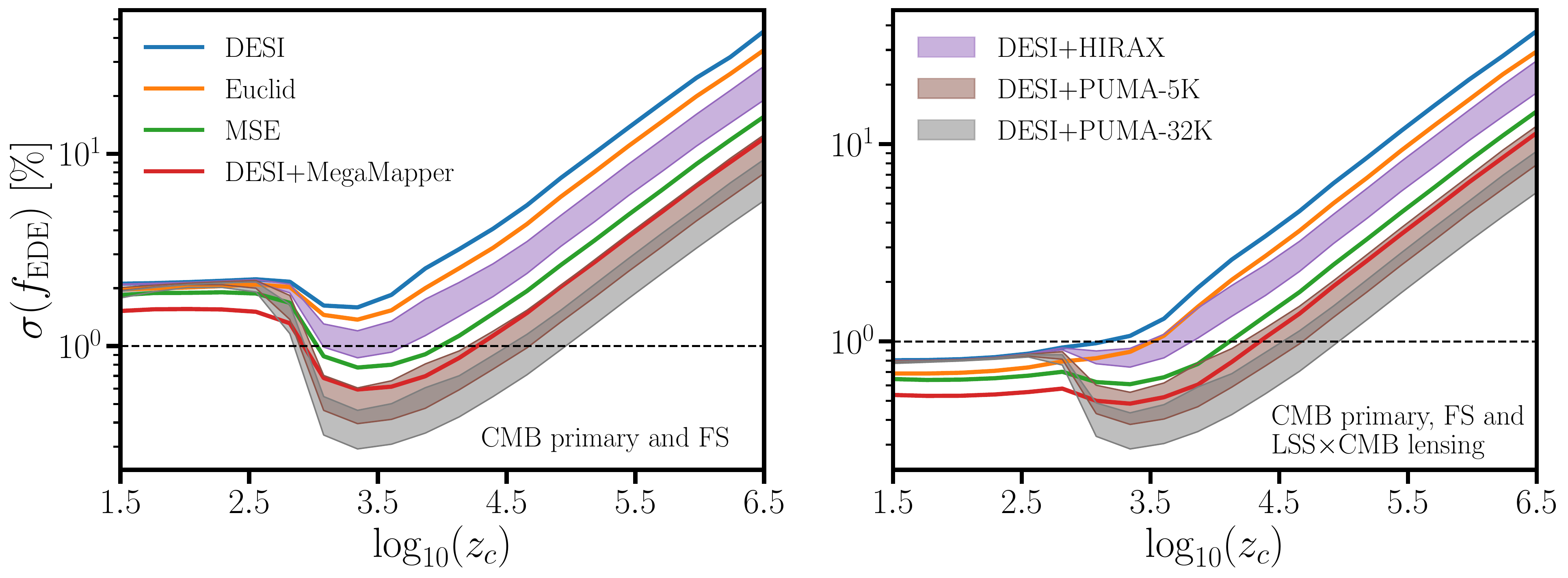}
\caption{
Constraints on the maximum amplitude of early dark energy ($f_\text{EDE}$) as a function of the time at which EDE peaks $z_c$, assuming $\theta_i=2.83$. We include a Planck+SO prior on $\Lambda$CDM for all experiments. In the left panel we show constraints from full shape (FS) measurements only, while in the right panel we include a prior on $\Lambda$CDM and the nuisance parameters in Table~\ref{tab:marg_params} from SO lensing and cross-correlations with the respective galaxy surveys, following the convention in \S\ref{sec:combined fisher}.
}
\label{fig:EDE_constraints}
\end{figure}

We give an illustrative example in Fig.~\ref{fig:percent_change_in_pk_from_EDE}, which considers the EDE scenario of ref.~\cite{Poulin_2019}, in which a light scalar $\phi$ with potential $V(\phi)\propto [1-\cos(\phi/f)]^3$ is initially displaced from its minimum. The field is parameterized by its maximum energy contribution, $f_\text{EDE}\equiv \text{max}[\rho_\text{EDE}(z)/\rho_\text{tot}(z)]$, which occurs at redshift $z_c$, and the initial value of the field $\theta_i \equiv \phi_i/f$.  Values of $z_c\gtrsim z_\star$, where $z_\star\sim1100$ is recombination, would help to relieve the Hubble tension, but we show that future large-scale structure surveys can probe even small values of $f_\text{EDE}$ over a very broad range of redshifts.   Using the publicly available code \verb|CLASS_EDE| \cite{Hill20} to produce the initial power spectra and \verb|velocileptors| to treat the damping of features from non-linear evolution and biasing \cite{Chen20b}, we forecast constraints on $f_\text{EDE}$ for values of $z_c$ ranging from $\sim 30$ to $10^6$. In all forecasts $z_c$ and $\theta_i=2.83$ are held fixed, and we marginalize over the parameters listed in Table~\ref{tab:marg_params}. The results are shown in Fig.~\ref{fig:EDE_constraints}. 
  
We see that near-term spectroscopic surveys are capable of constraining the amplitude of early dark energy at the few percent level out to redshift $10^4$, while PUMA would be capable of achieving subpercent level constraints out to redshift $10^5$. These are order of magnitude improvements over the $\mathcal{O}(10\%)$ constraints found by refs.\ \cite{smith2020early, Hill20} from a combination of current BOSS and Planck data. As shown in Fig.~\ref{fig:percent_change_in_pk_from_EDE}, EDE that peaks prior to recombination produces a phase shift in the BAO feature. From Fig.~\ref{fig:percent_change_in_pk_from_EDE} we see that this effect peaks at around $z_c\sim 10^{3.5}$, which sensibly corresponds to the minimum in Fig.~\ref{fig:EDE_constraints}. However, EDE that peaks after recombination induces no phase shift in the BAO signature. In this case, EDE damps the linear power spectrum in a scale-independent way (for $k > 0.01\,\, h\text{Mpc}^{-1}$) that is nearly degenerate with $A_s,\,b$ and $T_b$. Thus all of the information about post-recombination EDE comes from low $k$. This puts 21-cm observations at a disadvantage, as low $k$ modes are excluded by the foreground wedge. We see this effect in Fig.~\ref{fig:EDE_constraints}. PUMA-32K offers no improvement over DESI for $z_c<z_\star$, while MegaMapper improves over DESI by more than a factor of two. We note that constraints in the $z_c<z_\star$ regime benefit significantly from the addition of lensing cross-correlations (MegaMapper constraints decrease by more than a factor of $\sim2$), which soften the $f_\text{EDE}-b$ degeneracy. 

\subsection{Modified gravity}
\label{sec:MG}

Modifications to GR have been proposed as an alternative explanation for the observed accelerated expansion of the Universe at low redshift \cite{Jain10,Joyce16,EUCLID18,Slosar19c}.  The phenomenology of such models can be very rich, but a common ingredient is that the responses of the Newtonian potential $\Psi$ and spatial curvature $\Phi$ to matter and energy (which are equal in the low redshift Universe in GR) can differ. 
 
To estimate the sensitivity of future surveys to deviations from GR we examine the gravitational slip parameter $\gamma = \Phi/\Psi$ \cite{Clifton_2012,Will14,Joyce15,Joyce16,EUCLID18,Slosar19c}. We follow the method of ref.~\cite{Chen19a}, treating $\gamma$ as a scale-independent factor, and consider a scenario in which Poisson's equation for $\Phi$ is modified but $\Psi$ is left unchanged. This amounts to making the substitution $\delta_\kappa  = c_\kappa \delta_m$ in Eq.~\ref{eqn:Limber}, where $c_\kappa = (1+\gamma)/2$, since lensing probes the sum of the two potentials $\Phi+\Psi$. We approximate $c_\kappa$ as a piecewise constant linear Eulerian bias, whose value in each redshift bin is treated as an independent parameter.   

\begin{figure}[!h]
\centering
\includegraphics[width=0.6\linewidth]{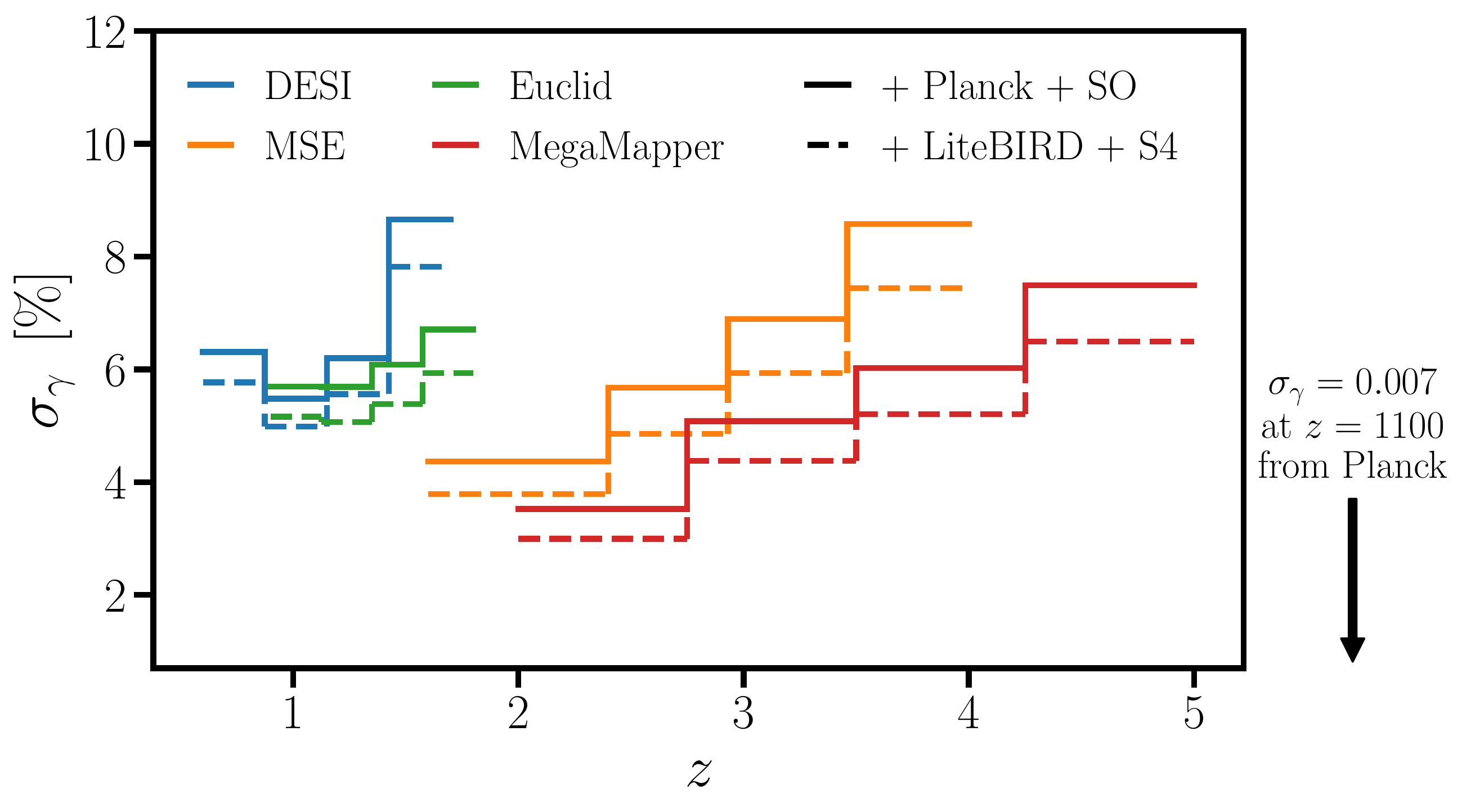}
\caption{
Error on the gravitational slip parameter $\gamma$ from LSS$\times$CMB lensing cross-correlations, LSS full shape data, and the primary CMB. For comparison, we show the constraint on $\gamma$ from Planck \cite{PlanckLegacy18} at $z\approx 1100$ (recombination) to the right of the plot.
}
\label{fig:gravitational_slip}
\end{figure}

In Fig.~\ref{fig:gravitational_slip} we show constraints on $\gamma_i$, where $\gamma_i$ is the value of the gravitational slip in the $i^{\rm th}$ redshift bin. In these constraints we include LSS$\times$CMB lensing, full shape clustering data, the primary CMB, and marginalize over all parameters listed in Table~\ref{tab:marg_params}. The direct constraints on $\gamma_i$ come only from $C^{\kappa\kappa}_\ell$ and $C^{\kappa g_i}_\ell$, while the full shape and primary CMB data serve as priors on $\Lambda$CDM and the nuisance parameters. Thus the constraints in Fig.~\ref{fig:gravitational_slip} should not be interpreted as independent measurements. Instead, they should be interpreted as a survey's sensitivity to $\gamma_i$ as measured by the full dataset (lensing cross-correlations and full shape clustering across the entire redshift range of the survey), assuming $\Lambda$CDM as the fiducial cosmology.

From Fig.~\ref{fig:gravitational_slip} we see that spectroscopic surveys are capable of measuring $\gamma(z)$ to better than $10\%$ out to redshift $6$ with current CMB data. The improvement from upgrading to LiteBIRD $+$ CMB-S4 from Planck $+$ SO is $\sim 15\%$ for all LSS surveys at all redshifts of interest. We see that MegaMapper improves over MSE by $20 - 40\%$ across $2<z<4$. These results are in rough agreement with ref.~\cite{Chen19a}, who considered an LBG survey covering $1000$ square degrees and found $0.02 <\sigma_\gamma < 0.20$ for $2<z<5$ after marginalizing over $\sigma_8$ and $b_\text{LBG}$. If we only marginalize over $\ln(A_s)$ and $b_\text{LBG}$, and reduce MegaMapper's sky coverage to $f_\text{sky}=0.024$, mimicking the approach of ref.~\cite{Chen19a}, we find $0.09 <\sigma_\gamma < 0.23$ for $2<z<5$. The disagreement at low redshift is likely due to different binning schemes (we use four redshift bins, while ref.~\cite{Chen19a} uses three), and differences in the assumed LBG number densities.

\subsection{Extended base model and multi-extension fits}
\label{sec:Extending_the_base_model}

Our discussion so far focused on one parameter extensions of our base $\Lambda$CDM model. We now investigate how our constraints degrade as we extend our base model by a single parameter, or as we simultaneously vary multiple extensions.

We first consider the constraints on the $\Lambda$CDM parameters when the base model is extended by a single parameter. In Table~\ref{tab:extended_base_model} we show the percent increase to PSD + MegaMapper (using full shape data only, without lensing) constraints as we extend the base model to include $M_\nu,\,N_\text{eff},\,\Omega_k$ or $f_\text{EDE}$ (with $\log_{10}(z_c)=3.61$). In particular, we find that constraints on $h$, $\omega_c$ and $\omega_b$ worsen by roughly a factor of $2$ when $N_\text{eff}$ is allowed to vary. When the base model is extended to include $M_\nu$ or $\Omega_k$, all $\Lambda$CDM constraints increase by less than $40\%$. Allowing $f_\text{EDE}$ to vary, however, has a negligible (few percent) impact on $\Lambda$CDM constraints. 

In Table~\ref{tab:extended_base_model} we also consider the impact of extending the base model on several one-parameter extensions. In particular, we find that $\{M_\nu,\,N_\text{eff},\,\Omega_k\}$ are not internally degenerate $-$ allowing any two of these three extensions to vary simultaneously worsens the constraints by less than $17\%$. We find a mild degeneracy between $f_\text{EDE}$ and $\Omega_k$; varying the two simultaneously worsens the constraints on both parameters by $\approx 30\%$.

\begin{table}[!h]
    \centering
    \begin{tabular}{c|cccccc|cccc}
         Extension & $h$ & $\ln(A_s)$ & $n_s$ & $\omega_c$ & $\omega_b$ & $\tau$ & $M_\nu$ & $N_\text{eff}$ & $\Omega_k$ & $f_\text{EDE}$\\
         \hline
         \hline 
         $M_\nu$ & 39.7 & 29.5 & 2.7 & 26.6 & 0.0 & 35.3                & -- & 11.5 & 12.5 & 16.2\\
         $N_\text{eff}$ & 138.0 & 0.0 &  46.0 &  67.0 &  73.0 & 1.0     & 11.5 & -- & 0.3 & 8.5\\
         $\Omega_k$ & 23.4 & 1.3 & 12.4 & 32.6 &  0.0 &  3.4            & 12.5 & 0.3 & -- & 31.8\\
         $f_\text{EDE}$ & 0.4 & 1.6 & 4.4 & 0.8 & 3.4 & 2.5             & 16.2 & 8.5 & 31.8 & -- \\
         \hline
    \end{tabular}
    \caption{Percent increase to PSD $+$ MegaMapper constraints on $\Lambda$CDM (columns 2-7) and several one-parameter extensions (columns 8-11) when the base model is extended by a single parameter. The constraint on $f_\text{EDE}$ is for fixed $\log_{10}(z_c)=3.61$ and $\theta_{i}=2.83$.}
    \label{tab:extended_base_model}
\end{table}

\begin{figure}[!h]
\centering
\includegraphics[width=\linewidth]{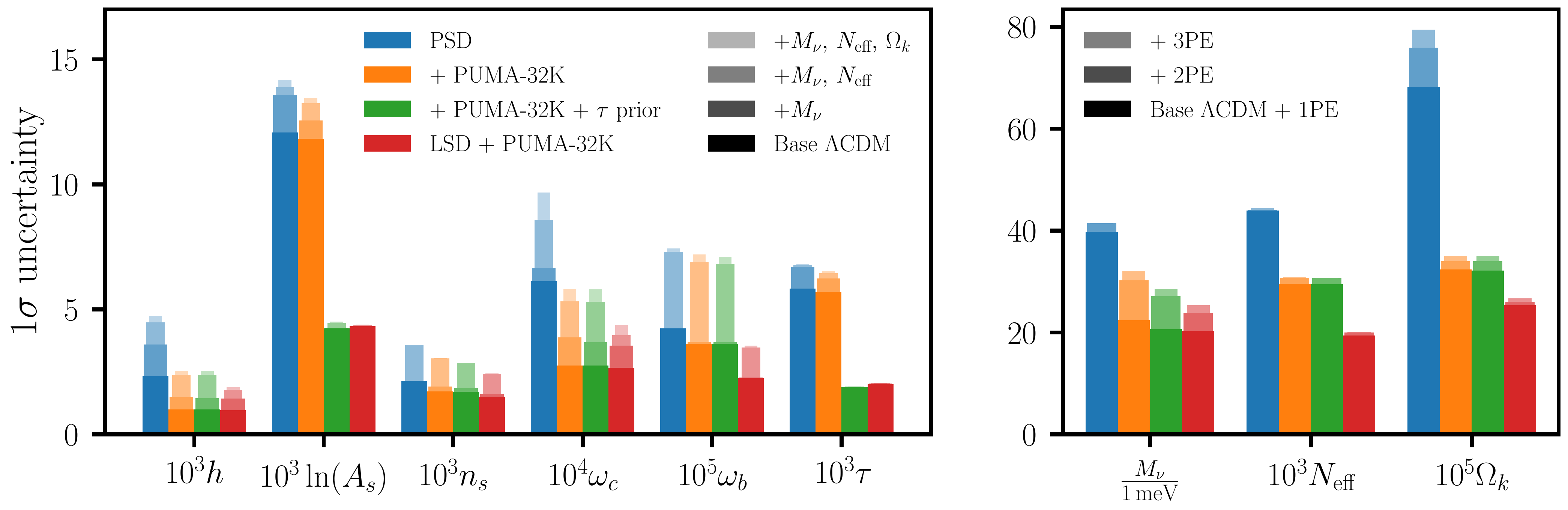}
\caption{Constraints on $\Lambda$CDM and extensions from four different combinations of experiments, using primary CMB and clustering data only. We include a $\sigma(\tau)=0.002$ prior for the green bars. \textit{Left}: The shading/width denotes which parameters are allowed to vary when fitting for $\Lambda$CDM (e.g.\ for Base $\Lambda$CDM+$M_\nu$ we keep $N_\text{eff}\,\text{and}\,\Omega_k$ fixed).
\textit{Right}: The shading denotes the number $N$ of Parameter Extensions ($N$PE) that are simultaneously allowed to vary in our fits, starting with the relevant parameter on the $x$-axis and continuing along the cycle $(M_\nu,N_\text{eff},\Omega_k)$. E.g.\ for $\Omega_k$, $2$PE is equivalent to $\Omega_k,\,M_\nu$. 
}
\label{fig:extensions_to_base_model}
\end{figure}

In Fig.~\ref{fig:extensions_to_base_model} we consider changes in our constraints as we vary more than one extension simultaneously. The blue bars in the left panels of Fig.~\ref{fig:extensions_to_base_model} shows the $1\,\sigma$ uncertainty on the $\Lambda$CDM parameters from PSD, which we compare to PSD + PUMA-32K, shown in orange (the analogous figure for MegaMapper looks nearly identical). Allowing for multiple extensions of the baseline model results in an overall degradation of the constraints, as shown by the different shading of the blue and orange bars. When the neutrino mass is marginalized over, this effect is the strongest on $h$, $A_s$ and $\tau$. This trend increases when simultaneously fitting for $N_{\rm eff}$ and $\Omega_k$. The net improvement over PSD brought by future surveys is also reduced as more extensions are added to the base $\Lambda$CDM model, indicating that the set of observables we consider is not powerful enough to break the new degeneracies among parameters. Fitting to more than one extension to $\Lambda$CDM simultaneously without significant degradation to $\Lambda$CDM requires next generation space-based CMB surveys, such as LiteBIRD, or other measurements which are capable of constraining the primordial amplitude and optical depth to extremely high precision. These constraints are shown by the red and green bars respectively. 
As seen in Fig.~\ref{fig:extensions_to_base_model}, the constraints on $\tau$ and $A_s$ from next generation CMB surveys are negligibly impacted when simultaneously fitting for multiple extensions. Thus no additional penalty is incurred to the remaining four $\Lambda$CDM parameters from the would-be worsened degeneracies with $\tau$ and $A_s$. 
 
On the other hand, the constraints to our selected extensions to $\Lambda$CDM (right panel of Fig.~\ref{fig:extensions_to_base_model}) remain relatively unchanged while fitting to more than one extension. Just as before (Table~\ref{tab:extended_base_model}), this suggests that $\{M_\nu, N_\text{eff}, \Omega_k\}$ are not internally degenerate.

\subsection{Linear vs. non-linear modeling}
\label{sec:lin vs non lin}
Currently, the vast majority of forecasts assume linear theory and scale-independent bias (though see \cite{Chudaykin19,Boyle20,chen2021precise} for notable exceptions), and model the effects of FoG-like contributions with a specific functional form (typically a Gaussian or Lorentzian). That is, the typical linear model (LM) for the redshift-space galaxy power spectrum takes the form \cite{White15}:
\begin{equation}
    P^\text{LM}_{gg}(\bm{k}) = \frac{(b+f\mu^2)^2}{1 + \sigma_v^2 (k \mu)^2 }P^\text{lin}(k) + N_0.
\label{eqn:linear_model}
\end{equation}

\begin{figure}[!h]
\centering
\includegraphics[width=\linewidth]{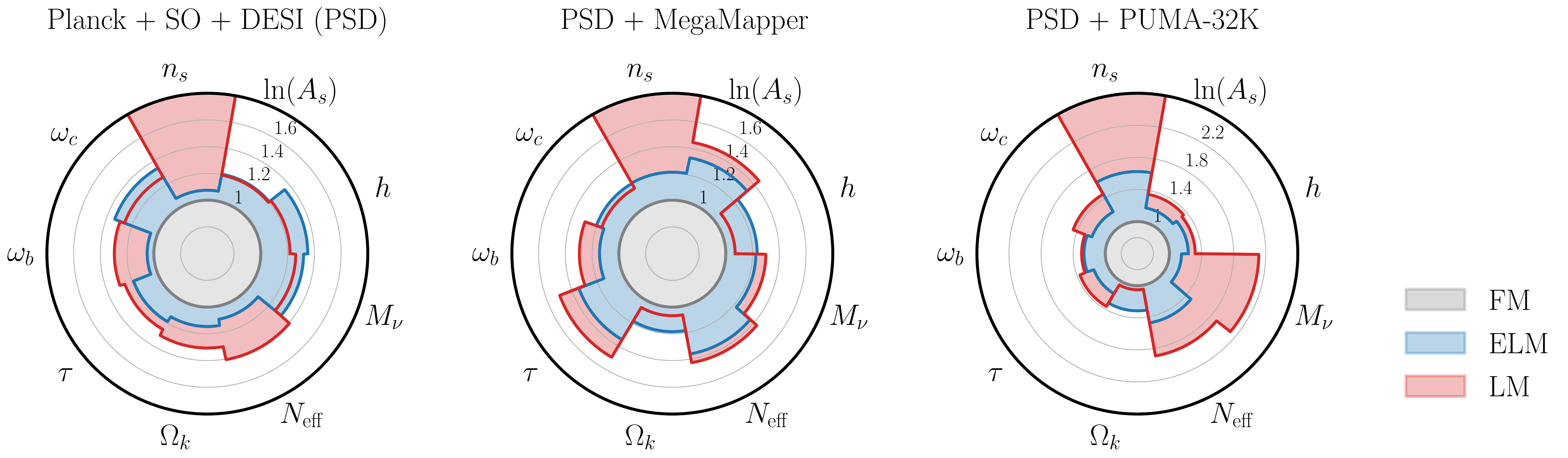}
\caption{Constraints on $\Lambda$CDM and several one-parameter extensions when using the full model (FM), relative to the linear model (LM, red) or extended linear model (ELM, blue). In this plot we only include the primary CMB and full shape clustering data. Note that the largest radius in the two leftmost panels is $1.8$, but $2.4$ in the right panel. For the linear model, the values for $n_s$ extend well beyond these limits: from left to right the ratios are 3.1, 6.8, and 11.6.
}
\label{fig:beyond_linear_params}
\end{figure}

To explore the difference between this simplified linear model and our fiducial analysis, we create forecasts for DESI, MegaMapper and PUMA-32K (with optimistic foregrounds) following the procedure of \S\ref{sec:full shape power spectrum} with Eq.~\eqref{eqn:linear_model} replacing the full model (FM), given by Eq.~\eqref{eqn:stochastic}. A comparison between the constraints on $\Lambda$CDM and a selection of single parameter extensions is shown in Fig.~\ref{fig:beyond_linear_params}. From Fig.~\ref{fig:beyond_linear_params} we see that the linear model generically predicts overly optimistic constraints: for both PSD and PSD + MegaMapper, the constraints using the full model are $20-40\%$ larger than those using the linear model for most parameters, with $n_s$ being a notable exception, where the linear and full models disagree by factors of $3$ and $6$ for the respective survey combinations. For PSD + PUMA-32K the disagreement is significantly worse: $\sigma(M_\nu)$ and $\sigma(N_\text{eff})$ from linear theory are overly optimistic by a factor of $\sim 2$, while the constraints on $n_s$ disagree by a factor of $12$. 

Leaving aside the fact that linear theory is clearly not sufficient for an unbiased estimate of cosmological parameters in the galaxy power spectrum, linear theory results in inaccurate forecasts for several reasons. Perhaps the most obvious is that the approximation of linear dynamics breaks down at intermediate to high $k$. Two beyond-linear effects are particularly important: the broadening of the BAO peak due to non-linear evolution and decorrelation of the evolved field with the initial conditions \cite{Bha96,ESW07,Mat08a,Mat08b,Crocce08,PWC09,Noh09,CLPT,TasZal12,McCSza12,White14,SenZal15,Schmittfull15,Baldauf15,Vlah16,McQuinn16,Blas16,Seo16}. In Fig.~\ref{fig:lin vs nonlin derivatives} we show a few logarithmic derivatives of the tracer power spectrum using both linear and non-linear theory at $z=2.25$. The non-linear scale at this redshift is roughly $0.46\,h\,{\rm Mpc}^{-1}$, and we see significant discrepancies (more than a few percent) by $k \sim 0.2\,h\,{\rm Mpc}^{-1} < 0.5\,k_{\rm nl}$. In addition to linear theory breaking down on small scales, linear forecasting is inaccurate since the true angular structure is more complex than the $b+f\mu^2$ form allows, the bias model has too few degrees of freedom, linear forecasting ignores contributions from small scale physics (i.e. counterterms) and the FoG model is overly simplified.

The linear model can be trivially extended to address the latter two shortfalls: the effects of small-scale physics can be mimicked with the counterterm-like contribution $\mu^{2n} k^2 P^\text{lin}$, and the functional dependence of the FoG-like contribution can be relaxed via a Taylor expansion. Doing so results in the ``extended linear model'' (ELM): 
\begin{equation}
    P^\text{ELM}_{gg} = (b+f\mu^2)^2 P^\text{lin} + (\alpha_0 + \alpha_2 \mu^2 + \alpha_4 \mu^4)(k/k_*)^2 P^\text{lin} + N_0 + N_2 (\mu k)^2 + N_4 (\mu k)^4,
\end{equation}
where we have suppressed the explicit $\bm{k}$ dependence.
A comparison of the ELM with the FM is shown in Fig.~\ref{fig:beyond_linear_params}. For PSD (PSD + MegaMapper) we find a minor improvement in the ELM over the LM, with the mean disagreement\footnote{ The ``mean disagreement" $\equiv\sum_\theta[\sigma^\text{FM}(\theta)/\sigma^\text{ELM}(\theta)-1]/8$, where $\theta\in \{h,\,\ln(A_s),\,\omega_c,\,\omega_b,\,\tau,\,\Omega_k,\,N_\text{eff},\,M_\nu\}$.} with the full model decreasing from $28\%$ ($28\%$) to $21\%$ ($25\%$). Most notable is the improved ELM prediction of $\sigma(n_s)$, which agrees with the full model to within $20\%$ for both PSD and PSD + MegaMapper. For PSD + PUMA-32K, we find that the mean disagreement with the full model decreases from 44\% to 25\%, which is primarily driven by the significantly more accurate ELM predictions of $\sigma(M_\nu)$ and $\sigma(N_\text{eff})$, disagreeing by $20\%$ and $40\%$ with the full model, as opposed to a factor of $2$. Likewise, the disagreement for $\sigma(n_s)$ decreases from a factor of 12 to $50\%$.

\section{Observational challenges}
\label{sec:issues}

In the previous section we neglected or delayed discussing the effects of several observational challenges that would arise in a more realistic treatment of future surveys. In this section we quantify some of these effects, summarizing the importance of the 21-cm foreground wedge, examining redshift uncertainties, the impact of a shallower $u$-band depth for LBG surveys, and the impact of a $k_\parallel$ cut on constraints of primordial non-Gaussianity.

Throughout \S\ref{sec:results} we presented both optimistic and pessimistic forecasts for 21-cm surveys, representing the ability to mitigate foreground contamination in modes with low $k_\parallel$. Here we summarize the main results. Foreground assumptions can impact the precision of distance measurements perpendicular to the line of sight by more than a factor of $2$, while measurements parallel to the line of sight are negligibly impacted. For PSD + PUMA-32K, the constraints on $h$, $\omega_c$, $A_\text{lin}$ and $\Omega_k$ vary by more than $50\%$ depending on the foreground assumptions, while the constraints on $N_\text{eff}$ and $M_\nu$ vary by more than $30\%$. For EDE the constraints can vary by more than a factor of $2$. Finally, while the 21-cm constraints of $f^\text{Loc}_\text{NL}$ under optimistic foreground assumptions will not be competitive with a high-$z$ galaxy survey, pessimistic foreground assumptions would inhibit a 21-cm measurement of $f^\text{Loc}_\text{NL}$ from scale-dependent bias altogether. 

\begin{figure}[!h]
\centering
\includegraphics[width=0.95\linewidth]{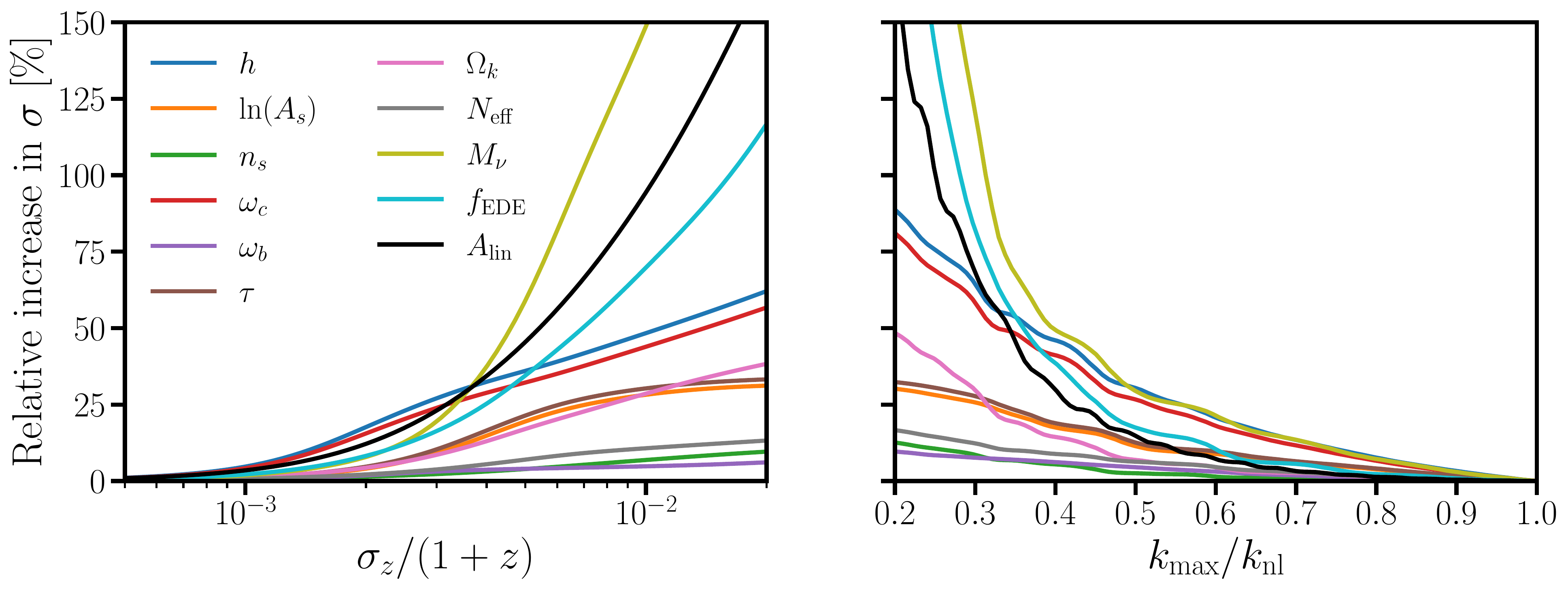}
\caption{\textit{Left}: Constraints on $\Lambda$CDM and several one-parameter extensions from MegaMapper full shape data vs redshift uncertainty, with $k_{\rm max}/k_{\rm nl}=1$. The constraints are measured relative to $\sigma_z=0$. \textit{Right}: Change in constraints as a function of $k_\text{max}/k_\text{nl}$, relative to $k_\text{max}/k_\text{nl}=1$. In both plots we include a prior on $\{\Lambda\text{CDM},N_\text{eff},\Omega_k,M_\nu\}$ from Planck $+$ SO. The constraint on $f_\text{EDE}$ is with fixed $\log_{10}(z_c)=3.61$, while the constraint on $A_\text{lin}$ is with fixed $\omega_\text{lin}=147\,\,h^{-1}\,\text{Mpc}$.
}
\label{fig:impact_of_redshift_error}
\end{figure}

As previously discussed in \S\ref{sec:redshift_uncertainties}, at high redshifts it becomes increasingly difficult to obtain small redshift errors, while the effects of redshift errors become increasingly more important due to the non-linear scale shifting to higher $k$. In Fig.~\ref{fig:impact_of_redshift_error} we show the impact of redshift errors on the constraints of $\Lambda$CDM and several one-parameter extensions from MegaMapper full shape data. We find that redshift errors negligibly impact our results for $\sigma_z/(1+z)<10^{-3}$, only degrading the constraints at the few percent level. However, when $\sigma_z/(1+z)$ nears $0.3\%$ we begin to see non-negligible ($\sim25$\%) effects on some parameters, most notably $M_\nu,\,f_\text{EDE}\text{ and }A_\text{lin}$. As shown in the right pannel of Fig.~\ref{fig:impact_of_redshift_error}, including redshift errors is roughly equivalent to enforcing a strict $k_\text{max}/k_\text{nl}$ cut, with $\sigma_z/(1+z)=0.01$ (0.003) corresponding to roughly $k_\text{max}/k_\text{nl} =  0.3$ (0.6) for our fiducial MegaMapper survey.

\begin{figure}[!h]
    \centering
    \includegraphics[width=0.48\linewidth]{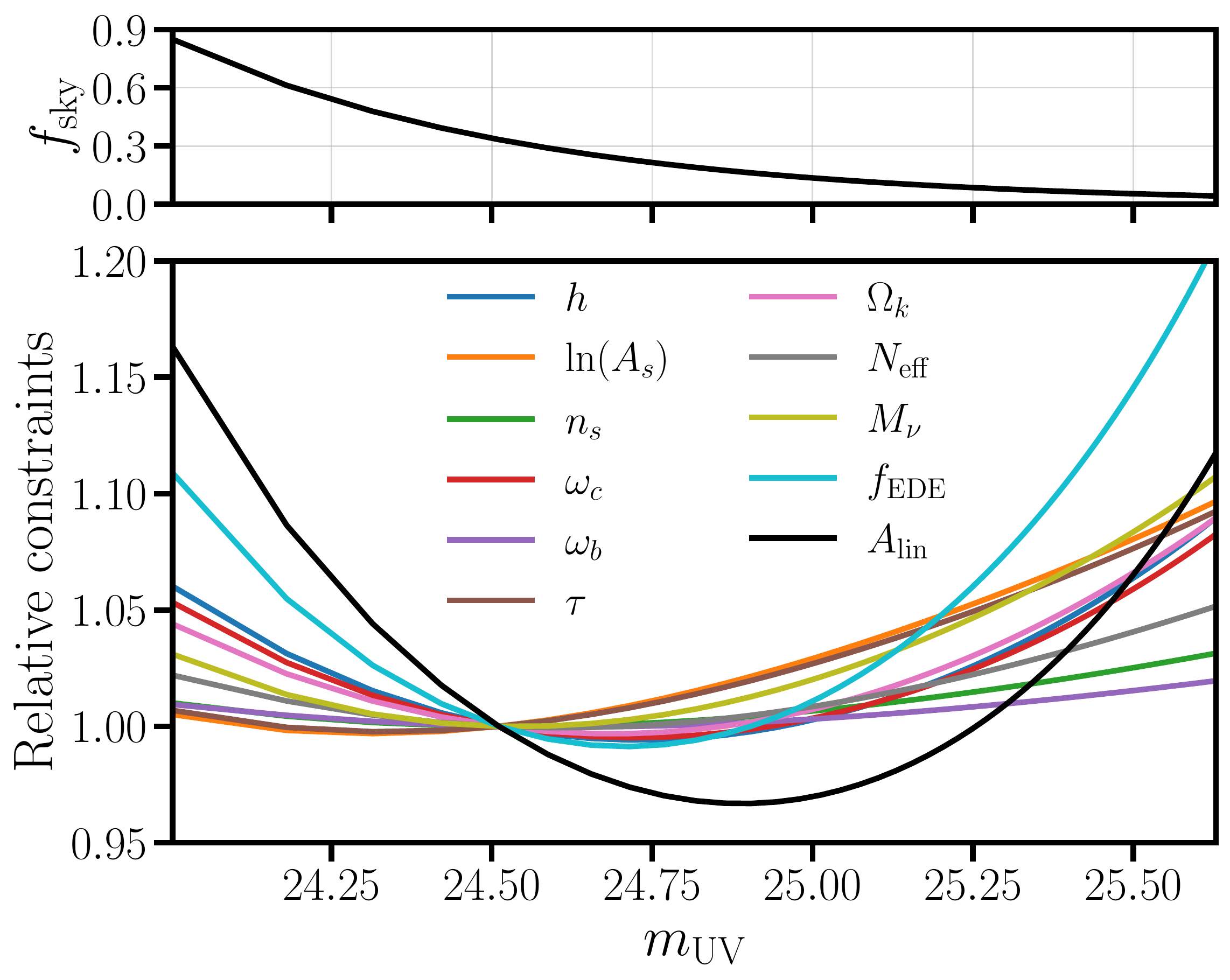}
    \includegraphics[width=0.4\linewidth]{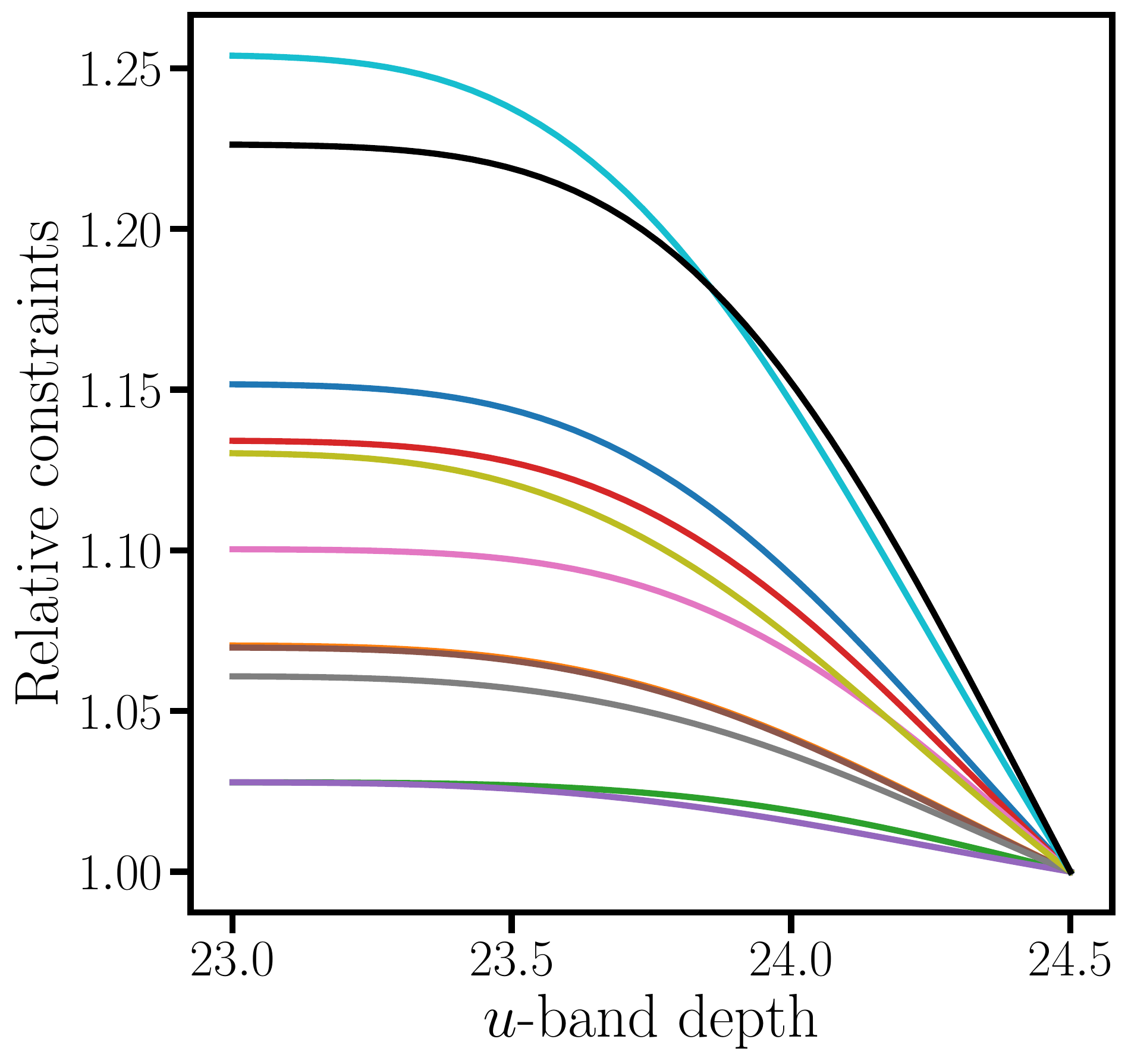}
    \caption{ \textit{Left:} Relative improvement to Planck + SO + MegaMapper-like parameters as (MegaMapper) area is traded for depth, for fixed observing time. We assume that $m_\text{UV}(t)=24.5 +2.5 \log_{10}( \sqrt{t})$ and $f_\text{sky}(t)=0.34/t$, and scan through $0.4<t<8$. 
    \textit{Right:} Degradation to Planck + SO + MegaMapper-like constraints with a shallower $u$-band depth. 
    Here the MegaMapper-like survey has a default 24.5 limiting magnitude, and detects $50\%$ of the LBGs for a given $m_\text{UV}$. 
    }
\label{fig:area_vs_depth}
\end{figure}

To estimate the impact of a lower $u$-band depth on MegaMapper constraints, we consider the idealized MegaMapper-like survey described in \S\ref{sec:surveys and probes}, which we split into four redshift bins spanning $2<z<5$. To mimic the effect of a lower $u$-band depth, we vary $m_\text{UV}$ in the two central redshift bins (spanning $2.75<z<4.25$) between $23$ and $24.5$. We leave $m_\text{UV}=24.5$ in the lowest and highest redshift bins unchanged, and assume a mean $50\%$ target efficiency ($\bar{n}/n_\text{LBG}=0.5$) in all redshift bins, regardless of $m_\text{UV}$. As shown in the right panel in Fig.~\ref{fig:area_vs_depth}, the constraints from Planck + SO + MegaMapper on $\Lambda$CDM and common one parameter extensions ($N_\text{eff},\,M_\nu,\,\Omega_k$) increase by less than $\approx15\%$ as the $u$-band depth is decreased from 24.5 to 23, while the constraints on $f_\text{EDE}$ (at fixed $\log_{10}(z_c)=3.61$) and $A_\text{lin}$ (at fixed $\omega_\text{lin}=147\,\,h^{-1}\,\text{Mpc}$) increase by less than $25\%$. 

We also perform a simplistic area vs depth trade-off study for the same MegaMapper-like survey, which we discuss here. For fixed observing time, we assume the sky coverage and the change to the limiting apparent magnitude scale as $f_\text{sky}\propto t^{-1}$ and $\delta m_\text{UV}\propto 2.5\log_{10}(\sqrt{t})$ respectively, which should be a very good approximation given that MegaMapper will be deep into the sky-noise limit. The fiducial survey corresponds to $t=1$. In principle, changing the limiting magnitude changes the bias of the LBG sample. We account for this effect by rescaling $n(m=24.5,z) \to n(m,z) b^2(m,z)/b^2(m=24.5,z)$ in our covariance. 

\begin{figure}[!h]
    \centering
    \includegraphics[width=0.95\linewidth]{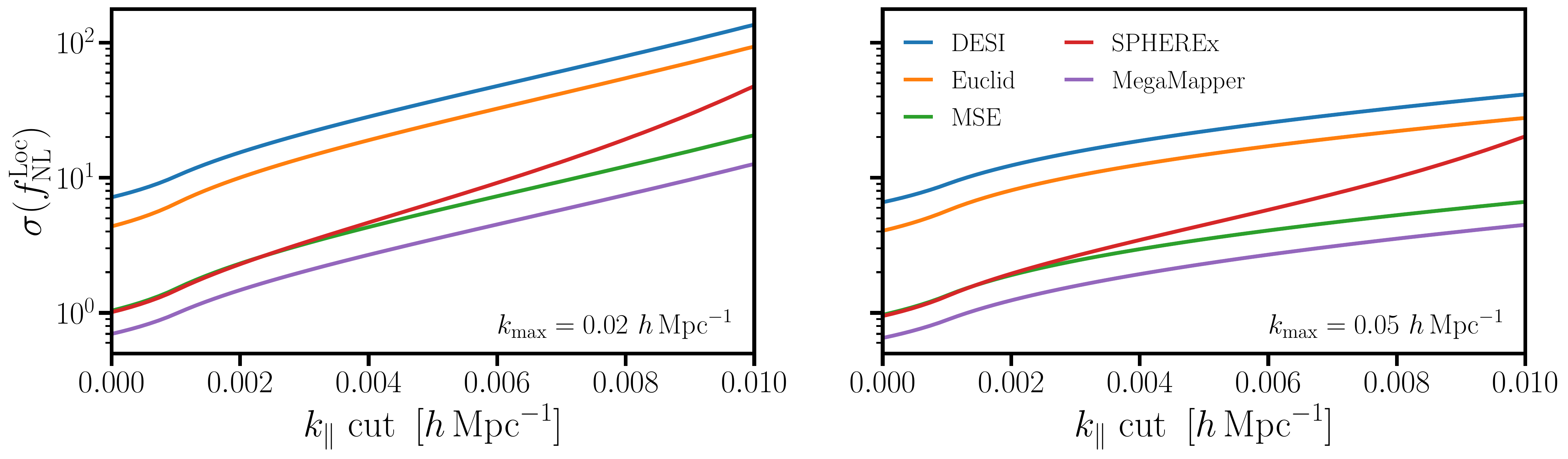}
    \caption{ Constraints on the amplitude of local primordial non-Gaussianity with a cut on $k_\parallel$ modes. In the left panel we set $k_\text{max} = 0.02\,\,h\,\text{Mpc}^{-1}$, while in the right panel we set $k_\text{max} = 0.05\,\,h\,\text{Mpc}^{-1}$. In both panels we take $k_\text{min}=0.001\,\,h\,\text{Mpc}^{-1}$}
\label{fig:fnl_vs_kparmin}
\end{figure}

As shown in the left panel of Fig.~\ref{fig:area_vs_depth}, the fiducial MegaMapper sky coverage ($f_\text{sky}=0.34$) is well optimised for the parameters considered in this work. In particular, we find that the fiducial MegaMapper sky coverage yields constraints that are within $5\%$ of their optimal values.

Finally we consider the impact of a $k_\parallel$ cut, which could be imposed to remove systematic errors, on constraints of local primordial non-Gaussianity. For our constraints we follow the method described in \S\ref{sec:PNG} with $k_\text{min}=0.001\,\,h\,\text{Mpc}^{-1}$, and remove modes whose component along the line of sight is smaller than some $k_\parallel$ cut, which we vary between $0$ and $0.01\,\,h\,\text{Mpc}^{-1}$. Our results are illustrated in Fig.~\ref{fig:fnl_vs_kparmin}. With a conservative $k_\text{max} = 0.02\,\,h\,\text{Mpc}^{-1}$, we find that MegaMapper constraints degrade from $\sigma(f^\text{Loc}_\text{NL})=0.7$ to $\sigma(f^\text{Loc}_\text{NL})=10$ as the $k_\parallel$ cut is increased from 0 to $0.01\,\,h\,\text{Mpc}^{-1}$. For a more optimistic $k_\text{max} = 0.05\,\,h\,\text{Mpc}^{-1}$ these constraints degrade from $0.7$ to $4$.
In addition, we find that the constraints from SPHEREx degrade more rapidly with a cut on $k_\parallel$ than for spectroscopic surveys due to its redshift errors (as in \S\ref{sec:PNG}, we assume $\sigma_z/(1+z)=0.05$). This emphasizes the importance of spectroscopy for diagnosing systematic errors without blowing up the error on $f^\text{Loc}_\text{NL}$.

\section{Conclusions}
\label{sec:conclusions}

Advances in experimental techniques make it possible to map the high redshift Universe at high fidelity in the near future. This will increase the observed volume by many-fold, while providing unprecedented access to very large scales, allowing the reconstruction of an order of magnitude more ``primordial modes''.  Such surveys would improve measurements of the expansion history and spatial curvature of the Universe, primordial non-Gaussianity, features in the power spectrum, models of inflation, dark energy and modified gravity, light relics and neutrino masses.  Recently developed theoretical models, well suited to the quasi-linear clustering of the early Universe, provide a firm basis for forecasting the science reach of such surveys including such complexities are non-linear evolution, redshift-space distortions and non-linear bias.  In this paper we have presented such forecasts for a handful of proposed experiments (\S\ref{sec:surveys and probes}) that illustrate the potential of next-generation large-scale structure surveys to inform our theories of fundamental physics and discussed some of the limitations and synergies with other probes.  We used a Fisher matrix formalism (\S\ref{sec:fisher}) to forecast errors on cosmological parameters from the experiments individually and in combination with other probes, such as primary CMB anisotropies and CMB lensing.  Baryon acoustic oscillations (BAO) are treated as a special case, as are methods for determining the amplitude of clustering using a fixed shape of the power spectrum (e.g.\ refs.\ \cite{BOSS_DR12,eBOSS21}).

Section \ref{sec:distance_measures} shows that such surveys could map the expansion history and spatial curvature of the Universe, in a relatively model independent fashion and with high precision, over more than $90\%$ of cosmic time ($z<6$).  Fig.~\ref{fig:distance_constraints} shows that MegaMapper could provide sub-percent measures of the angular diameter distance to $z\approx 5$ while a PUMA-like experiment could reduce the errors by a further factor of 2 under optimistic assumptions for foregrounds.  The line-of-sight distance, i.e.\ the expansion rate, can be determined with similar accuracy with a PUMA-like experiment providing $<0.5\%$ precision to $z\approx 6$ for a wide range of foreground assumptions.

Section \ref{sec:structure_growth} shows that future high redshift surveys will enable percent-level measurements of structure growth out to redshift $5$. Fig.~\ref{fig:fs8} shows that PUMA could achieve $5-10\%$ measurements of $\sigma_8(z)$ across $2<z<6$, while MegaMapper could reduce these errors to less than $1\%$ across $2<z<5$. 21-cm surveys are severely limited by a degeneracy with the mean 21-cm brightness temperature, while high redshift spectroscopic surveys benefit greatly from the addition of CMB lensing cross-correlations, which decrease the error on $\sigma_8(z)$ by a factor of $2$.

The $\Lambda$CDM model with minimal-mass neutrinos provides a highly predictive and rigid model that can explain the majority of cosmological observations \cite{PlanckLegacy18,eBOSS21} and thus makes a natural target for forecasts (\S\ref{sec:base_lcdm}).  Table \ref{tab:big_summary} and Figs.~\ref{fig:polar_plot_HiZ} and \ref{fig:extensions_to_base_model} present a summary of our findings.
Upcoming surveys like  DESI and Euclid, in combination with Planck, will improve current measurements of $\omega_c$ and $h$ by a factor of two, but will largely exhaust the information available from the low redshift power spectrum.  Adding high-$z$ redshift measurements allows further improvements by more than a factor of 3, exceeding the reach of even ambitious next-generation CMB experiments. The improvement in the determinant of the Fisher matrix for base $\Lambda$CDM in going from PSD to PSD + MegaMapper (PSD + PUMA-32K) is a factor of $2\times 10^4$ ($7\times 10^4$).

Massive neutrinos provide one of the best motivated extensions of the minimal $\Lambda$CDM model (\S\ref{sec:neutrinos}), and large-scale structure provides the tightest constraints on the sum of the neutrino masses \cite{PDG20}.  Table \ref{tab:big_summary} shows that constraints on $M_\nu$ tighten significantly with the inclusion of high-redshift LSS data.  While the forecasts depend sensitively on the smallest scales that can be used (\S\ref{sec:neutrinos}; Fig.~\ref{fig:mnu_vs_kmax}) the next-generation of CMB surveys coupled with a next-generation, high-redshift LSS survey should comfortably achieve $\sigma(M_\nu)=20$ meV.  This could be further improved by a better measurement of $\tau$ (from the CMB or other means) and the inclusion of CMB lensing, which aids in degeneracy breaking (see Fig.~\ref{fig:kk_vs_kg_gg}).

Light relics (\S\ref{sec:relics}) suppress structure on small scales and thus can also be well constrained by high-redshift large-scale structure surveys (Table \ref{tab:big_summary} and Fig.~\ref{fig:Neff}).  We find that both MegaMapper and PUMA are nearly cosmic variance limited for this measurement and capable of achieving $\sigma(N_{\rm eff})\simeq 0.02$ in combination with future CMB experiments (depending upon the $k_{\rm max}$ assumed: Table \ref{tab:big_summary} and Fig.~\ref{fig:Neff}).
Such a measurement could detect light vectors at $3\sigma$, regardless of their freeze-out temperature, and would be capable of detecting (or ruling out) all light relics with a freeze-out temperature smaller than the QCD phase transition ($\sim 0.2$ GeV) at $5\sigma$ significance.  Interestingly we find that a proper marginalization over nuisance parameters (such as non-linear biases and counterterms) is important for forecasting $\sigma(N_{\rm eff})$, with constraints underestimated by $\simeq 40\%$ when these parameters are held fixed (as is often done).

One window into the very early Universe and the physics of inflation is the study of primordial non-Gaussianity (\S\ref{sec:PNG}). The Planck constraints $f_{\rm NL}^{\rm loc}=-0.9\pm 5.1$ are within a factor of 3 of the cosmic variance limit, so pushing to $\sigma(f_{\rm NL}^{\rm loc})\lesssim 1$ requires LSS data.  We find a high-$z$ LSS survey would be capable of a 5-fold improvement over a cosmic variance limited CMB survey, with $\sigma(f_{\rm NL}^{\rm loc})\sim 0.8$, just from the power spectrum alone.  Further improvements could be possible by including information from the bispectrum or from including CMB lensing.

A second window into the physics of inflation from large-scale structure surveys is to search for primordial features (\S\ref{sec:primordial_features}) in the power spectrum.  The space of possible features is large, so as an example we show that a future LSS survey could constrain the amplitude of a sinusoidal modulation of $P(k)$ to better than $0.1\%$ over a broad range of frequencies (Fig.~\ref{fig:primodial_features}).
A precise measurement of the power spectrum shape can also be used to place constraints on (or detect) short-term deviations from matter or radiation domination, since such periods will change the shape of the power spectrum due to differential growth of modes.  Thus the shape of the matter power spectrum has the potential to detect sub-percent fractional amounts of Early Dark Energy to $z \sim 10^5$ (\S\ref{sec:EDE}), probing Dark Energy all the way to when the Universe was a only a few years old (Fig.~\ref{fig:EDE_constraints}).

The precision of these measurements, combined with CMB observations, also has the promise of tightening constraints on the gravitational slip, which is equal to one in General Relativity but generically departs from unity in other models.  For the simplest case of a scale-independent slip, we show that future high redshift experiments could constrain $\gamma$ to a few percent (Fig.~\ref{fig:gravitational_slip}) with the tightest constraints coming from MegaMapper near $z\approx 2-3$.

We find that inclusion of models that go beyond constant bias and linear theory is important for accurately assessing the potential of future experiments to constrain parameters.  Neglecting the complexities of bias or fixing too many nuisance parameters can artificially inflate the precision of parameter measurements by a factor of up to $12$ in $\Lambda$CDM models (\S\ref{sec:lin vs non lin} and Fig.~\ref{fig:beyond_linear_params}).

At higher redshift the non-linear scale shifts to higher $k$, and the impact of fingers of god on the redshift-space clustering also declines.  To make use of the extra information available along the line of sight requires an increase in redshift precision (\S\ref{sec:redshift_uncertainties}; Figs.~\ref{fig:integration_region} and \ref{fig:redshift_uncertainty}).  Achieving this precision observationally may be challenging if broad lines, or lines with significant radiative transfer effects, are used.  In \S\ref{sec:issues} we discuss the impact of redshift uncertainties on a MegaMapper-like survey, and find that our constraints degrade by less than $15\%$ for $\sigma_z/(1+z)<0.002$, which corresponds to a systematic velocity of $600$ km s$^{-1}$. 

Studies of linear or quasi-linear large-scale structure with redshift surveys and the CMB currently provide our tightest constraints on cosmology and fundamental physics.  Pushing the redshift and volume frontier will provide guaranteed, significant improvements in the state-of-the-art in a manner that can be reliably forecast and optimized.  This paper presents a first step in this direction, considering a number of proposed experiments that would map the Universe in the epochs before cosmic noon.

\acknowledgments
We would like to thank Arjun Dey and David Schlegel for helpful discussions on redshift errors for high redshift galaxies and Kyle Dawson for conversations and calculations regarding SpecTel. 
We also thank Shi-Fan Chen, Mikhail Ivanov, Marko Simonovic, Juna Kollmeier, Eric Linder, Emmanuel Schaan, Uro\v{s} Seljak, Leonardo Senatore, An\v{z}e Slosar and  Michael Wilson for useful discussions during the preparation of this manuscript.
N.S.~is supported by the NSF.
S.F.~is supported by the Physics Division of Lawrence Berkeley National Laboratory.
M.W.~is supported by the DOE and the NSF.
This research has made use of NASA's Astrophysics Data System and the arXiv preprint server.
This research used resources of the National Energy Research Scientific Computing Center (NERSC), a U.S. Department of Energy Office of Science User Facility operated under Contract No.\ DE-AC02-05CH11231.

\appendix

\section{Primordial figure of merit ($N_\text{modes}$)}
\label{sec:fom}

Here we describe a new figure of merit for  ``counting reconstructable linear modes.'' This is similar to the effective volume, but integrated to a $k_\text{max}$ that incorporates the decorrelation of the observed nonlinear field with the initial conditions at high $k$, and hence loss of primordial information on small scales. This loss of information can be modeled as a Gaussian ``propagator'' $G(\bm{k})$. That is, we split the observed nonlinear overdensities $\delta_F(\bm{k}) = G(\bm{k})\delta_L(\bm{k}) + d(\bm{k})$ into two pieces: one which is correlated with the linear overdensities, $\delta_L(\bm{k})$, and one which isn't $\langle\delta_L(\bm{k}) d(\bm{k}')\rangle=0$. The Gaussian propogator is given by:
\begin{equation}
    G(k,\mu) \equiv\frac{\langle\delta_F\delta_L\rangle}{\langle\delta_L\delta_L\rangle}
    \simeq \exp\left[-\frac{1}{2}\left(k_\perp^2+k_\parallel^2\{1+f\}^2 \right)\Sigma^2\right],
\end{equation}
where $\Sigma$ is set by the rms displacement within the Zel'dovich approximation, as in Eq.~\eqref{eqn:Sigma2}.

In the sample variance limit, the relative uncertainty of the linear power spectrum's amplitude $A$ is given by $\sqrt{2/N_\text{modes}}$. We can calculate this uncertainty from the Fisher matrix:
\begin{equation}
    F_{ii}
    =
    \frac{f_\text{sky}}{2}
    \int_{z_\text{min}}^{z_\text{max}} dz
    \frac{dV}{dz}
    \int_{k-\text{wedge}(z)}^\infty
    \frac{d^3 \bm{k}}{(2\pi)^3}
    \left(
    \frac{
    \partial_{\theta_i} P_F(\bm{k},z)}{
    P_F(\bm{k},z) + N(z)}
    \right)^2
    ,
\end{equation}
where $k-\text{wedge}(z)$ defines the lower limit of integration. The uncertainty on the amplitude is $\delta A/A = \delta P_L(\bm{k})/P_L(\bm{k}) = 1/\sqrt{F_{AA}}$, so that $N_\text{modes}=2F_{AA}$. This defines the figure of merit $N_\text{modes}$.

Let $\hat{A}=P_L(\bm{k})/\hat{P}_L(\bm{k})$ be an estimator for the amplitude of the linear power spectrum $P_L(\bm{k}) = (b+f\mu^2)^2 P_m(k)$, where $\hat{P}_L(\bm{k})$ is any (unbiased) estimator for the linear galaxy power spectrum in redshift space, and $P_m(k)$ is the linear matter power spectrum. Since $P_F(\bm{k}) = G^2(\bm{k}) P_L(\bm{k}) + P_d(\bm{k})$ does not explicitly depend on $\hat{P}_L(\bm{k})$, the derivative of the non-linear power spectrum with respect to $\hat{A}$ is simply\footnote{This is not true in a strict mathematical sense. In reality the decorrelated piece $P_d(\bm{k})$ is still dependent on the initial conditions, although the manner in which it depends on $P_L(\bm{k})$ in highly non-trivial and subdominant, so we neglect it in our derivative.} $G^2(\bm{k}) \hat{P}_L(\bm{k})$. This derivative should be evaluated at the fiducial cosmology, so $\partial_{A} P_F(\bm{k}) = G^2(\bm{k}) P_L(\bm{k})$. Putting this together gives:
\begin{equation}
    N_\text{modes}
    =
    f_\text{sky}
    \int_{z_\text{min}}^{z_\text{max}} dz
    \frac{dV}{dz}
    \int_{k-\text{wedge}(z)}^\infty
    \frac{d^3 \bm{k}}{(2\pi)^3}
    \left(
    \frac{
    G^2(\bm{k},z)P_L(\bm{k},z)}{
    P_F(\bm{k},z) + N(z)}
    \right)^2.
\label{eqn:nmodes}
\end{equation}

The calculation of the CMB equivalent of $N_\text{modes}$ is almost identical. We include the TT, TE, and EE power spectra $\bm{\mu}_\ell = \left(C^\text{TT}_\ell, C^\text{TE}_\ell,C^\text{EE}_\ell\right)^T$ in our Fisher forecast, which has covariance matrix:
\begin{equation}
\footnotesize{
    \bm{C}_{\ell,\ell'}
    =
    \frac{2\delta^K_{\ell,\ell'}}{(2\ell+1)f_\text{sky}}
    \begin{pmatrix}
    \left(C^\text{TT}_\ell + N_\ell\right)^2
    &
    \left(C^\text{TT}_\ell + N_\ell\right) C^\text{TE}_\ell
    &
    \left(C^\text{TE}_\ell\right)^2
    \\
    \left(C^\text{TT}_\ell + N_\ell\right) C^\text{TE}_\ell
    &
    \frac{1}{2}\left[\left(C^\text{TE}_\ell\right)^2 + \left(C^\text{TT}_\ell+N_\ell\right) \left(C^\text{EE}_\ell+2N_\ell\right)\right]
    &
    \left(C^\text{EE}_\ell + 2N_\ell\right) C^\text{TE}_\ell
    \\
    \left(C^\text{TE}_\ell\right)^2
    &
    \left(C^\text{EE}_\ell + 2N_\ell\right) C^\text{TE}_\ell
    &
    \left(C^\text{EE}_\ell + 2N_\ell\right)^2
    \end{pmatrix},
}
\label{eqn:CMB_covariance}
\end{equation}
where
\begin{equation}
N_{\ell} =  \Delta^2_{T} \exp{\left( \frac{\theta^2_{\rm FWHM} \ell^2}{8 \ln2}\right) }
\end{equation}
accounts for the instrumental noise. Here $\Delta_{T}$ is the noise level of the experiment and $\theta_{\rm FWHM}$ is the full-width at half-maximum (FWHM) of the beam in radians. Given $\bm{\mu}_\ell$ and $\bm{C}_{\ell,\ell'}$, the Fisher matrix is simply:
\begin{equation}
\begin{aligned}
    F_{ij} 
    &= 
    \frac{1}{2}
    \sum_{\ell,\ell_1,\ell_2,\ell_3}
    \text{Tr}\left[
    \bm{C}^{-1}_{\ell,\ell_1}
    \frac{\partial\bm{C}_{\ell_1,\ell_2}}{\partial p_i}
    \bm{C}^{-1}_{\ell_2,\ell_3}
    \frac{\partial\bm{C}_{\ell_3,\ell}}{\partial p_j}
    \right]
    +
    \sum_{\ell,\ell'}
    \frac{\partial\bm{\mu}^T_{\ell}}{\partial p_i}
    \bm{C}^{-1}_{\ell,\ell'}
    \frac{\partial\bm{\mu}_{\ell'}}{\partial p_j}.
\end{aligned}
\end{equation}
The second term dominates since the summand is weighted by the number of modes, whereas in the first term all factors of $2\ell+1$ cancel. Just as before, we take $N_\text{modes}=2F_\text{AA}$:
\begin{equation}
    N_\text{modes} \simeq 2 \sum_{\ell,\ell'} \bm{\mu}^T_{\ell} \bm{C}^{-1}_{\ell,\ell'} \bm{\mu}_{\ell'}.
\end{equation}
Due to the bias from foreground contamination, $C^\text{TT}_\ell$ will realistically be measurable out to $\ell_\text{max}\simeq 3000$. We enforce this in our calculation of $N_\text{modes}$ by multiplying $C^\text{TT}_\ell$ by $10^6$ in our covariance matrix when $\ell>3000$.

Our definition of $N_\text{modes}$ can be easily generalized to represent a survey's ability to measure the amplitude of the linear matter power spectrum (at $z=0$) in a specific $k$-bin. Following the same argument as before, let $\hat{A}(k) = P_m(k)/\hat{P}_m(k)$ be an estimator for the amplitude, where $P_m(k)$ is the linear matter power spectrum at $z=0$. Then $\partial_{\hat{A}(k')} P_F(\bm{k},z) =G^2(\bm{k},z)T^2(k,z) (b+f\mu^2)^2 \hat{P}_m(k)\, \Theta(\Delta k-|k-k'|)$, where $T(k,z)$ is the transfer function, $\Theta(x)$ is the Heaviside step function, and $\Delta k$ is the width of the $k$-bin. When evaluated at the fiducial cosmology, this yields $\partial_{\hat{A}(k')} P_F(\bm{k},z) =G^2(\bm{k},z)P_L(\bm{k},z) \Theta(\Delta k-|k-k'|)$. Just as before, we interpret $2 F_{A(k)A(k)}$ as the effective number of sample variance limited modes that measure the amplitude:
\begin{equation}
    N_\text{modes}(k)
    =
    f_\text{sky}
    \int_{z_\text{min}}^{z_\text{max}} dz
    \frac{dV}{dz}
    \frac{k^2 \Delta k}{2\pi^2}
    \left(
    \frac{
    G^2(\bm{k},z)P_L(\bm{k},z)}{
    P_F(\bm{k},z) + N(z)}
    \right)^2.
\end{equation}
We make use this expression in Fig.~\ref{fig:marius_plot}, where the error in each bin is $P_m(k)\sqrt{2/N_\text{modes}(k)}$.

One note on the interpretation of $N_\text{modes}$ is in order here: we have defined $N_\text{modes}$ to be the number of modes recoverable that are correlated with the initial conditions.  Therefore it is a figure of merit for how well we can constrain the primordial potential or other statistical properties of the initial conditions. $N_\text{modes}$ as defined here does not take into account the sensitivity of a given observable to the parameters of interest, so it should not be taken as the only comparison metric, especially between high and low-redshift probes (such as CMB vs LSS). For example, it is very possible for an LSS experiment with lower $N_\text{modes}$ than the CMB to be a lot more sensitive to Dark Energy or neutrino masses, since it performs measurements on a range of redshifts where they are dynamically important. 

For this reason, while a direct comparison between CMB and LSS experiments based on $N_\text{modes}$ alone might be misleading, we believe that it provides a useful metric to compare different survey designs covering a similar redshift range. Since it represents a proxy for ``primordial information'' available, survey optimization aimed at maximizing it can ensure better sensitivity to the physics of the early Universe.

\section{Covariance approximations}
\label{sec:cov_approx}

\begin{figure}[!h]
\centering
\includegraphics[width=\linewidth]{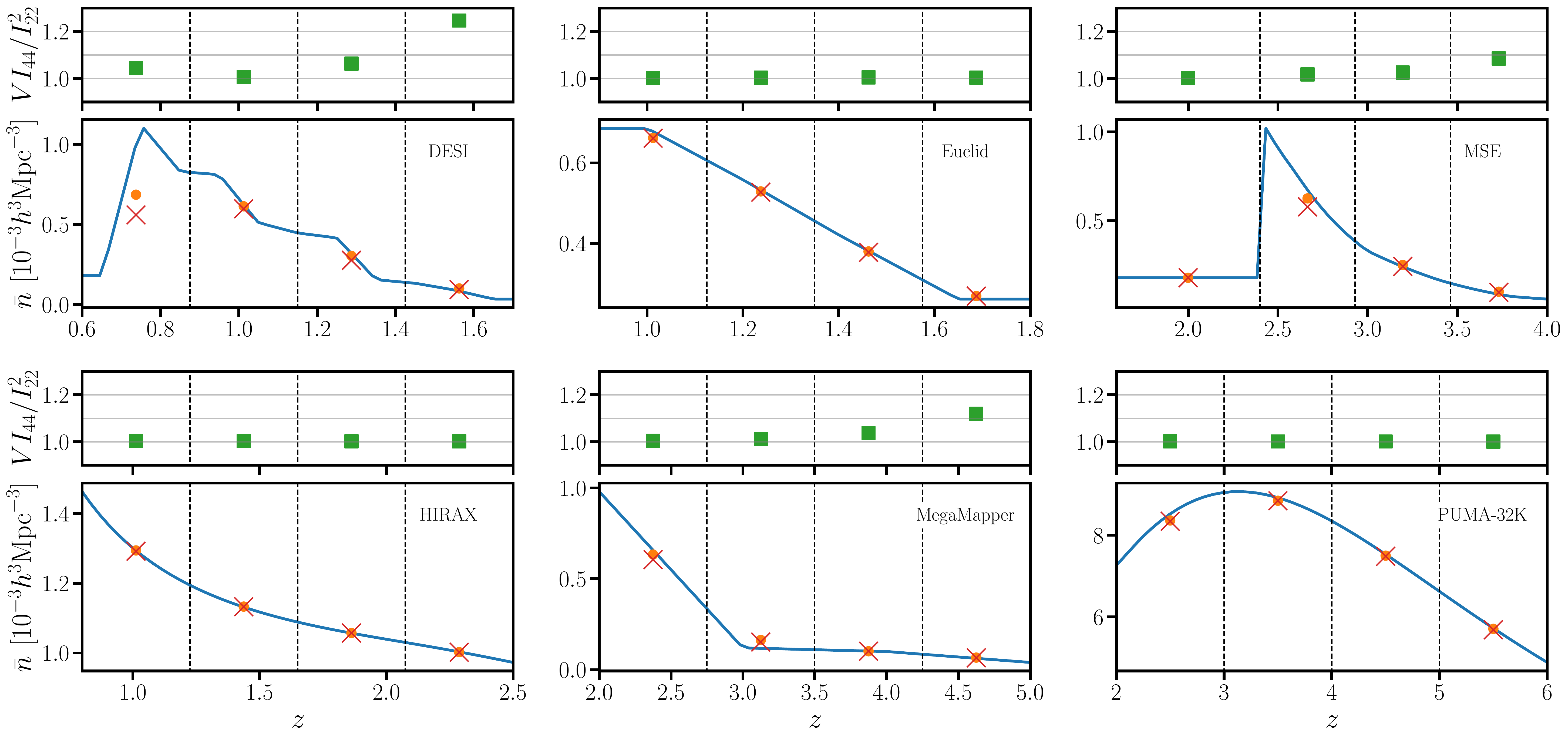}
\caption{Corrections to the number densities and volume from the effects considered by ref.~\cite{Wadekar20} for six of the surveys considered in this work, when $P_\text{fid}=10000\,\,h^{-3}\text{Mpc}^3$. In blue is the ``true'' number density $\bar{n}(z)$ of each survey. In orange we plot $I_{44}/I_{34}$ at the center of each redshift bin. In red is $\sqrt{I_{44}/I_{24}}$. In green we plot the ratio $V I_{44}/I_{22}^2$.}
\label{fig:cov_corrections}
\end{figure}

In the main text we have followed the ``usual'' approach of approximating the covariance matrix by the Gaussian (disconnected) piece. In Eq.~\eqref{eqn:partial_fisher} we also assume that the survey has a uniform number density in each redshift bin. As emphasized by ref.~\cite{Wadekar20}, a more realistic approach would include the impact of window functions, galaxy weights, etc. Inclusion of those effects leads to a covariance with the same functional form 
\begin{equation}
    C \propto (P^2 + 2 P\bar{n}^{-1} + \bar{n}^{-2})/V
\end{equation}
with the substitutions 
\begin{equation}
    V\to I^2_{22}/I_{44}, \quad \bar{n}^{-1} \to I_{34}/I_{44}, \quad \bar{n}^{-2} \to I_{24}/I_{44},
\end{equation}
where the $I_{ij}$'s are integrals over the number density \cite{Wadekar20}:
\begin{equation}
  I_{ij} = \int d^3x\ \bar{n}^i(\mathbf{x}) w^j(\mathbf{x})  
  \qquad\mathrm{with}\quad
  w(\mathbf{x}) = \frac{1}{1+\bar{n}(\mathbf{x})P_{\rm fid}}
\end{equation}
and $P_{\rm fid}$ a fiducial level of power. Ref.~\cite{Wadekar20} argues that for ``realistic'' surveys, the ratios of $I_{ij}$ can be numerically quite different than a typical ``effective volume'' or ``constant number density'' approximation might lead you to believe. We find that these corrections are negligible for the majority of our forecasts, as shown in Fig.~\ref{fig:cov_corrections}. The corrections are larger for surveys with smaller number densities. For DESI, the volume correction in the highest redshift bin is $25\%$, while the number density correction in the lowest redshift bin is $30\%$. For all other surveys, the correction to the volume is less that $13\%$, and the correction to the number density is less than $12\%$. We thus expect these corrections to have a small ($\lesssim 10\%$) impact on our parameter constraints.

\section{BAO forecasting}
\label{sec:bao_comparison}

In \S\ref{sec:bao} we outlined our procedure for inferring distance constraints from the BAO feature in the power spectrum. In this section we compare our method to the standard procedure\footnote{Implemented using: \hyperlink{https://www.cfa.harvard.edu/~deisenst/acousticpeak/bao_forecast.c}{https://www.cfa.harvard.edu/deisenst/acousticpeak/bao\_forecast.c}} of ref.~\cite{Seo_2007} or \cite{DESI,DESI2013}.  In the standard procedure it is assumed that the BAO signal is exponentially damped, but to avoid any information coming from the damping these parameters are not varied in the Fisher matrix.  Rather the A-P parameters, $\alpha_\parallel$ and $\alpha_\perp$ are allowed to vary so as to ``shift'' the BAO template.  A scale-independent bias is assumed.  For an isotropic shift, $s$, the Fisher matrix would look schematically like \cite{Seo_2007}
\begin{equation}
    F_{ss} \approx \frac{V_{\rm surv}}{2} \int\frac{d^3k}{(2\pi)^3}\left(\frac{\partial P_{\rm osc}e^{-k^2\Sigma^2/2}}{\partial\ln s}\right)^2 \frac{1}{(P_L+\bar{n}^{-1})^2}
\end{equation}
where in the numerator $P_{\rm osc}$ represents only the oscillatory, or BAO, piece of the power spectrum and the exponential indicates the damping due to non-linear structure formation.  The 2D model is only slightly more complex, with different damping across and along the line of sight and redshift-space distortions included assuming linear theory.  To include the effects of reconstruction the damping parameters ($\Sigma$) are reduced by $\approx 0.5$.  By contrast our method uses the Zeldovich-based model of ref.~\cite{Chen20c}.  This model automatically accounts for the differences in damping of the different components, more complex bias, non-linear redshift-space distortions and the manner in which reconstruction reduces the damping.  By using a forward model of the reconstruced power spectrum, our forecast is of the standard Fisher form wherein a parameter error is forecast through the theory derivative rather than requiring any modification of the procedure.

\begin{figure}[!h]
\centering
\resizebox{\columnwidth}{!}{\includegraphics{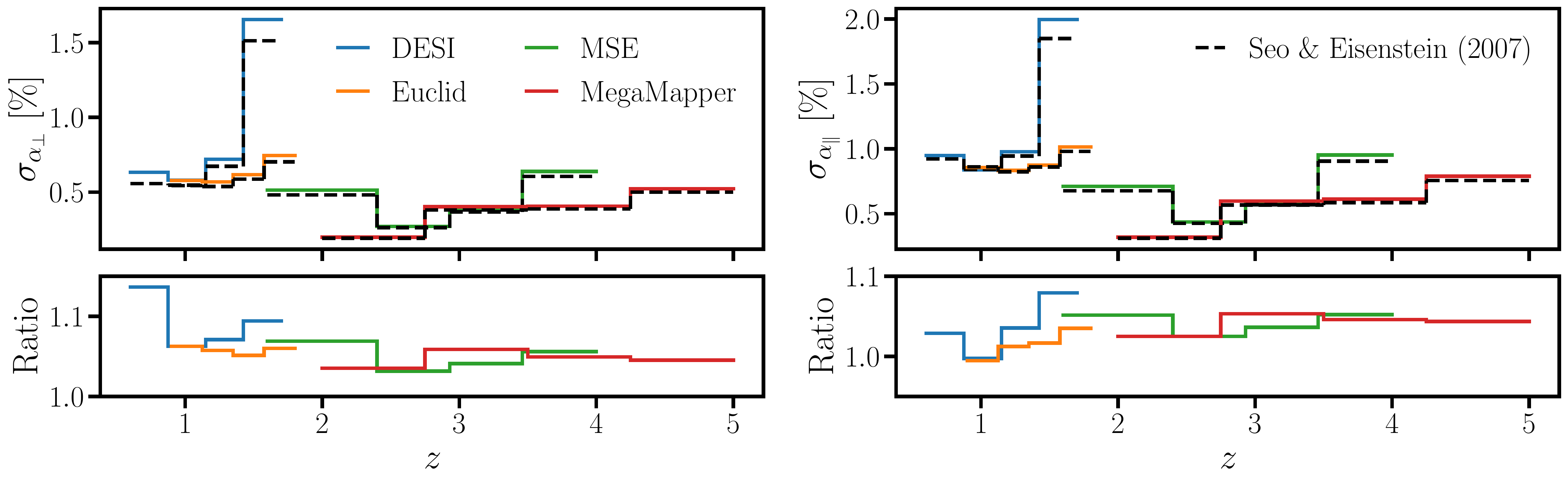}}
\caption{
In the top two pannels we compare our distance constraints (solid), calculated using the method of \S\ref{sec:bao}, to those found using the standard procedure of ref.~\cite{Seo_2007} (dashed). The ratio of the two methods is shown in the bottom two panels.
}
\label{fig:bao_comparison}
\end{figure}

However, our method is actually quite close (within $\sim15\%$) to the standard approach in practice, as shown in Fig.~\ref{fig:bao_comparison}. The damping of the BAO signal by non-linear evolution is very close to Gaussian, as assumed in the standard approach, and while reconstruction affects each of the damping terms differently along and across the line of sight the major impact is to reduce the scale of the damping by approximately a factor of $2$ (indeed it was a perturbative analysis of BAO reconstruction that suggested this approximation). 

\section{Lensing auto- vs. cross-correlation}
\label{app:lensing_confusion}

\begin{figure}[!h]
\centering
\resizebox{\columnwidth}{!}{\includegraphics{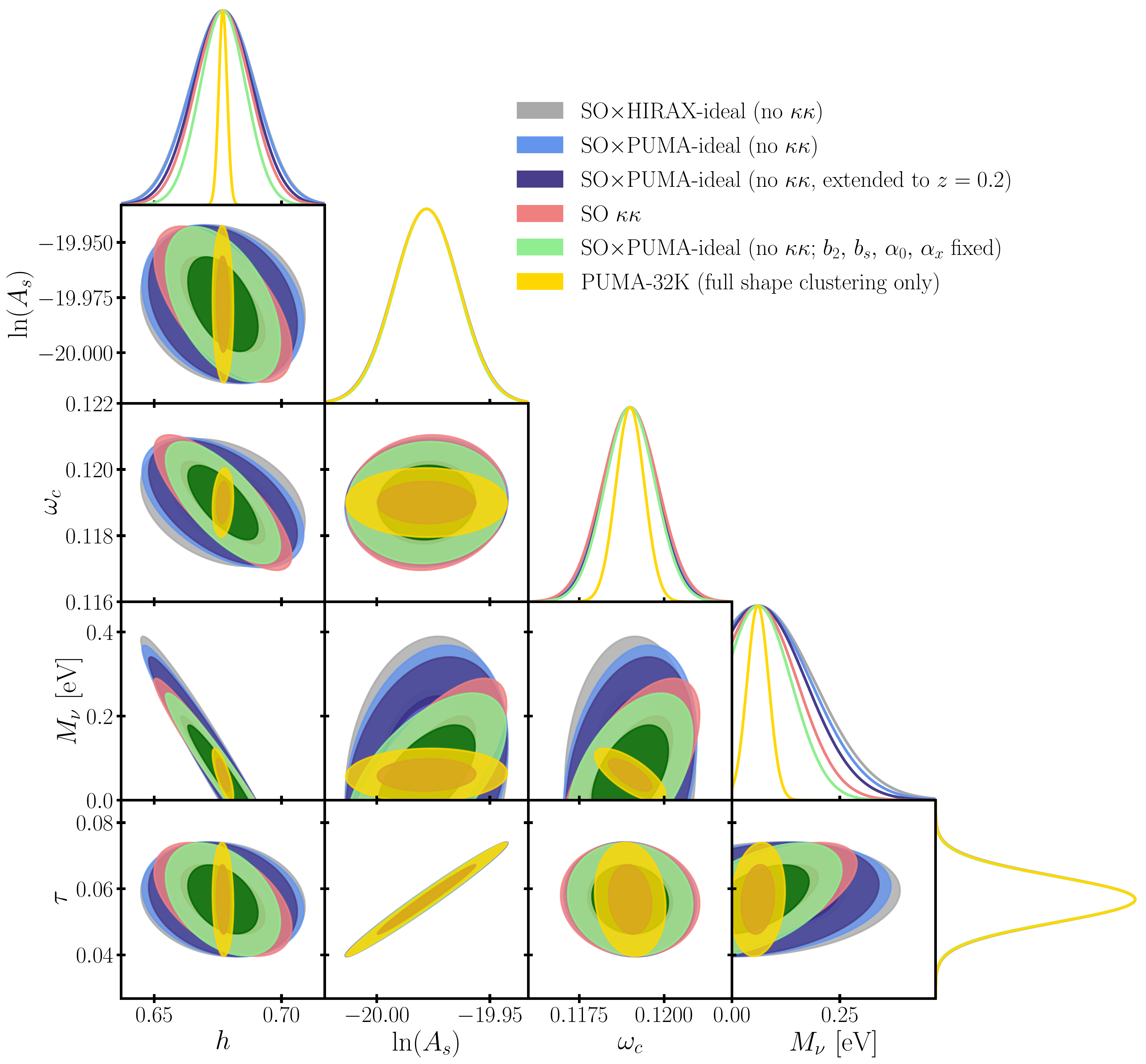}}
\caption{Constraints from cross-correlations of SO lensing with HIRAX-ideal (greay), PUMA-ideal (light blue), a version of PUMA-ideal that extends down to $z=0.2$ (dark blue), and PUMA-ideal with non-linear nuisance terms held fixed (green). In red we plot the constraints from the SO lensing convergence autospectrum. To illustrate how the degeneracies from lensing and full shape measurements differ, we plot the constraints from PUMA-32K full shape data in yellow. All contours share the same Planck + SO prior on $\Lambda$CDM+$M_\nu$.}
\label{fig:kk_vs_kg_gg}
\end{figure}

Here we comment on the information content of the CMB lensing auto power spectrum, vs the cross-correlation with galaxies. Given a spectroscopic survey with sufficiently high number densities that covers the majority of the redshift range where the CMB lensing kernel is large, one can in principle reconstruct $C^{\kappa\kappa}_\ell$ from the galaxy auto-correlation and galaxy-convergence cross-correlation. This is easily seen in linear theory for a noiseless galaxy survey, so that the number density $n\to\infty$. In this limit we can make the redshift bins arbitrarily small, and approximate the integral expression for $C^{\kappa g_i}_\ell$ and $C^{g_i g_i}_\ell$ (Eq.~\eqref{eqn:Limber}) at a single comoving distance $\chi_i$. Under these approximations, $C^{\kappa\kappa}_\ell$ is related to $C^{\kappa g_i}_\ell$ and $C^{g_i g_i}_\ell$ via
\begin{equation}
    C^{\kappa\kappa}_\ell = \sum_i \left(C^{\kappa g_i}_\ell\right)^2/C^{g_i g_i}_\ell,
\label{eq:kk_from_kg_gg}
\end{equation}
where $i$ runs over redshift bins covering $0<z<z_\star$. Thus $C^{\kappa g_i}_\ell,\,C^{g_i g_i}_\ell$ carries at least the same information as $C^{\kappa\kappa}_\ell$ in this idealized case. Since $C^{\kappa g_i}_\ell,\,C^{g_i g_i}_\ell$ can in principle be measured to a higher signal-to-noise than $C^{\kappa\kappa}_\ell$, and can be used to measure time evolution, one might naively expect that parameter constraints are driven by $C^{\kappa g_i}_\ell,\,C^{g_i g_i}_\ell$. However, in our forecasts we find this not to be the case, as illustrated in Fig.~\ref{fig:kk_vs_kg_gg}. The contours from an SO measurement of the convergence auto-correlation (red) are significantly tighter than the contours from galaxy auto- and cross-correlations, as measured by SO and our idealized PUMA-32K survey (light blue). 

There are multiple ways in which this simplistic argument goes wrong: (1) the assumption that $C^{\kappa g_i}_\ell,\, C^{g_ig_i}_\ell$ can be measured to a high signal-to-noise ratio depends on the details of the survey, (2) a redshift survey might not cover the entire region where the lensing kernel is large, and (3) forecasts that include galaxy auto- and cross-correlations marginalize over galaxy biases, counterterms, and a Poisson-like stochastic contribution; however, including $C^{\kappa\kappa}_\ell$ in a forecast introduces no new nuisance parameters. This is by no means an exhaustive list, since  Eq.~\eqref{eq:kk_from_kg_gg} is no longer true for a survey with finite noise, redshift bins of finite extent, when magnification bias is included, or when modeling the tracer beyond linear theory, etc.

To test which of these effects causes the largest degradation to constraints from $C^{\kappa g_i}_\ell$ and $C^{g_i g_i}_\ell$, we consider idealized versions of the HIRAX and PUMA-32K surveys which have no foreground wedge and are assumed to measure $T_b$ to infinite precision (so that it is held fixed), dubbed HIRAX-ideal and PUMA-ideal, enabling cross-correlations with CMB lensing. We take HIRAX-ideal and PUMA-ideal as representative (toy-model) low noise surveys at high redshift. As illustrated in Fig.~\ref{fig:kk_vs_kg_gg}, we see a negligible improvement to parameter constraints when upgrading from HIRAX-ideal (grey) to PUMA-ideal (light blue). Thus (1) is not the problem, and PUMA-ideal has sufficiently low noise levels for high fidelity measurements of $C^{\kappa g_i}_\ell$ and $C^{g_i g_i}_\ell$. We also see that extending PUMA-ideal to $z=0.2$ (light vs. dark blue) has a small impact on the constraints, suggesting that PUMA's fiducial redshift range $(2<z<6)$ sufficiently covers the region where the CMB lensing kernel is large, eliminating (2). Instead, the primary culprit is (3). From Fig.~\ref{fig:kk_vs_kg_gg} we see that the covariance from the convergence auto-correlation (red) is nearly identical to the covariance from the galaxy auto- and cross-correlation (green) when the nuisance terms are fixed.  

We should stress that the value of cross-correlations extends considerably beyond constraining standard cosmological parameters. While $C_\ell^{\kappa \kappa}$ depends on the integrated matter fluctuations over a wide redshift range, as discussed in Sec.~\ref{sec:structure_growth}, they allow for a tomographic reconstruction of the amplitude of structure $\sigma_8(z)$ that only depends on the local amplitude.  This can provide important model-independent tests beyond the rigidity of $\Lambda$CDM, considerably opening up the discovery space.

\section{Empirical  checks}
\label{app:checks}
\begin{figure}[!h]
    \centering
    \includegraphics[width=0.7\linewidth]{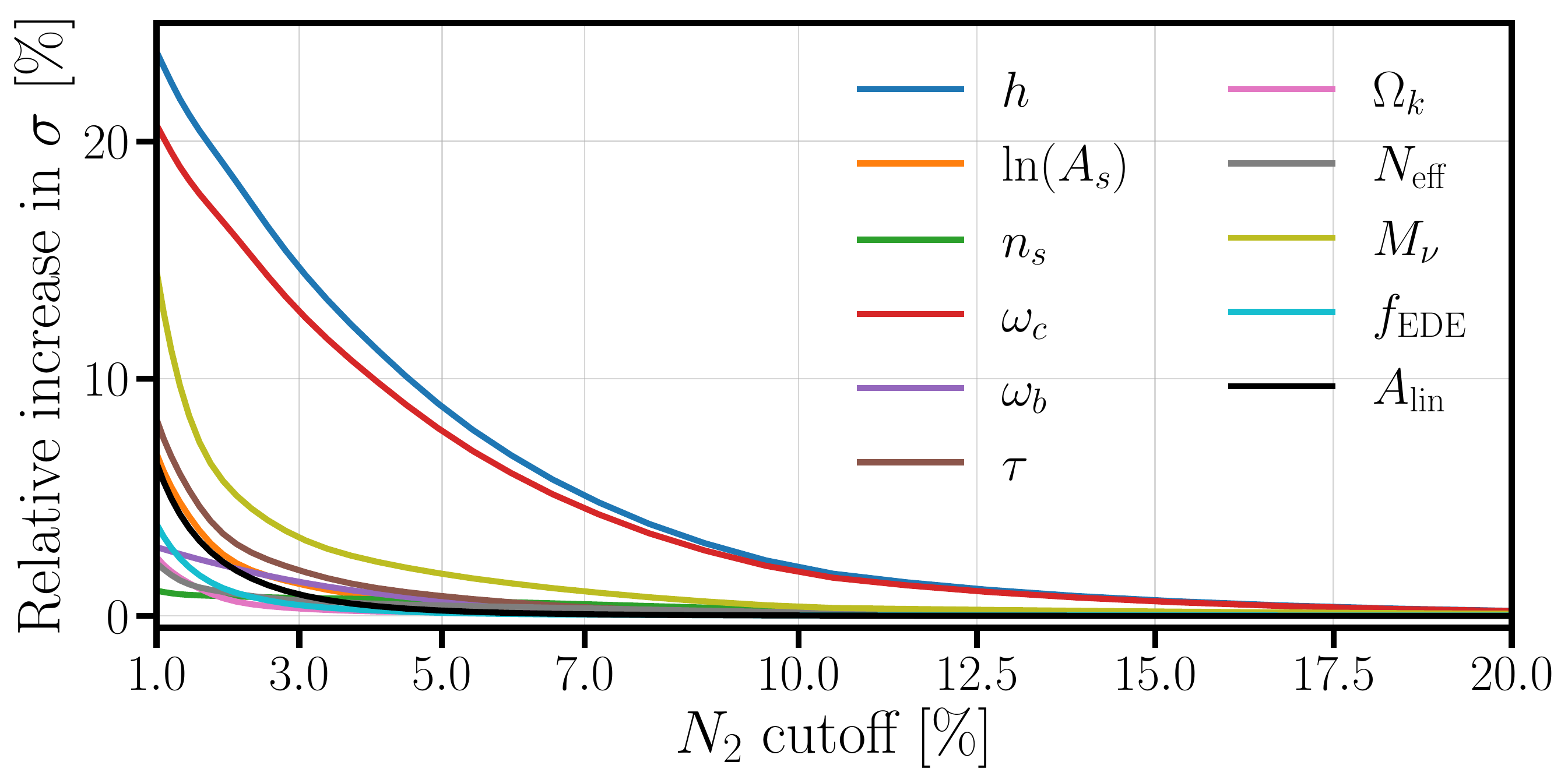}
    \caption{ Planck $+$ SO $+$ MegaMapper (full shape only) constraints as the $N_2$ cutoff is varied. This is with $\log_{10}(z_c)=3.61$ and $\omega_\text{lin}=147\,\,h^{-1}\,\text{Mpc}$. 
    }
\label{fig:N2_cutoff}
\end{figure}

As discussed in \S\ref{sec:full shape power spectrum}, we only integrate over modes where the $N_2$ term ($N_0 \sigma_v^2 (k\mu)^2$) is less than $20\%$ of the fiducial power. This ``$N_2$ cut'' is only relevant for MSE and MegaMapper, whose number densities are sufficiently low enough for the $N_2$ term to contribute a non-negligible amount to the fiducial power for $k<k_\text{nl}$. As we show in Fig.~\ref{fig:N2_cutoff}, our results are largely insensitive to the choice of $N_2$ cutoff. For any cutoff above $4\%$ the constraints from Planck $+$ SO $+$ MegaMapper change by at most $10\%$. 

To quantify the importance of nuisance parameters' fiducial values, we consider a ``MiniMapper'' survey that observes LBGs in a single redshift bin covering $2<z<3$ with $f_\text{sky}=0.34$. We assume that MiniMapper has $100\%$ efficiency for a limiting magnitude $m_\text{UV}=24.5$. The default values of MiniMapper's nuisance terms (following the conventions in \S\ref{sec:full shape power spectrum}) are listed in the first row of Table~\ref{tab:new_vs_old}. The constraints from Planck + MiniMapper on $\{h,\omega_c,\tau,\Omega_k,N_\text{eff},M_\nu\}$ with these default values are shown in the last row of Table~\ref{tab:different_nuisance}. 
\begin{table}
\centering
\begin{tabular}{c|ccccccc}
\hline
Parameter & $b_2$ & $b_s$ & $\alpha_0$ & $\alpha_2$ & $\alpha_4$ & $N_2$  & $N_4$ \\
\hline
\hline
Default value & 0.93 & -0.70 & -8.65 & 0 & 0 & -728 & 0\\
New value & 1.8 & -1.4 & 0 & 5 & 5 & 0 & 318\\
\hline
\end{tabular}
\caption{Default and new fiducial values for the nuisance parameters of the MiniMapper survey.
}
\label{tab:new_vs_old}
\end{table} 
The remaining rows of Table~\ref{tab:different_nuisance} show the constraints when the fiducial value of each of the seven nuisance terms is individually varied, with the new fiducial values listed in the second row of Table~\ref{tab:new_vs_old}. We find that varying the values of $\{\alpha_0,\alpha_2,N_2,\alpha_4\}$ has a negligible ($<6\%$) impact on our results, while varying the fiducial values of $\{b_2,b_s\}$ has less than a $27\%$ impact.  We should be able to refine our fiducial values as we learn more about the clustering of these high redshift samples.
\begin{table}
\centering
\begin{tabular}{c|ccccccccc}
\hline
Parameter changed & $10^4 h$ & $10^4\omega_c$ & $10^3 \tau$ &  $10^3 \Omega_k$ & $10^3N_{\rm eff}$ & $M_\nu\,\,[{\rm meV}]$\\
\hline
\hline
$b_2$ & $25.1$ & $5.82$ & $6.12$ & $0.71$ & $72.35$ & $34.02$ \\
$b_s$ & $22.77$ & $5.42$ & $4.96$ & $0.69$ & $77.17$ & $28.23$ \\
$\alpha_0$ & $19.58$ & $4.79$ & $5.35$ & $0.69$ & $81.02$ & $31.96$ \\
$\alpha_2$ & $19.4$ & $4.75$ & $5.34$ & $0.69$ & $80.38$ & $31.85$ \\
$\alpha_4$ & $19.23$ & $4.71$ & $5.34$ & $0.69$ & $79.88$ & $31.83$ \\
$N_2$ & $19.75$ & $4.83$ & $5.32$ & $0.69$ & $81.35$ & $31.73$ \\
$N_4$ & $19.69$ & $4.82$ & $5.31$ & $0.69$ & $81.23$ & $31.7$ \\
\hline
Default & $19.68$ & $4.82$ & $5.31$ & $0.69$ & $81.23$ & $31.7$ \\
\hline
\end{tabular}
\caption{Constraints on selected $\Lambda$CDM parameters and extensions for Planck $+$ MiniMapper. In each row we change the relevant nuisance parameter's fiducial value to the new value listed in Table~\ref{tab:new_vs_old}. The fiducial values of the remaining nuisance parameters are left unchanged. In the last row we show the constraints when all of the nuisance terms fiducial values' are set to the default values in Table~\ref{tab:new_vs_old}.
}
\label{tab:different_nuisance}
\end{table} 

We have also checked the sensitivity of Early Dark Energy constraints to the fiducial value of $f_\text{EDE}$. We find that the constraint on $f_\text{EDE}$ (for $\log_{10}(z_c)=3.5\text{ and }\theta_i=2.83$) from Planck $+$ MiniMapper changes by less than $3\%$ when the fiducial value of $f_\text{EDE}$ is increased from $0$ to $0.02$.

\section{Python implementation}
\label{app:code}

The Python code used to create these forecasts (\verb|FishLSS|) is publicly available\footnote{\href{https://github.com/NoahSailer/FishLSS}{https://github.com/NoahSailer/FishLSS}}. We include an example Jupyter notebook that walks the user through using the code in detail, and provide a brief overview of the organizational structure below. 

The primary module is \verb|fisherForecast.py|, which takes a cosmology (an instance of \verb|CLASS|) and experiment (an instance of \verb|experiment.py|) as inputs, and creates a Fisher forecast for a user-specified set of parameters and observables. \verb|fisherForecast.py| contains functions for calculating derivatives of power spectra, computing the 21-cm foreground wedge (under optimistic or pessimistic assumptions), computing Fisher matrices for both 3D and 2D clustering measurements, combining the Fisher matrices from two uncorrelated measurements, and calculating both $N_\text{modes}$ and $N_\text{modes}(k)$. \verb|fisherForecast.py| has three helper modules $-$ \verb|experiment.py|, \verb|twoPoint.py|, and \verb|twoPointNoise.py| $-$ which we describe below.

\verb|experiment.py| specifies the experimental configuration of the survey, including the redshift range, redshift bins, linear bias, nonlinear biases (optional), redshift errors, velocity dispersion for FoG-like contributions, and sky coverage. For 21-cm surveys, additional details of the interferometric array are stored in \verb|experiment.py|, such as the number of detectors, fill factor and observing time. We include a set of flags (\verb|LBG|, \verb|HI|, \verb|Halpha|, \verb|ELG|, \verb|Euclid|, \verb|MSE|) which may be switched to \verb|True| should the user wish to use the default biases and number densities of a specific survey described in \S\ref{sec:surveys and probes}.

Power spectra are computed with \verb|twoPoint.py|. In particular, \verb|twoPoint.py|  has several functions for computing the fiducial linear biases for the surveys and samples considered in this work, as well as functions for computing both the matter and galaxy power spectra in redshift space ($P_{mm}(\bm{k}),\,P_{gg}(\bm{k})$); the real space galaxy-matter cross spectrum $(P_{gm}(k))$; the CMB lensing auto-spectrum, cross-correlation with galaxies, and the projected galaxy auto-spectrum ($C^{\kappa\kappa}_\ell,\,C^{\kappa g}_\ell, C^{gg}_\ell$); and the reconstructed power spectrum. These functions require the installation of both \verb|CLASS|\footnote{\href{https://github.com/lesgourg/class_public}{https://github.com/lesgourg/class\_public}} \cite{CLASS} and \verb|velocileptors|\footnote{\href{https://github.com/sfschen/velocileptors}{https://github.com/sfschen/velocileptors}} \cite{Chen19a}, which in turn requires \verb|pyFFTW|\footnote{\href{https://hgomersall.github.io/pyFFTW/}{https://hgomersall.github.io/pyFFTW/}}. 

We calculate covariances (at the $\bm{k}$ or $\ell$-level) with \verb|twoPointNoise.py|. In particular, \verb|twoPointNoise.py| contains functions for computing the Poisson shot noise of each of the samples considered in this work, as well as functions for computing Eqs. \eqref{eqn:partial_fisher} and \eqref{eq:angular_covariance}. 

While we do not include our code for computing the CMB priors, the Fisher matrices used for Planck, SO, S4, and LiteBIRD can be found in the \verb|input| folder. 

\bibliographystyle{JHEP}
\bibliography{main}

\providecommand{\href}[2]{#2}\begingroup\raggedright\begin{thebibliography}{100}

\bibitem{Slosar19c}
A.~{Slosar}, R.~{Mandelbaum}, and D.~{Eisenstein}, {\it {Dark Energy and
  Modified Gravity}},  {\em \baas} {\bf 51} (May, 2019) 97,
  [\href{http://arxiv.org/abs/1903.12016}{{\tt arXiv:1903.12016}}].

\bibitem{Takada:2015mma}
M.~Takada and O.~Dore, {\it {Geometrical Constraint on Curvature with BAO
  experiments}},  {\em Phys. Rev. D} {\bf 92} (2015), no.~12 123518,
  [\href{http://arxiv.org/abs/1508.02469}{{\tt arXiv:1508.02469}}].

\bibitem{Meerburg19}
P.~D. {Meerburg}, D.~{Green}, R.~{Flauger}, et~al., {\it {Primordial
  Non-Gaussianity}},  {\em \baas} {\bf 51} (May, 2019) 107,
  [\href{http://arxiv.org/abs/1903.04409}{{\tt arXiv:1903.04409}}].

\bibitem{Alonso:2015uua}
D.~Alonso, P.~Bull, P.~G. Ferreira, et~al., {\it {Ultra large-scale cosmology
  in next-generation experiments with single tracers}},  {\em Astrophys. J.}
  {\bf 814} (2015), no.~2 145, [\href{http://arxiv.org/abs/1505.07596}{{\tt
  arXiv:1505.07596}}].

\bibitem{Slosar19b}
A.~{Slosar}, X.~{Chen}, C.~{Dvorkin}, et~al., {\it {Scratches from the Past:
  Inflationary Archaeology through Features in the Power Spectrum of Primordial
  Fluctuations}},  {\em \baas} {\bf 51} (May, 2019) 98,
  [\href{http://arxiv.org/abs/1903.09883}{{\tt arXiv:1903.09883}}].

\bibitem{Hill20}
J.~C. {Hill}, E.~{McDonough}, M.~W. {Toomey}, and S.~{Alexander}, {\it {Early
  Dark Energy Does Not Restore Cosmological Concordance}},  {\em arXiv
  e-prints} (Mar., 2020) arXiv:2003.07355,
  [\href{http://arxiv.org/abs/2003.07355}{{\tt arXiv:2003.07355}}].

\bibitem{Ivanov20}
M.~M. {Ivanov}, E.~{McDonough}, J.~C. {Hill}, et~al., {\it {Constraining Early
  Dark Energy with Large-Scale Structure}},  {\em arXiv e-prints} (June, 2020)
  arXiv:2006.11235, [\href{http://arxiv.org/abs/2006.11235}{{\tt
  arXiv:2006.11235}}].

\bibitem{DAmico20b}
G.~{D'Amico}, L.~{Senatore}, P.~{Zhang}, and H.~{Zheng}, {\it {The Hubble
  Tension in Light of the Full-Shape Analysis of Large-Scale Structure Data}},
  {\em arXiv e-prints} (June, 2020) arXiv:2006.12420,
  [\href{http://arxiv.org/abs/2006.12420}{{\tt arXiv:2006.12420}}].

\bibitem{Klypin20}
A.~{Klypin}, V.~{Poulin}, F.~{Prada}, et~al., {\it {Clustering and Halo
  Abundances in Early Dark Energy Cosmological Models}},  {\em arXiv e-prints}
  (June, 2020) arXiv:2006.14910, [\href{http://arxiv.org/abs/2006.14910}{{\tt
  arXiv:2006.14910}}].

\bibitem{Chen20b}
S.-F. {Chen}, Z.~{Vlah}, and M.~{White}, {\it {Modeling features in the
  redshift-space halo power spectrum with perturbation theory}},  {\em arXiv
  e-prints} (July, 2020) arXiv:2007.00704,
  [\href{http://arxiv.org/abs/2007.00704}{{\tt arXiv:2007.00704}}].

\bibitem{ferraro2019inflation}
S.~Ferraro et~al., {\it {Inflation and Dark Energy from Spectroscopy at $z >
  2$}},  \href{http://arxiv.org/abs/1903.09208}{{\tt arXiv:1903.09208}}.

\bibitem{Castorina20}
E.~{Castorina}, S.~{Foreman}, D.~{Karagiannis}, et~al., {\it {Packed
  Ultra-wideband Mapping Array (PUMA): Astro2020 RFI Response}},  {\em arXiv
  e-prints} (Feb., 2020) arXiv:2002.05072,
  [\href{http://arxiv.org/abs/2002.05072}{{\tt arXiv:2002.05072}}].

\bibitem{PlanckInf18}
{Planck Collaboration}, Y.~{Akrami}, F.~{Arroja}, et~al., {\it {Planck 2018
  results. X. Constraints on inflation}},  {\em arXiv e-prints} (July, 2018)
  arXiv:1807.06211, [\href{http://arxiv.org/abs/1807.06211}{{\tt
  arXiv:1807.06211}}].

\bibitem{PDE15}
{Planck Collaboration}, P.~A.~R. {Ade}, N.~{Aghanim}, et~al., {\it {Planck 2015
  results. XX. Constraints on inflation}},  {\em \aap} {\bf 594} (Sept., 2016)
  A20, [\href{http://arxiv.org/abs/1502.02114}{{\tt arXiv:1502.02114}}].

\bibitem{Slosar19a}
A.~{Slosar}, Z.~{Ahmed}, D.~{Alonso}, et~al., {\it {Packed Ultra-wideband
  Mapping Array (PUMA): A Radio Telescope for Cosmology and Transients}},  in
  {\em \baas}, vol.~51, p.~53, Sep, 2019.
\newblock \href{http://arxiv.org/abs/1907.12559}{{\tt arXiv:1907.12559}}.

\bibitem{CMBS4}
K.~N. {Abazajian}, P.~{Adshead}, Z.~{Ahmed}, et~al., {\it {CMB-S4 Science Book,
  First Edition}},  {\em ArXiv e-prints} (Oct., 2016)
  [\href{http://arxiv.org/abs/1610.02743}{{\tt arXiv:1610.02743}}].

\bibitem{Dvorkin19}
C.~{Dvorkin}, M.~{Gerbino}, D.~{Alonso}, et~al., {\it {Neutrino Mass from
  Cosmology: Probing Physics Beyond the Standard Model}},  {\em \baas} {\bf 51}
  (May, 2019) 64, [\href{http://arxiv.org/abs/1903.03689}{{\tt
  arXiv:1903.03689}}].

\bibitem{Green19}
D.~{Green}, M.~A. {Amin}, J.~{Meyers}, et~al., {\it {Messengers from the Early
  Universe: Cosmic Neutrinos and Other Light Relics}},  {\em \baas} {\bf 51}
  (May, 2019) 159, [\href{http://arxiv.org/abs/1903.04763}{{\tt
  arXiv:1903.04763}}].

\bibitem{Jain10}
B.~{Jain} and J.~{Khoury}, {\it {Cosmological tests of gravity}},  {\em Annals
  of Physics} {\bf 325} (Jul, 2010) 1479--1516,
  [\href{http://arxiv.org/abs/1004.3294}{{\tt arXiv:1004.3294}}].

\bibitem{Joyce16}
A.~{Joyce}, L.~{Lombriser}, and F.~{Schmidt}, {\it {Dark Energy Versus Modified
  Gravity}},  {\em Annual Review of Nuclear and Particle Science} {\bf 66}
  (Oct., 2016) 95--122, [\href{http://arxiv.org/abs/1601.06133}{{\tt
  arXiv:1601.06133}}].

\bibitem{EUCLID18}
L.~{Amendola}, S.~{Appleby}, A.~{Avgoustidis}, et~al., {\it {Cosmology and
  fundamental physics with the Euclid satellite}},  {\em Living Reviews in
  Relativity} {\bf 21} (Apr., 2018) 2,
  [\href{http://arxiv.org/abs/1606.00180}{{\tt arXiv:1606.00180}}].

\bibitem{Castorina19}
E.~{Castorina} and M.~{White}, {\it {Measuring the growth of structure with
  intensity mapping surveys}},  {\em \jcap} {\bf 2019} (June, 2019) 025,
  [\href{http://arxiv.org/abs/1902.07147}{{\tt arXiv:1902.07147}}].

\bibitem{PlanckLegacy18}
{Planck Collaboration}, Y.~{Akrami}, F.~{Arroja}, et~al., {\it {Planck 2018
  results. I. Overview and the cosmological legacy of Planck}},  {\em arXiv
  e-prints} (July, 2018) arXiv:1807.06205,
  [\href{http://arxiv.org/abs/1807.06205}{{\tt arXiv:1807.06205}}].

\bibitem{Chabanier_2019}
S.~Chabanier, M.~Millea, and N.~Palanque-Delabrouille, {\it {Matter power
  spectrum: from Ly$\alpha$ forest to CMB scales}},  {\em Mon. Not. Roy.
  Astron. Soc.} {\bf 489} (2019), no.~2 2247--2253,
  [\href{http://arxiv.org/abs/1905.08103}{{\tt arXiv:1905.08103}}].

\bibitem{DESI}
{DESI Collaboration}, A.~{Aghamousa}, J.~{Aguilar}, et~al., {\it {The DESI
  Experiment Part I: Science,Targeting, and Survey Design}},  {\em ArXiv
  e-prints} (Oct., 2016) [\href{http://arxiv.org/abs/1611.00036}{{\tt
  arXiv:1611.00036}}].

\bibitem{Euclid2020prep}
A.~Blanchard, S.~Camera, C.~Carbone, et~al., {\it Euclid preparation},  {\em
  \aa} {\bf 642} (Oct, 2020) A191, [\href{http://arxiv.org/abs/1910.09273}{{\tt
  arXiv:1910.09273}}].

\bibitem{Ilic:2021dwx}
S.~Ili\'c et~al., {\it {$Euclid$ preparation: XV. Forecasting cosmological
  constraints for the $Euclid$ and CMB joint analysis}},
  \href{http://arxiv.org/abs/2106.08346}{{\tt arXiv:2106.08346}}.

\bibitem{Dore14}
O.~{Dor{\'e}}, J.~{Bock}, M.~{Ashby}, et~al., {\it {Cosmology with the SPHEREX
  All-Sky Spectral Survey}},  {\em arXiv e-prints} (Dec., 2014)
  arXiv:1412.4872, [\href{http://arxiv.org/abs/1412.4872}{{\tt
  arXiv:1412.4872}}].

\bibitem{MegaMapper}
D.~{Schlegel}, J.~A. {Kollmeier}, and S.~{Ferraro}, {\it {The MegaMapper: a
  $z>2$ spectroscopic instrument for the study of Inflation and Dark Energy}},
  in {\em \baas}, vol.~51, p.~229, Sept., 2019.
\newblock \href{http://arxiv.org/abs/1907.11171}{{\tt arXiv:1907.11171}}.

\bibitem{Wilson19}
M.~J. {Wilson} and M.~{White}, {\it {Cosmology with dropout selection:
  straw-man surveys \&amp; CMB lensing}},  {\em \jcap} {\bf 2019} (Oct., 2019)
  015, [\href{http://arxiv.org/abs/1904.13378}{{\tt arXiv:1904.13378}}].

\bibitem{MSE}
{The MSE Science Team}, C.~{Babusiaux}, M.~{Bergemann}, et~al., {\it {The
  Detailed Science Case for the Maunakea Spectroscopic Explorer, 2019
  edition}},  {\em arXiv e-prints} (Apr., 2019) arXiv:1904.04907,
  [\href{http://arxiv.org/abs/1904.04907}{{\tt arXiv:1904.04907}}].

\bibitem{percival2019cosmology}
W.~J. Percival et~al., {\it {Cosmology with the MaunaKea Spectroscopic
  Explorer}},  \href{http://arxiv.org/abs/1903.03158}{{\tt arXiv:1903.03158}}.

\bibitem{Blanchard:2021ffq}
A.~Blanchard et~al., {\it {Gravitation And the Universe from large
  Scale-Structures: The GAUSS mission concept. Mapping the cosmic web up to the
  reionization era}},  \href{http://arxiv.org/abs/2102.03931}{{\tt
  arXiv:2102.03931}}.

\bibitem{SpecTel}
R.~{Ellis} and K.~{Dawson}, {\it {SpecTel: A 10-12 meter class Spectroscopic
  Survey Telescope}},  in {\em \baas}, vol.~51, p.~45, Sept., 2019.
\newblock \href{http://arxiv.org/abs/1907.06797}{{\tt arXiv:1907.06797}}.

\bibitem{dawson2018cosmic}
K.~Dawson, J.~Frieman, K.~Heitmann, et~al., {\it {Cosmic Visions Dark Energy:
  Small Projects Portfolio}},  \href{http://arxiv.org/abs/1802.07216}{{\tt
  arXiv:1802.07216}}.

\bibitem{Roman}
D.~{Spergel}, N.~{Gehrels}, C.~{Baltay}, et~al., {\it {Wide-Field InfrarRed
  Survey Telescope-Astrophysics Focused Telescope Assets WFIRST-AFTA 2015
  Report}},  {\em arXiv e-prints} (Mar., 2015) arXiv:1503.03757,
  [\href{http://arxiv.org/abs/1503.03757}{{\tt arXiv:1503.03757}}].

\bibitem{CHIME}
K.~{Bandura}, G.~E. {Addison}, M.~{Amiri}, et~al., {\it {Canadian Hydrogen
  Intensity Mapping Experiment (CHIME) pathfinder}},  in {\em Ground-based and
  Airborne Telescopes V}, vol.~9145 of {\em \procspie}, p.~914522, July, 2014.
\newblock \href{http://arxiv.org/abs/1406.2288}{{\tt arXiv:1406.2288}}.

\bibitem{Tianlai}
X.~{Chen}, {\it {The Tianlai Project: a 21CM Cosmology Experiment}},  in {\em
  International Journal of Modern Physics Conference Series}, vol.~12 of {\em
  International Journal of Modern Physics Conference Series}, pp.~256--263,
  March, 2012.
\newblock \href{http://arxiv.org/abs/1212.6278}{{\tt arXiv:1212.6278}}.

\bibitem{CHORD}
K.~{Vanderlinde}, A.~{Liu}, B.~{Gaensler}, et~al., {\it {The Canadian Hydrogen
  Observatory and Radio-transient Detector (CHORD)}},  in {\em Canadian Long
  Range Plan for Astronomy and Astrophysics White Papers}, vol.~2020, p.~28,
  Oct., 2019.
\newblock \href{http://arxiv.org/abs/1911.01777}{{\tt arXiv:1911.01777}}.

\bibitem{Newburgh_2016}
L.~B. Newburgh et~al., {\it {HIRAX: A Probe of Dark Energy and Radio
  Transients}},  {\em Proc. SPIE Int. Soc. Opt. Eng.} {\bf 9906} (2016) 99065X,
  [\href{http://arxiv.org/abs/1607.02059}{{\tt arXiv:1607.02059}}].

\bibitem{SKACosmo}
F.~B. {Abdalla}, P.~{Bull}, S.~{Camera}, et~al., {\it {Cosmology from HI galaxy
  surveys with the SKA}},  {\em Advancing Astrophysics with the Square
  Kilometre Array (AASKA14)} (April, 2015) 17,
  [\href{http://arxiv.org/abs/1501.04035}{{\tt arXiv:1501.04035}}].

\bibitem{collaboration2018inflation}
{\bf Cosmic Visions 21 cm} Collaboration, R.~Ansari et~al., {\it {Inflation and
  Early Dark Energy with a Stage II Hydrogen Intensity Mapping experiment}},
  \href{http://arxiv.org/abs/1810.09572}{{\tt arXiv:1810.09572}}.

\bibitem{VN18}
F.~{Villaescusa-Navarro}, S.~{Genel}, E.~{Castorina}, et~al., {\it {Ingredients
  for 21 cm Intensity Mapping}},  {\em \apj} {\bf 866} (Oct, 2018) 135,
  [\href{http://arxiv.org/abs/1804.09180}{{\tt arXiv:1804.09180}}].

\bibitem{Chen19a}
S.-F. {Chen}, E.~{Castorina}, M.~{White}, and A.~{Slosar}, {\it {Synergies
  between radio, optical and microwave observations at high redshift}},  {\em
  \jcap} {\bf 2019} (July, 2019) 023,
  [\href{http://arxiv.org/abs/1810.00911}{{\tt arXiv:1810.00911}}].

\bibitem{Sheth_2001}
R.~K. Sheth, H.~J. Mo, and G.~Tormen, {\it Ellipsoidal collapse and an improved
  model for the number and spatial distribution of dark matter haloes},  {\em
  Monthly Notices of the Royal Astronomical Society} {\bf 323} (May, 2001)
  1–12.

\bibitem{Bull2014}
P.~Bull, P.~G. Ferreira, P.~Patel, and M.~G. Santos, {\it {Late-time cosmology
  with 21cm intensity mapping experiments}},  {\em Astrophys. J.} {\bf 803}
  (2015), no.~1 21, [\href{http://arxiv.org/abs/1405.1452}{{\tt
  arXiv:1405.1452}}].

\bibitem{Seo2010}
H.-J. Seo, S.~Dodelson, J.~Marriner, et~al., {\it A {GROUND}-{BASED} 21 cm
  {BARYON} {ACOUSTIC} {OSCILLATION} {SURVEY}},  {\em The Astrophysical Journal}
  {\bf 721} (aug, 2010) 164--173.

\bibitem{Castorina_2017}
E.~Castorina and F.~Villaescusa-Navarro, {\it On the spatial distribution of
  neutral hydrogen in the universe: bias and shot-noise of the h{\sc i} power
  spectrum},  {\em Monthly Notices of the Royal Astronomical Society} {\bf 471}
  (Jun, 2017) 1788–1796.

\bibitem{Obuljen2017}
A.~Obuljen, E.~Castorina, F.~Villaescusa-Navarro, and M.~Viel, {\it
  {High-redshift post-reionization cosmology with 21cm intensity mapping}},
  {\em JCAP} {\bf 05} (2018) 004, [\href{http://arxiv.org/abs/1709.07893}{{\tt
  arXiv:1709.07893}}].

\bibitem{Liu20}
A.~{Liu} and J.~R. {Shaw}, {\it {Data Analysis for Precision 21 cm Cosmology}},
   {\em \pasp} {\bf 132} (June, 2020) 062001,
  [\href{http://arxiv.org/abs/1907.08211}{{\tt arXiv:1907.08211}}].

\bibitem{Bagley20}
M.~B. {Bagley}, C.~{Scarlata}, V.~{Mehta}, et~al., {\it {HST Grism-derived
  Forecasts for Future Galaxy Redshift Surveys}},  {\em arXiv e-prints} (June,
  2020) arXiv:2006.03602, [\href{http://arxiv.org/abs/2006.03602}{{\tt
  arXiv:2006.03602}}].

\bibitem{Dore16}
O.~{Dor{\'e}}, M.~W. {Werner}, M.~{Ashby}, et~al., {\it {Science Impacts of the
  SPHEREx All-Sky Optical to Near-Infrared Spectral Survey: Report of a
  Community Workshop Examining Extragalactic, Galactic, Stellar and Planetary
  Science}},  {\em arXiv e-prints} (June, 2016) arXiv:1606.07039,
  [\href{http://arxiv.org/abs/1606.07039}{{\tt arXiv:1606.07039}}].

\bibitem{SimonsObs}
N.~{Galitzki}, A.~{Ali}, K.~S. {Arnold}, et~al., {\it {The Simons Observatory:
  instrument overview}},  in {\em Millimeter, Submillimeter, and Far-Infrared
  Detectors and Instrumentation for Astronomy IX}, vol.~10708 of {\em Society
  of Photo-Optical Instrumentation Engineers (SPIE) Conference Series},
  p.~1070804, July, 2018.
\newblock \href{http://arxiv.org/abs/1808.04493}{{\tt arXiv:1808.04493}}.

\bibitem{LiteBIRD}
H.~Sugai, P.~A.~R. Ade, Y.~Akiba, et~al., {\it Updated design of the cmb
  polarization experiment satellite litebird},  {\em Journal of Low Temperature
  Physics} {\bf 199} (Jan, 2020) 1107–1117.

\bibitem{Dunkley:2013vu}
J.~Dunkley et~al., {\it {The Atacama Cosmology Telescope: likelihood for
  small-scale CMB data}},  {\em JCAP} {\bf 07} (2013) 025,
  [\href{http://arxiv.org/abs/1301.0776}{{\tt arXiv:1301.0776}}].

\bibitem{McCarthy:2021lfp}
F.~McCarthy, J.~C. Hill, and M.~S. Madhavacheril, {\it {Baryonic feedback
  biases on fundamental physics from lensed CMB power spectra}},
  \href{http://arxiv.org/abs/2103.05582}{{\tt arXiv:2103.05582}}.

\bibitem{Pico}
{\bf NASA PICO} Collaboration, S.~Hanany et~al., {\it {PICO: Probe of Inflation
  and Cosmic Origins}},  \href{http://arxiv.org/abs/1902.10541}{{\tt
  arXiv:1902.10541}}.

\bibitem{PCP18}
{Planck Collaboration}, N.~{Aghanim}, Y.~{Akrami}, et~al., {\it {Planck 2018
  results. VI. Cosmological parameters}},  {\em arXiv e-prints} (July, 2018)
  arXiv:1807.06209, [\href{http://arxiv.org/abs/1807.06209}{{\tt
  arXiv:1807.06209}}].

\bibitem{DESI2013}
A.~Font-Ribera, P.~McDonald, N.~Mostek, et~al., {\it {DESI and other dark
  energy experiments in the era of neutrino mass measurements}},  {\em JCAP}
  {\bf 05} (2014) 023, [\href{http://arxiv.org/abs/1308.4164}{{\tt
  arXiv:1308.4164}}].

\bibitem{Schaan2020}
E.~Schaan, S.~Ferraro, and U.~Seljak, {\it {Photo-z outlier self-calibration in
  weak lensing surveys}},  {\em JCAP} {\bf 12} (2020) 001,
  [\href{http://arxiv.org/abs/2007.12795}{{\tt arXiv:2007.12795}}].

\bibitem{Yu2018}
B.~Yu, R.~Z. Knight, B.~D. Sherwin, et~al., {\it {Towards Neutrino Mass from
  Cosmology without Optical Depth Information}},
  \href{http://arxiv.org/abs/1809.02120}{{\tt arXiv:1809.02120}}.

\bibitem{Schmittfull17}
M.~{Schmittfull}, T.~{Baldauf}, and M.~{Zaldarriaga}, {\it {Iterative initial
  condition reconstruction}},  {\em \prd} {\bf 96} (July, 2017) 023505,
  [\href{http://arxiv.org/abs/1704.06634}{{\tt arXiv:1704.06634}}].

\bibitem{Brinckmann2018}
T.~Brinckmann, D.~C. Hooper, M.~Archidiacono, et~al., {\it {The promising
  future of a robust cosmological neutrino mass measurement}},  {\em JCAP} {\bf
  01} (2019) 059, [\href{http://arxiv.org/abs/1808.05955}{{\tt
  arXiv:1808.05955}}].

\bibitem{Vagnozzi2018}
S.~Vagnozzi, T.~Brinckmann, M.~Archidiacono, et~al., {\it {Bias due to
  neutrinos must not uncorrect'd go}},  {\em JCAP} {\bf 09} (2018) 001,
  [\href{http://arxiv.org/abs/1807.04672}{{\tt arXiv:1807.04672}}].

\bibitem{LSST}
{\bf LSST Dark Energy Science} Collaboration, D.~Alonso et~al., {\it {The LSST
  Dark Energy Science Collaboration (DESC) Science Requirements Document}},
  \href{http://arxiv.org/abs/1809.01669}{{\tt arXiv:1809.01669}}.

\bibitem{Chudaykin19}
A.~{Chudaykin} and M.~M. {Ivanov}, {\it {Measuring neutrino masses with
  large-scale structure: Euclid forecast with controlled theoretical error}},
  {\em \jcap} {\bf 2019} (Nov, 2019) 034,
  [\href{http://arxiv.org/abs/1907.06666}{{\tt arXiv:1907.06666}}].

\bibitem{Boyle20}
A.~{Boyle} and F.~{Schmidt}, {\it {Neutrino mass constraints beyond linear
  order: cosmology dependence and systematic biases}},  {\em arXiv e-prints}
  (Nov., 2020) arXiv:2011.10594, [\href{http://arxiv.org/abs/2011.10594}{{\tt
  arXiv:2011.10594}}].

\bibitem{chen2021precise}
S.-F. Chen, H.~Lee, and C.~Dvorkin, {\it Precise and accurate cosmology with
  cmbxlss power spectra and bispectra},  2021.

\bibitem{Mead20}
A.~{Mead} and L.~{Verde}, {\it {Including beyond-linear halo bias in halo
  models}},  {\em arXiv e-prints} (Nov., 2020) arXiv:2011.08858,
  [\href{http://arxiv.org/abs/2011.08858}{{\tt arXiv:2011.08858}}].

\bibitem{Chen20c}
S.-F. {Chen}, Z.~{Vlah}, E.~{Castorina}, and M.~{White}, {\it {Redshift-Space
  Distortions in Lagrangian Perturbation Theory}},  {\em arXiv e-prints} (Dec.,
  2020) arXiv:2012.04636, [\href{http://arxiv.org/abs/2012.04636}{{\tt
  arXiv:2012.04636}}].

\bibitem{CLASS}
D.~{Blas}, J.~{Lesgourgues}, and T.~{Tram}, {\it {The Cosmic Linear Anisotropy
  Solving System (CLASS). Part II: Approximation schemes}},  {\em \jcap} {\bf
  7} (July, 2011) 034, [\href{http://arxiv.org/abs/1104.2933}{{\tt
  arXiv:1104.2933}}].

\bibitem{Chan2012}
K.~C. Chan, R.~Scoccimarro, and R.~K. Sheth, {\it {Gravity and Large-Scale
  Non-local Bias}},  {\em Phys. Rev. D} {\bf 85} (2012) 083509,
  [\href{http://arxiv.org/abs/1201.3614}{{\tt arXiv:1201.3614}}].

\bibitem{Baldauf2012}
T.~Baldauf, U.~Seljak, V.~Desjacques, and P.~McDonald, {\it Evidence for
  quadratic tidal tensor bias from the halo bispectrum},  {\em Phys. Rev. D}
  {\bf 86} (Oct, 2012) 083540.

\bibitem{Takahashi12}
R.~{Takahashi}, M.~{Sato}, T.~{Nishimichi}, et~al., {\it {Revising the Halofit
  Model for the Nonlinear Matter Power Spectrum}},  {\em \apj} {\bf 761} (Dec.,
  2012) 152, [\href{http://arxiv.org/abs/1208.2701}{{\tt arXiv:1208.2701}}].

\bibitem{Villaescusa-Navarro2018}
F.~Villaescusa-Navarro et~al., {\it {Ingredients for 21 cm Intensity Mapping}},
   {\em Astrophys. J.} {\bf 866} (2018), no.~2 135,
  [\href{http://arxiv.org/abs/1804.09180}{{\tt arXiv:1804.09180}}].

\bibitem{Field59}
G.~B. {Field}, {\it {An Attempt to Observe Neutral Hydrogen Between the
  Galaxies.}},  {\em \apj} {\bf 129} (May, 1959) 525.

\bibitem{Crighton2015}
N.~H.~M. Crighton et~al., {\it {The neutral hydrogen cosmological mass density
  at z = 5}},  {\em Mon. Not. Roy. Astron. Soc.} {\bf 452} (2015), no.~1
  217--234, [\href{http://arxiv.org/abs/1506.02037}{{\tt arXiv:1506.02037}}].

\bibitem{Villaescusa-Navarro2013}
F.~Villaescusa-Navarro, F.~Marulli, M.~Viel, et~al., {\it {Cosmology with
  massive neutrinos I: towards a realistic modeling of the relation between
  matter, haloes and galaxies}},  {\em JCAP} {\bf 03} (2014) 011,
  [\href{http://arxiv.org/abs/1311.0866}{{\tt arXiv:1311.0866}}].

\bibitem{Castorina2013}
E.~Castorina, E.~Sefusatti, R.~K. Sheth, et~al., {\it {Cosmology with massive
  neutrinos II: on the universality of the halo mass function and bias}},  {\em
  JCAP} {\bf 02} (2014) 049, [\href{http://arxiv.org/abs/1311.1212}{{\tt
  arXiv:1311.1212}}].

\bibitem{Castorina2015}
E.~Castorina, C.~Carbone, J.~Bel, et~al., {\it {DEMNUni: The clustering of
  large-scale structures in the presence of massive neutrinos}},  {\em JCAP}
  {\bf 07} (2015) 043, [\href{http://arxiv.org/abs/1505.07148}{{\tt
  arXiv:1505.07148}}].

\bibitem{Villaescusa-Navarro2017}
F.~Villaescusa-Navarro, A.~Banerjee, N.~Dalal, et~al., {\it {The imprint of
  neutrinos on clustering in redshift-space}},  {\em Astrophys. J.} {\bf 861}
  (2018), no.~1 53, [\href{http://arxiv.org/abs/1708.01154}{{\tt
  arXiv:1708.01154}}].

\bibitem{LoVerde2014}
M.~LoVerde, {\it {Halo bias in mixed dark matter cosmologies}},  {\em Phys.
  Rev. D} {\bf 90} (2014), no.~8 083530,
  [\href{http://arxiv.org/abs/1405.4855}{{\tt arXiv:1405.4855}}].

\bibitem{Munoz2018}
J.~B. Mu\~noz and C.~Dvorkin, {\it {Efficient Computation of Galaxy Bias with
  Neutrinos and Other Relics}},  {\em Phys. Rev. D} {\bf 98} (2018), no.~4
  043503, [\href{http://arxiv.org/abs/1805.11623}{{\tt arXiv:1805.11623}}].

\bibitem{Fidler2018}
C.~Fidler, N.~Sujata, and M.~Archidiacono, {\it {Relativistic bias in neutrino
  cosmologies}},  {\em JCAP} {\bf 06} (2019) 035,
  [\href{http://arxiv.org/abs/1812.09266}{{\tt arXiv:1812.09266}}].

\bibitem{Aviles:2020cax}
A.~Aviles and A.~Banerjee, {\it {A Lagrangian Perturbation Theory in the
  presence of massive neutrinos}},  {\em JCAP} {\bf 10} (2020) 034,
  [\href{http://arxiv.org/abs/2007.06508}{{\tt arXiv:2007.06508}}].

\bibitem{Wadekar20}
D.~{Wadekar}, M.~M. {Ivanov}, and R.~{Scoccimarro}, {\it {Cosmological
  constraints from BOSS with analytic covariance matrices}},  {\em \prd} {\bf
  102} (Dec., 2020) 123521, [\href{http://arxiv.org/abs/2009.00622}{{\tt
  arXiv:2009.00622}}].

\bibitem{LEWIS_2006}
A.~LEWIS and A.~CHALLINOR, {\it Weak gravitational lensing of the cmb},  {\em
  Physics Reports} {\bf 429} (Jun, 2006) 1–65.

\bibitem{Bermejo-Climent:2021jxf}
J.~R. Bermejo-Climent, M.~Ballardini, F.~Finelli, et~al., {\it {Cosmological
  parameter forecasts by a joint 2D tomographic approach to CMB and galaxy
  clustering}},  {\em Phys. Rev. D} {\bf 103} (2021), no.~10 103502,
  [\href{http://arxiv.org/abs/2106.05267}{{\tt arXiv:2106.05267}}].

\bibitem{LoVerde_2008}
M.~LoVerde and N.~Afshordi, {\it Extended limber approximation},  {\em Physical
  Review D} {\bf 78} (Dec, 2008).

\bibitem{LovAfs08}
M.~{Loverde} and N.~{Afshordi}, {\it {Extended Limber approximation}},  {\em
  \prd} {\bf 78} (Dec., 2008) 123506,
  [\href{http://arxiv.org/abs/0809.5112}{{\tt arXiv:0809.5112}}].

\bibitem{Krause2010}
{Krause, E.} and {Hirata, C. M.}, {\it Weak lensing power spectra for precision
  cosmology - multiple-deflection, reduced shear, and lensing bias
  corrections},  {\em A\&A} {\bf 523} (2010) A28.

\bibitem{Pratten2016}
G.~Pratten and A.~Lewis, {\it {Impact of post-Born lensing on the CMB}},  {\em
  JCAP} {\bf 08} (2016) 047, [\href{http://arxiv.org/abs/1605.05662}{{\tt
  arXiv:1605.05662}}].

\bibitem{Marozzi2016}
G.~Marozzi, G.~Fanizza, E.~Di~Dio, and R.~Durrer, {\it {CMB-lensing beyond the
  Born approximation}},  {\em JCAP} {\bf 09} (2016) 028,
  [\href{http://arxiv.org/abs/1605.08761}{{\tt arXiv:1605.08761}}].

\bibitem{Fabbian2017}
G.~Fabbian, M.~Calabrese, and C.~Carbone, {\it {CMB weak-lensing beyond the
  Born approximation: a numerical approach}},  {\em JCAP} {\bf 02} (2018) 050,
  [\href{http://arxiv.org/abs/1702.03317}{{\tt arXiv:1702.03317}}].

\bibitem{Fabbian2019}
G.~Fabbian, A.~Lewis, and D.~Beck, {\it {CMB lensing reconstruction biases in
  cross-correlation with large-scale structure probes}},  {\em JCAP} {\bf 10}
  (2019) 057, [\href{http://arxiv.org/abs/1906.08760}{{\tt arXiv:1906.08760}}].

\bibitem{Boehm2019}
V.~B\"ohm, C.~Modi, and E.~Castorina, {\it {Lensing corrections on
  galaxy-lensing cross correlations and galaxy-galaxy auto correlations}},
  {\em JCAP} {\bf 03} (2020) 045, [\href{http://arxiv.org/abs/1910.06722}{{\tt
  arXiv:1910.06722}}].

\bibitem{Modi_2017}
C.~Modi, M.~White, and Z.~Vlah, {\it Modeling cmb lensing cross correlations
  with cleft},  {\em Journal of Cosmology and Astroparticle Physics} {\bf 2017}
  (Aug, 2017) 009–009.

\bibitem{Foreman2015}
S.~Foreman and L.~Senatore, {\it {The EFT of Large Scale Structures at All
  Redshifts: Analytical Predictions for Lensing}},  {\em JCAP} {\bf 04} (2016)
  033, [\href{http://arxiv.org/abs/1503.01775}{{\tt arXiv:1503.01775}}].

\bibitem{Braganca2020}
D.~P.~L. Bragan\c{c}a, M.~Lewandowski, D.~Sekera, et~al., {\it {Baryonic
  effects in the Effective Field Theory of Large-Scale Structure and an
  analytic recipe for lensing in CMB-S4}},
  \href{http://arxiv.org/abs/2010.02929}{{\tt arXiv:2010.02929}}.

\bibitem{McCarth2020}
F.~McCarthy, S.~Foreman, and A.~van Engelen, {\it {Avoiding baryonic feedback
  effects on neutrino mass measurements from CMB lensing}},
  \href{http://arxiv.org/abs/2011.06582}{{\tt arXiv:2011.06582}}.

\bibitem{Chung2019}
E.~Chung, S.~Foreman, and A.~van Engelen, {\it {Baryonic effects on CMB lensing
  and neutrino mass constraints}},  {\em Phys. Rev. D} {\bf 101} (2020), no.~6
  063534, [\href{http://arxiv.org/abs/1910.09565}{{\tt arXiv:1910.09565}}].
  [Erratum: Phys.Rev.D 102, 109903 (2020)].

\bibitem{Chung_2020}
E.~Chung, S.~Foreman, and A.~van Engelen, {\it Baryonic effects on cmb lensing
  and neutrino mass constraints},  {\em Physical Review D} {\bf 101} (Mar,
  2020).

\bibitem{AP}
C.~{Alcock} and B.~{Paczynski}, {\it {An evolution free test for non-zero
  cosmological constant}},  {\em \nat} {\bf 281} (Oct., 1979) 358.

\bibitem{Seo_2003}
H.~Seo and D.~J. Eisenstein, {\it Probing dark energy with baryonic acoustic
  oscillations from future large galaxy redshift surveys},  {\em The
  Astrophysical Journal} {\bf 598} (Dec, 2003) 720–740.

\bibitem{Wei13}
D.~H. {Weinberg}, M.~J. {Mortonson}, D.~J. {Eisenstein}, et~al., {\it
  {Observational probes of cosmic acceleration}},  {\em \physrep} {\bf 530}
  (Sept., 2013) 87--255, [\href{http://arxiv.org/abs/1201.2434}{{\tt
  arXiv:1201.2434}}].

\bibitem{ESSS07}
D.~J. {Eisenstein}, H.-J. {Seo}, E.~{Sirko}, and D.~N. {Spergel}, {\it
  {Improving Cosmological Distance Measurements by Reconstruction of the Baryon
  Acoustic Peak}},  {\em \apj} {\bf 664} (Aug., 2007) 675--679,
  [\href{http://xxx.lanl.gov/abs/astro-ph/0604362}{{\tt astro-ph/0604362}}].

\bibitem{Pad12}
N.~{Padmanabhan}, X.~{Xu}, D.~J. {Eisenstein}, et~al., {\it {A 2 per cent
  distance to z = 0.35 by reconstructing baryon acoustic oscillations - I.
  Methods and application to the Sloan Digital Sky Survey}},  {\em \mnras} {\bf
  427} (Dec., 2012) 2132--2145, [\href{http://arxiv.org/abs/1202.0090}{{\tt
  arXiv:1202.0090}}].

\bibitem{Chen19b}
S.-F. {Chen}, Z.~{Vlah}, and M.~{White}, {\it {The reconstructed power spectrum
  in the Zeldovich approximation}},  {\em \jcap} {\bf 2019} (Sept., 2019) 017,
  [\href{http://arxiv.org/abs/1907.00043}{{\tt arXiv:1907.00043}}].

\bibitem{BOSS_DR12}
S.~{Alam}, M.~{Ata}, S.~{Bailey}, et~al., {\it {The clustering of galaxies in
  the completed SDSS-III Baryon Oscillation Spectroscopic Survey: cosmological
  analysis of the DR12 galaxy sample}},  {\em \mnras} {\bf 470} (Sept., 2017)
  2617--2652, [\href{http://arxiv.org/abs/1607.03155}{{\tt arXiv:1607.03155}}].

\bibitem{eBOSS21}
{eBOSS Collaboration}, S.~{Alam}, M.~{Aubert}, et~al., {\it {The Completed
  SDSS-IV extended Baryon Oscillation Spectroscopic Survey: Cosmological
  Implications from two Decades of Spectroscopic Surveys at the Apache Point
  observatory}},  {\em arXiv e-prints} (July, 2020) arXiv:2007.08991,
  [\href{http://arxiv.org/abs/2007.08991}{{\tt arXiv:2007.08991}}].

\bibitem{Seo_2007}
H.~Seo and D.~J. Eisenstein, {\it Improved forecasts for the baryon acoustic
  oscillations and cosmological distance scale},  {\em The Astrophysical
  Journal} {\bf 665} (Aug, 2007) 14–24.

\bibitem{Shapley03}
A.~E. {Shapley}, C.~C. {Steidel}, M.~{Pettini}, and K.~L. {Adelberger}, {\it
  {Rest-Frame Ultraviolet Spectra of z\raisebox{-0.5ex}\textasciitilde3 Lyman
  Break Galaxies}},  {\em \apj} {\bf 588} (May, 2003) 65--89,
  [\href{http://xxx.lanl.gov/abs/astro-ph/0301230}{{\tt astro-ph/0301230}}].

\bibitem{2018MNRAS.478L..60V}
A.~{Verhamme}, T.~{Garel}, E.~{Ventou}, et~al., {\it {Recovering the systemic
  redshift of galaxies from their Lyman alpha line profile}},  {\em \mnras}
  {\bf 478} (July, 2018) L60--L65, [\href{http://arxiv.org/abs/1804.01883}{{\tt
  arXiv:1804.01883}}].

\bibitem{Obuljen2016}
A.~Obuljen, F.~Villaescusa-Navarro, E.~Castorina, and M.~Viel, {\it {Baryon
  Acoustic Oscillations reconstruction with pixels}},  {\em JCAP} {\bf 09}
  (2017) 012, [\href{http://arxiv.org/abs/1610.05768}{{\tt arXiv:1610.05768}}].

\bibitem{Philcox2020}
O.~H.~E. Philcox, M.~M. Ivanov, M.~Simonovi\'c, and M.~Zaldarriaga, {\it
  {Combining Full-Shape and BAO Analyses of Galaxy Power Spectra: A
  1.6\textbackslash{}\% CMB-independent constraint on H$_0$}},  {\em JCAP} {\bf
  05} (2020) 032, [\href{http://arxiv.org/abs/2002.04035}{{\tt
  arXiv:2002.04035}}].

\bibitem{Modi:2019hnu}
C.~Modi, M.~White, A.~Slosar, and E.~Castorina, {\it {Reconstructing
  large-scale structure with neutral hydrogen surveys}},  {\em JCAP} {\bf 11}
  (2019) 023, [\href{http://arxiv.org/abs/1907.02330}{{\tt arXiv:1907.02330}}].

\bibitem{Modi:2021okf}
C.~Modi, M.~White, E.~Castorina, and A.~Slosar, {\it {Mind the gap: the power
  of combining photometric surveys with intensity mapping}},
  \href{http://arxiv.org/abs/2102.08116}{{\tt arXiv:2102.08116}}.

\bibitem{PDG20}
{\bf Particle Data Group} Collaboration, P.~A. Zyla et~al., {\it {Review of
  Particle Physics}},  {\em PTEP} {\bf 2020} (2020), no.~8 083C01.

\bibitem{planckcollaboration2019planck}
P.~Collaboration, Y.~Akrami, F.~Arroja, et~al., {\it Planck 2018 results. ix.
  constraints on primordial non-gaussianity},  2019.

\bibitem{2014SPIE.9153E..1IE}
T.~{Essinger-Hileman}, A.~{Ali}, M.~{Amiri}, et~al., {\it {CLASS: the cosmology
  large angular scale surveyor}},  in {\em Millimeter, Submillimeter, and
  Far-Infrared Detectors and Instrumentation for Astronomy VII} (W.~S.
  {Holland} and J.~{Zmuidzinas}, eds.), vol.~9153 of {\em Society of
  Photo-Optical Instrumentation Engineers (SPIE) Conference Series}, p.~91531I,
  July, 2014.
\newblock \href{http://arxiv.org/abs/1408.4788}{{\tt arXiv:1408.4788}}.

\bibitem{Ferraro:2018izc}
S.~Ferraro and K.~M. Smith, {\it {Characterizing the epoch of reionization with
  the small-scale CMB: Constraints on the optical depth and duration}},  {\em
  Phys. Rev. D} {\bf 98} (2018), no.~12 123519,
  [\href{http://arxiv.org/abs/1803.07036}{{\tt arXiv:1803.07036}}].

\bibitem{Alvarez:2020gvl}
M.~A. Alvarez, S.~Ferraro, J.~C. Hill, et~al., {\it {Mitigating the optical
  depth degeneracy using the kinematic Sunyaev-Zel'dovich effect with CMB-S4}},
   {\em Phys. Rev. D} {\bf 103} (2021), no.~6 063518,
  [\href{http://arxiv.org/abs/2006.06594}{{\tt arXiv:2006.06594}}].

\bibitem{Meerburg:2017lfh}
P.~D. Meerburg, J.~Meyers, K.~M. Smith, and A.~van Engelen, {\it
  {Reconstructing CMB fluctuations and the mean reionization optical depth}},
  {\em Phys. Rev. D} {\bf 95} (2017), no.~12 123538,
  [\href{http://arxiv.org/abs/1701.06992}{{\tt arXiv:1701.06992}}].

\bibitem{Liu:2015txa}
A.~Liu, J.~R. Pritchard, R.~Allison, et~al., {\it {Eliminating the optical
  depth nuisance from the CMB with 21 cm cosmology}},  {\em Phys. Rev. D} {\bf
  93} (2016), no.~4 043013, [\href{http://arxiv.org/abs/1509.08463}{{\tt
  arXiv:1509.08463}}].

\bibitem{billings2021extracting}
T.~S. Billings, P.~L. Plante, and J.~E. Aguirre, {\it Extracting the optical
  depth to reionization $\tau$ from 21 cm data using machine learning
  techniques},  2021.

\bibitem{Mangano_2005}
G.~Mangano, G.~Miele, S.~Pastor, et~al., {\it Relic neutrino decoupling
  including flavour oscillations},  {\em Nuclear Physics B} {\bf 729} (Nov,
  2005) 221–234.

\bibitem{Bashinsky:2003tk}
S.~Bashinsky and U.~Seljak, {\it {Neutrino perturbations in CMB anisotropy and
  matter clustering}},  {\em Phys. Rev. D} {\bf 69} (2004) 083002,
  [\href{http://xxx.lanl.gov/abs/astro-ph/0310198}{{\tt astro-ph/0310198}}].

\bibitem{Nico:2004ie}
J.~S. Nico et~al., {\it {Measurement of the neutron lifetime by counting
  trapped protons in a cold neutron beam}},  {\em Phys. Rev. C} {\bf 71} (2005)
  055502, [\href{http://xxx.lanl.gov/abs/nucl-ex/0411041}{{\tt
  nucl-ex/0411041}}].

\bibitem{Pattie:2017vsj}
R.~W. Pattie, Jr. et~al., {\it {Measurement of the neutron lifetime using a
  magneto-gravitational trap and in situ detection}},  {\em Science} {\bf 360}
  (2018), no.~6389 627--632, [\href{http://arxiv.org/abs/1707.01817}{{\tt
  arXiv:1707.01817}}].

\bibitem{Baumann:2015rya}
D.~Baumann, D.~Green, J.~Meyers, and B.~Wallisch, {\it {Phases of New Physics
  in the CMB}},  {\em JCAP} {\bf 01} (2016) 007,
  [\href{http://arxiv.org/abs/1508.06342}{{\tt arXiv:1508.06342}}].

\bibitem{Baumann:2017lmt}
D.~Baumann, D.~Green, and M.~Zaldarriaga, {\it {Phases of New Physics in the
  BAO Spectrum}},  {\em JCAP} {\bf 11} (2017) 007,
  [\href{http://arxiv.org/abs/1703.00894}{{\tt arXiv:1703.00894}}].

\bibitem{Simha:2008zj}
V.~Simha and G.~Steigman, {\it {Constraining The Early-Universe Baryon Density
  And Expansion Rate}},  {\em JCAP} {\bf 06} (2008) 016,
  [\href{http://arxiv.org/abs/0803.3465}{{\tt arXiv:0803.3465}}].

\bibitem{Kneller:2004jz}
J.~P. Kneller and G.~Steigman, {\it {BBN for pedestrians}},  {\em New J. Phys.}
  {\bf 6} (2004) 117, [\href{http://xxx.lanl.gov/abs/astro-ph/0406320}{{\tt
  astro-ph/0406320}}].

\bibitem{Hou:2011ec}
Z.~Hou, R.~Keisler, L.~Knox, et~al., {\it {How Massless Neutrinos Affect the
  Cosmic Microwave Background Damping Tail}},  {\em Phys. Rev. D} {\bf 87}
  (2013) 083008, [\href{http://arxiv.org/abs/1104.2333}{{\tt
  arXiv:1104.2333}}].

\bibitem{Maldacena_2003}
J.~Maldacena, {\it Non-gaussian features of primordial fluctuations in single
  field inflationary models},  {\em Journal of High Energy Physics} {\bf 2003}
  (May, 2003) 013–013.

\bibitem{Creminelli04}
P.~Creminelli and M.~Zaldarriaga, {\it {Single field consistency relation for
  the 3-point function}},  {\em JCAP} {\bf 10} (2004) 006,
  [\href{http://xxx.lanl.gov/abs/astro-ph/0407059}{{\tt astro-ph/0407059}}].

\bibitem{Schmidt_2010}
F.~Schmidt and M.~Kamionkowski, {\it Halo clustering with nonlocal
  non-gaussianity},  {\em Physical Review D} {\bf 82} (Nov, 2010).

\bibitem{Castorina2019}
E.~Castorina et~al., {\it {Redshift-weighted constraints on primordial
  non-Gaussianity from the clustering of the eBOSS DR14 quasars in Fourier
  space}},  {\em JCAP} {\bf 09} (2019) 010,
  [\href{http://arxiv.org/abs/1904.08859}{{\tt arXiv:1904.08859}}].

\bibitem{Slosar:2008}
A.~Slosar, C.~Hirata, U.~Seljak, et~al., {\it {Constraints on local primordial
  non-Gaussianity from large scale structure}},  {\em JCAP} {\bf 08} (2008)
  031, [\href{http://arxiv.org/abs/0805.3580}{{\tt arXiv:0805.3580}}].

\bibitem{Biagetti:2019bnp}
M.~Biagetti, {\it {The Hunt for Primordial Interactions in the Large Scale
  Structures of the Universe}},  {\em Galaxies} {\bf 7} (2019), no.~3 71,
  [\href{http://arxiv.org/abs/1906.12244}{{\tt arXiv:1906.12244}}].

\bibitem{Biagetti2016}
M.~Biagetti, T.~Lazeyras, T.~Baldauf, et~al., {\it {Verifying the consistency
  relation for the scale-dependent bias from local primordial
  non-Gaussianity}},  {\em Mon. Not. Roy. Astron. Soc.} {\bf 468} (2017), no.~3
  3277--3288, [\href{http://arxiv.org/abs/1611.04901}{{\tt arXiv:1611.04901}}].

\bibitem{Barreira2020}
A.~Barreira, {\it {On the impact of galaxy bias uncertainties on primordial
  non-Gaussianity constraints}},  {\em JCAP} {\bf 12} (2020) 031,
  [\href{http://arxiv.org/abs/2009.06622}{{\tt arXiv:2009.06622}}].

\bibitem{MoradinezhadDizgah2020}
A.~Moradinezhad~Dizgah, M.~Biagetti, E.~Sefusatti, et~al., {\it {Primordial
  Non-Gaussianity from Biased Tracers: Likelihood Analysis of Real-Space Power
  Spectrum and Bispectrum}},  \href{http://arxiv.org/abs/2010.14523}{{\tt
  arXiv:2010.14523}}.

\bibitem{Cunnington:2020wdu}
S.~Cunnington, S.~Camera, and A.~Pourtsidou, {\it {The degeneracy between
  primordial non-Gaussianity and foregrounds in 21 cm intensity mapping
  experiments}},  {\em Mon. Not. Roy. Astron. Soc.} {\bf 499} (2020), no.~3
  4054--4067, [\href{http://arxiv.org/abs/2007.12126}{{\tt arXiv:2007.12126}}].

\bibitem{dePutter:2018jqk}
R.~de~Putter, {\it {Primordial physics from large-scale structure beyond the
  power spectrum}},  \href{http://arxiv.org/abs/1802.06762}{{\tt
  arXiv:1802.06762}}.

\bibitem{Karagiannis:2018jdt}
D.~Karagiannis, A.~Lazanu, M.~Liguori, et~al., {\it {Constraining primordial
  non-Gaussianity with bispectrum and power spectrum from upcoming optical and
  radio surveys}},  {\em Mon. Not. Roy. Astron. Soc.} {\bf 478} (2018), no.~1
  1341--1376, [\href{http://arxiv.org/abs/1801.09280}{{\tt arXiv:1801.09280}}].

\bibitem{Gleyzes_2017}
J.~Gleyzes, R.~de~Putter, D.~Green, and O.~Doré, {\it Biasing and the search
  for primordial non-gaussianity beyond the local type},  {\em Journal of
  Cosmology and Astroparticle Physics} {\bf 2017} (Apr, 2017) 002–002.

\bibitem{Kitanidis:2019rzi}
E.~Kitanidis et~al., {\it {Imaging Systematics and Clustering of DESI Main
  Targets}},  {\em Mon. Not. Roy. Astron. Soc.} {\bf 496} (2020), no.~2
  2262--2291, [\href{http://arxiv.org/abs/1911.05714}{{\tt arXiv:1911.05714}}].

\bibitem{2013PASP..125..705P}
A.~R. {Pullen} and C.~M. {Hirata}, {\it {Systematic Effects in Large-Scale
  Angular Power Spectra of Photometric Quasars and Implications for
  Constraining Primordial Non-Gaussianity}},  {\em \pasp} {\bf 125} (June,
  2013) 705, [\href{http://arxiv.org/abs/1212.4500}{{\tt arXiv:1212.4500}}].

\bibitem{Leistedt:2014zqa}
B.~Leistedt, H.~V. Peiris, and N.~Roth, {\it {Constraints on Primordial
  Non-Gaussianity from 800 000 Photometric Quasars}},  {\em Phys. Rev. Lett.}
  {\bf 113} (2014), no.~22 221301, [\href{http://arxiv.org/abs/1405.4315}{{\tt
  arXiv:1405.4315}}].

\bibitem{Agarwal:2013ajb}
N.~Agarwal et~al., {\it {Characterizing unknown systematics in large scale
  structure surveys}},  {\em JCAP} {\bf 04} (2014) 007,
  [\href{http://arxiv.org/abs/1309.2954}{{\tt arXiv:1309.2954}}].

\bibitem{2013MNRAS.428.1116R}
A.~J. {Ross}, W.~J. {Percival}, A.~{Carnero}, et~al., {\it {The clustering of
  galaxies in the SDSS-III DR9 Baryon Oscillation Spectroscopic Survey:
  constraints on primordial non-Gaussianity}},  {\em \mnras} {\bf 428} (Jan.,
  2013) 1116--1127, [\href{http://arxiv.org/abs/1208.1491}{{\tt
  arXiv:1208.1491}}].

\bibitem{Weaverdyck:2020mff}
N.~Weaverdyck and D.~Huterer, {\it {Mitigating contamination in LSS surveys: a
  comparison of methods}},  {\em Mon. Not. Roy. Astron. Soc.} {\bf 503} (2021),
  no.~4 5061--5084, [\href{http://arxiv.org/abs/2007.14499}{{\tt
  arXiv:2007.14499}}].

\bibitem{2013MNRAS.432.2945H}
D.~{Huterer}, C.~E. {Cunha}, and W.~{Fang}, {\it {Calibration errors unleashed:
  effects on cosmological parameters and requirements for large-scale structure
  surveys}},  {\em \mnras} {\bf 432} (July, 2013) 2945--2961,
  [\href{http://arxiv.org/abs/1211.1015}{{\tt arXiv:1211.1015}}].

\bibitem{Schmittfull18}
M.~{Schmittfull} and U.~{Seljak}, {\it {Parameter constraints from
  cross-correlation of CMB lensing with galaxy clustering}},  {\em \prd} {\bf
  97} (June, 2018) 123540, [\href{http://arxiv.org/abs/1710.09465}{{\tt
  arXiv:1710.09465}}].

\bibitem{Baumann09}
D.~{Baumann}, {\it {TASI Lectures on Inflation}},  {\em arXiv e-prints} (July,
  2009) arXiv:0907.5424, [\href{http://arxiv.org/abs/0907.5424}{{\tt
  arXiv:0907.5424}}].

\bibitem{Adams_2001}
J.~Adams, B.~Cresswell, and R.~Easther, {\it Inflationary perturbations from a
  potential with a step},  {\em Physical Review D} {\bf 64} (Nov, 2001).

\bibitem{Ballardini:2016hpi}
M.~Ballardini, F.~Finelli, C.~Fedeli, and L.~Moscardini, {\it {Probing
  primordial features with future galaxy surveys}},  {\em JCAP} {\bf 10} (2016)
  041, [\href{http://arxiv.org/abs/1606.03747}{{\tt arXiv:1606.03747}}].
  [Erratum: JCAP 04, E01 (2018)].

\bibitem{Beutler_2019}
F.~Beutler, M.~Biagetti, D.~Green, et~al., {\it Primordial features from linear
  to nonlinear scales},  {\em Physical Review Research} {\bf 1} (Dec, 2019).

\bibitem{Ballardini:2019tuc}
M.~Ballardini, R.~Murgia, M.~Baldi, et~al., {\it {Non-linear damping of
  superimposed primordial oscillations on the matter power spectrum in galaxy
  surveys}},  {\em JCAP} {\bf 04} (2020), no.~04 030,
  [\href{http://arxiv.org/abs/1912.12499}{{\tt arXiv:1912.12499}}].

\bibitem{EricLinderRise}
E.~V. Linder, {\it {The Rise of Dark Energy}},
  \href{http://arxiv.org/abs/2106.09581}{{\tt arXiv:2106.09581}}.

\bibitem{Efstathiou20}
G.~{Efstathiou}, {\it {A Lockdown Perspective on the Hubble Tension (with
  comments from the SH0ES team)}},  {\em arXiv e-prints} (July, 2020)
  arXiv:2007.10716, [\href{http://arxiv.org/abs/2007.10716}{{\tt
  arXiv:2007.10716}}].

\bibitem{Poulin_2019}
V.~Poulin, T.~L. Smith, T.~Karwal, and M.~Kamionkowski, {\it Early dark energy
  can resolve the hubble tension},  {\em Physical Review Letters} {\bf 122}
  (Jun, 2019).

\bibitem{smith2020early}
T.~L. Smith, V.~Poulin, J.~L. Bernal, et~al., {\it Early dark energy is not
  excluded by current large-scale structure data},  2020.

\bibitem{Clifton_2012}
T.~Clifton, P.~G. Ferreira, A.~Padilla, and C.~Skordis, {\it Modified gravity
  and cosmology},  {\em Physics Reports} {\bf 513} (Mar, 2012) 1–189.

\bibitem{Will14}
C.~M. {Will}, {\it {The Confrontation between General Relativity and
  Experiment}},  {\em Living Reviews in Relativity} {\bf 17} (Dec., 2014) 4,
  [\href{http://arxiv.org/abs/1403.7377}{{\tt arXiv:1403.7377}}].

\bibitem{Joyce15}
A.~{Joyce}, B.~{Jain}, J.~{Khoury}, and M.~{Trodden}, {\it {Beyond the
  cosmological standard model}},  {\em \physrep} {\bf 568} (Mar, 2015) 1--98,
  [\href{http://arxiv.org/abs/1407.0059}{{\tt arXiv:1407.0059}}].

\bibitem{White15}
M.~{White}, B.~{Reid}, C.-H. {Chuang}, et~al., {\it {Tests of redshift-space
  distortions models in configuration space for the analysis of the BOSS final
  data release}},  {\em \mnras} {\bf 447} (Feb, 2015) 234--245,
  [\href{http://arxiv.org/abs/1408.5435}{{\tt arXiv:1408.5435}}].

\bibitem{Bha96}
S.~{Bharadwaj}, {\it {The Evolution of Correlation Functions in the Zeldovich
  Approximation and Its Implications for the Validity of Perturbation Theory}},
   {\em \apj} {\bf 472} (Nov., 1996) 1--+,
  [\href{http://arxiv.org/abs/astro-ph/9}{{\tt astro-ph/9}}].

\bibitem{ESW07}
D.~J. {Eisenstein}, H.-J. {Seo}, and M.~{White}, {\it {On the Robustness of the
  Acoustic Scale in the Low-Redshift Clustering of Matter}},  {\em \apj} {\bf
  664} (Aug., 2007) 660--674,
  [\href{http://xxx.lanl.gov/abs/astro-ph/0604361}{{\tt astro-ph/0604361}}].

\bibitem{Mat08a}
T.~{Matsubara}, {\it {Resumming cosmological perturbations via the Lagrangian
  picture: One-loop results in real space and in redshift space}},  {\em \prd}
  {\bf 77} (Mar., 2008) 063530, [\href{http://arxiv.org/abs/0711.2521}{{\tt
  arXiv:0711.2521}}].

\bibitem{Mat08b}
T.~{Matsubara}, {\it {Nonlinear perturbation theory with halo bias and
  redshift-space distortions via the Lagrangian picture}},  {\em \prd} {\bf 78}
  (Oct., 2008) 083519, [\href{http://arxiv.org/abs/0807.1733}{{\tt
  arXiv:0807.1733}}].

\bibitem{Crocce08}
M.~{Crocce} and R.~{Scoccimarro}, {\it {Nonlinear evolution of baryon acoustic
  oscillations}},  {\em \prd} {\bf 77} (Jan, 2008) 023533,
  [\href{http://arxiv.org/abs/0704.2783}{{\tt arXiv:0704.2783}}].

\bibitem{PWC09}
N.~{Padmanabhan}, M.~{White}, and J.~D. {Cohn}, {\it {Reconstructing baryon
  oscillations: A Lagrangian theory perspective}},  {\em \prd} {\bf 79} (Mar.,
  2009) 063523, [\href{http://arxiv.org/abs/0812.2905}{{\tt arXiv:0812.2905}}].

\bibitem{Noh09}
Y.~{Noh}, M.~{White}, and N.~{Padmanabhan}, {\it {Reconstructing baryon
  oscillations}},  {\em \prd} {\bf 80} (Dec., 2009) 123501,
  [\href{http://arxiv.org/abs/0909.1802}{{\tt arXiv:0909.1802}}].

\bibitem{CLPT}
J.~{Carlson}, B.~{Reid}, and M.~{White}, {\it {Convolution Lagrangian
  perturbation theory for biased tracers}},  {\em \mnras} {\bf 429} (Feb.,
  2013) 1674--1685, [\href{http://arxiv.org/abs/1209.0780}{{\tt
  arXiv:1209.0780}}].

\bibitem{TasZal12}
S.~{Tassev} and M.~{Zaldarriaga}, {\it {Towards an optimal reconstruction of
  baryon oscillations}},  {\em \jcap} {\bf 10} (Oct., 2012) 006,
  [\href{http://arxiv.org/abs/1203.6066}{{\tt arXiv:1203.6066}}].

\bibitem{McCSza12}
N.~{McCullagh} and A.~S. {Szalay}, {\it {Nonlinear Behavior of Baryon Acoustic
  Oscillations from the Zel'dovich Approximation Using a Non-Fourier
  Perturbation Approach}},  {\em \apj} {\bf 752} (June, 2012) 21,
  [\href{http://arxiv.org/abs/1202.1306}{{\tt arXiv:1202.1306}}].

\bibitem{White14}
M.~{White}, {\it {The Zel'dovich approximation}},  {\em \mnras} {\bf 439}
  (Apr., 2014) 3630--3640, [\href{http://arxiv.org/abs/1401.5466}{{\tt
  arXiv:1401.5466}}].

\bibitem{SenZal15}
L.~{Senatore} and M.~{Zaldarriaga}, {\it {The IR-resummed Effective Field
  Theory of Large Scale Structures}},  {\em \jcap} {\bf 2} (Feb., 2015) 13,
  [\href{http://arxiv.org/abs/1404.5954}{{\tt arXiv:1404.5954}}].

\bibitem{Schmittfull15}
M.~{Schmittfull}, Y.~{Feng}, F.~{Beutler}, et~al., {\it {Eulerian BAO
  reconstructions and N -point statistics}},  {\em \prd} {\bf 92} (Dec., 2015)
  123522, [\href{http://arxiv.org/abs/1508.06972}{{\tt arXiv:1508.06972}}].

\bibitem{Baldauf15}
T.~{Baldauf}, M.~{Mirbabayi}, M.~{Simonovi{\'c}}, and M.~{Zaldarriaga}, {\it
  {Equivalence principle and the baryon acoustic peak}},  {\em \prd} {\bf 92}
  (Aug., 2015) 043514, [\href{http://arxiv.org/abs/1504.04366}{{\tt
  arXiv:1504.04366}}].

\bibitem{Vlah16}
Z.~{Vlah}, U.~{Seljak}, M.~{Yat Chu}, and Y.~{Feng}, {\it {Perturbation theory,
  effective field theory, and oscillations in the power spectrum}},  {\em
  \jcap} {\bf 2016} (Mar, 2016) 057,
  [\href{http://arxiv.org/abs/1509.02120}{{\tt arXiv:1509.02120}}].

\bibitem{McQuinn16}
M.~{McQuinn} and M.~{White}, {\it {Cosmological perturbation theory in 1+1
  dimensions}},  {\em \jcap} {\bf 1} (Jan., 2016) 043,
  [\href{http://arxiv.org/abs/1502.07389}{{\tt arXiv:1502.07389}}].

\bibitem{Blas16}
D.~{Blas}, M.~{Garny}, M.~M. {Ivanov}, and S.~{Sibiryakov}, {\it {Time-sliced
  perturbation theory II: baryon acoustic oscillations and infrared
  resummation}},  {\em \jcap} {\bf 2016} (July, 2016) 028,
  [\href{http://arxiv.org/abs/1605.02149}{{\tt arXiv:1605.02149}}].

\bibitem{Seo16}
H.-J. {Seo}, F.~{Beutler}, A.~J. {Ross}, and S.~{Saito}, {\it {Modeling the
  reconstructed BAO in Fourier space}},  {\em \mnras} {\bf 460} (Aug., 2016)
  2453--2471, [\href{http://arxiv.org/abs/1511.00663}{{\tt arXiv:1511.00663}}].

\end{thebibliography}\endgroup
\end{document}